\documentclass[onecolumn, draftcls]{IEEEtran}

\usepackage{ifpdf}

\usepackage[noadjust]{cite}

\ifCLASSINFOpdf
  \usepackage[pdftex]{graphicx}
\else
  \usepackage[dvips]{graphicx}
\fi

\usepackage[cmex10]{amsmath}
\usepackage{array}

\ifCLASSOPTIONcompsoc
  \usepackage[caption=false,font=normalsize,labelfont=sf,textfont=sf]{subfig}
\else
  \usepackage[caption=false,font=footnotesize]{subfig}
\fi

\usepackage{fixltx2e}

\usepackage{url}

\usepackage[utf8]{inputenc} 
\usepackage[T1]{fontenc}
\usepackage{ifthen}

\usepackage{mathtools}
\usepackage{amssymb}
\usepackage{relsize}
\usepackage{bbm}
\usepackage{xfrac}
\usepackage{amsthm}
\theoremstyle{plain}
\newtheorem{definition}{Definition}
\newtheorem{lemma}{Lemma}

\newtheorem{theorem}{Theorem}
\newtheorem{corollary}{Corollary}

\newtheorem{example}{Example}

\usepackage[varg, cmbraces]{newtxmath}

\usepackage{xcolor}
\definecolor{burgundy}{rgb}{0.545098,0,0}
\definecolor{navyblue}{rgb}{0.0, 0.0, 0.5}
\definecolor{leafgreen}{rgb}{0.290196, 0.470588, 0.0}
\definecolor{bluegreen}{rgb}{0, 0.470588, 0.415686}
\definecolor{zuhl}{rgb}{0.1875, 0.26171875, 0.46484375}
\definecolor{orange}{rgb}{1, 0.6470588235, 0}
\definecolor{red}{rgb}{1, 0, 0}

\usepackage{overpic}
\usepackage{pict2e}
\usepackage{booktabs}

\newcommand{\bvec}[1]{\boldsymbol{#1}}

\newcommand{\lcm}{\operatorname{lcm}}

\newcommand{\tabref}[1]{Table~\ref{#1}}
\newcommand{\figref}[1]{Fig.~\ref{#1}}

\newcommand{\lemref}[1]{Lemma~\ref{#1}}

\newcommand{\thref}[1]{Theorem~\ref{#1}}
\newcommand{\defref}[1]{Definition~\ref{#1}}
\newcommand{\corref}[1]{Corollary~\ref{#1}}
\newcommand{\sectref}[1]{Section~\ref{#1}}

\newcommand{\exref}[1]{Example~\ref{#1}}

\allowdisplaybreaks[3]

\usepackage{array}
\usepackage[linesnumbered, algoruled, boxed, lined, vlined]{algorithm2e}
\SetKwRepeat{Do}{do}{while}

\usepackage{hyperref}
\hypersetup{
    colorlinks,
    linkcolor={red!60!black},
    citecolor={blue!60!black},
    urlcolor={blue!50!black}
}
\usepackage{microtype}

\begin{document}

\title{Asymptotic Distribution of \\ Multilevel Channel Polarization for \\ a Certain Class of Erasure Channels}

\IEEEoverridecommandlockouts

\author{%
\IEEEauthorblockN{%
Yuta~Sakai,~\IEEEmembership{Student Member,~IEEE,}
Ken-ichi~Iwata,~\IEEEmembership{Member,~IEEE,} \\
and Hiroshi~Fujisaki,~\IEEEmembership{Member,~IEEE,}%
\thanks{This work was supported by JSPS KAKENHI Grant Numbers 17K06422 and 17J11247.}%
\thanks{Y.~Sakai and K.~Iwata are with Graduate School of Engineering,
University of Fukui,
Japan,
Email: \{y-sakai, k-iwata\}@u-fukui.ac.jp}%
\thanks{H.~Fujisaki is with Graduate School of Natural Science and Technology,
Kanazawa University,
Japan,
Email: fujisaki@ec.t.kanazawa-u.ac.jp}
}%
}

\maketitle

\begin{abstract}
This study examines multilevel channel polarization for a certain class of erasure channels that the input alphabet size is an arbitrary composite number.
We derive asymptotic proportions of partially noiseless channels for such a class.
The results of this study are proved by an argument of convergent sequences, inspired by Alsan and Telatar's simple proof of polarization \cite{alsan3}, and without martingale convergence theorems for polarization process.
\end{abstract}

\IEEEpeerreviewmaketitle

\section{Introduction}

Ar{\i}kan \cite{arikan} proposed binary polar codes as a class of provable symmetric capacity achieving codes with deterministic constructions and low encoding/decoding complexity for binary-input discrete memoryless channels (DMCs).

In non-binary polar codes, there are two types of channel polarization: strong polarization \cite{mori, sasoglu} and multilevel polarization \cite{ergodic1, ergodic2, quasigroup, park, sahebi}.
Strong polarization asymptotically makes similar extremal channels to binary cases, i.e., either noiseless or pure noisy.
On the other hand, multilevel polarization allows to converse several types of \emph{partially noiseless} channels.
It was independently shown in \cite{mori, sasoglu, ergodic1, ergodic2, quasigroup, park, sahebi} that both strong and multilevel channel polarization can achieve the symmetric capacity by showing rate of polarization for the Bhattacharyya parameters.
Although the asymptotic distributions of strong polarization are fully and simply characterized by the symmetric capacity, the asymptotic distribution of multilevel channel polarization is, however, still an open problem.

Recently, the authors \cite{itw2016} proposed a certain class of erasure channels together with the recursive formulas of the polar transforms for such a class.
In addition, we \cite{itw2016} also clarified the asymptotic distribution of multilevel channel polarization for such a class when the input alphabet size $q$ is a prime power.
In this paper, we examine further the asymptotic distribution for general composite numbers $q$.

\section{Preliminaries}

\subsection{Basic Notations of DMCs and Polar Transforms}
\label{sect:notations}

In this study, discrete memoryless channels are given as follows:
The input alphabet of a DMC is denoted by a finite set $\mathcal{X}$ having two or more elements;
and the output alphabet of a DMC is denoted by a nonempty and countable set $\mathcal{Y}$.
The transition probability of a DMC from an input symbol $x \in \mathcal{X}$ to an output symbol $y \in \mathcal{Y}$ is denoted by $W(y \mid x)$.
Let $W : \mathcal{X} \to \mathcal{Y}$, or simply $W$, be a shorthand for such a DMC.
We shall denote by $q = |\mathcal{X}|$ the input alphabet size of a DMC $W$, where $|\cdot|$ denotes the cardinality of a finite set.
The symmetric capacity of a DMC $W$ is given by
\begin{align}
I( W )
\coloneqq
\sum_{y \in \mathcal{Y}} \sum_{x \in \mathcal{X}} \frac{ 1 }{ q } W(y \mid x) \log \frac{ W(y \mid x) }{ \sum_{x^{\prime} \in \mathcal{X}} (1/q) W(y \mid x^{\prime}) } ,
\end{align}
where the base of logarithms is $q$.

We now introduce non-binary polar transforms with a quasigroup operation%
\footnote{%
A quasigroup is the pair $(\mathcal{Q}, \ast)$ of a nonempty set $\mathcal{Q}$ and a closed binary operation $\ast$ on $\mathcal{Q}$ satisfying the divisibility: for any $a, b \in \mathcal{Q}$, there exist unique $c, d \in \mathcal{Q}$ such that $a = b \ast c$ and $a = d \ast b$.%
}
$\ast$ on the input alphabet $\mathcal{X}$.
For a given DMC $W : \mathcal{X} \to \mathcal{Y}$, the polar transform makes two synthetic channels%
\footnote{%
The terms \emph{worse} and \emph{better} come from the inequalities $I(W^{-}) \le I(W) \le I(W^{+})$ under arbitrary quasigroup operation $\ast$ (cf. \cite{quasigroup}).
}%
: a worse channel $W^{-} : \mathcal{X} \to \mathcal{Y}^{2}$ defined by
\begin{align}
W^{-}(y_{1}, y_{2} \mid u_{1})
\coloneqq
\sum_{u_{2}^{\prime} \in \mathcal{X}} \frac{ 1 }{ q } \, W(y_{1} \mid u_{1} \ast u_{2}^{\prime}) \, W(y_{2} \mid u_{2}^{\prime}) ;
\label{def:minus}
\end{align}
and a better channel $W^{+} : \mathcal{X} \to \mathcal{Y}^{2} \times \mathcal{X}$ defined by
\begin{align}
W^{+}(y_{1}, y_{2}, u_{1} \mid u_{2})
\coloneqq
\frac{ 1 }{ q } \, W(y_{1} \mid u_{1} \ast u_{2}) \, W(y_{2} \mid u_{2}) .
\label{def:plus}
\end{align}
After the $n$-step polar transforms, $n \in \mathbb{N}$, the synthetic channel $W^{\bvec{s}} : \mathcal{X} \to \mathcal{Y}^{2^{n}} \times \mathcal{X}^{w(\bvec{s})}$ is created by
\begin{align}
W^{\bvec{s}}
& \coloneqq
( \cdots (W^{s_{1}})^{s_{2}} \cdots )^{s_{n}}
\label{def:n-step}
\end{align}
for each $\bvec{s} = s_{1}s_{2} \cdots s_{n} \in \{ -, + \}^{n}$, where the function%
\footnote{%
The set $\mathbb{N}_{0} \coloneqq \mathbb{N} \cup \{ 0 \}$ consists of all nonnegative integers.
}
$w : \{ -, + \}^{\ast} \to \mathbb{N}_{0}$ is recursively defined by%
\footnote{%
For example, $w(+, -, +) = 2 \, w(+, -) + 1 = 2 \cdot 2 \, w(+) + 1 = 2 \cdot 2 \cdot 1 + 1 = 5$.
As $w( \cdot )$ seems binary expansions by replacing $(-, +)$ with $(0, 1)$, the map $w : \{ -, + \}^{n} \to \{ 0, 1, \dots, 2^{n}-1 \}$ is bijective.
}
\begin{align}
w( s_{1}, \dots, s_{n} )
\coloneqq
\begin{cases}
2 \, w( s_{1}, \dots, s_{n-1} )
& \text{if} \ n \ge 1 \ \mathrm{and} \ s_{n} = - ,
\\
2 \, w( s_{1}, \dots, s_{n-1} ) + 1
& \text{if} \ n \ge 1 \ \mathrm{and} \ s_{n} = + ,
\\
0
& \mathrm{otherwise} ,
\end{cases}
\end{align}
and
$\{ -, + \}^{\ast} \coloneqq \{ \epsilon, -, +, --, -+, +-, ++, \dots \}$ denotes the set of $\{ -, + \}$-valued finite-length sequences containing the empty sequence $\epsilon$.
Namely, the output alphabet size $|\mathcal{Y}^{2^{n}} \times \mathcal{X}^{w(\bvec{s})}|$ of the synthetic channel $W^{\bvec{s}}$ grows double-exponentially as the number $n$ of polar transforms increases.
Difficulties of constructing and analyzing polar codes are due to this issue.

\subsection{Strong Polarization}
\label{sect:strong}

When the input alphabet size $q$ is a prime number, \c{S}a\c{s}o\u{g}lu et~al. \cite{sasoglu_etal} showed that for any $q$-ary input DMC $W: \mathcal{X} \to \mathcal{Y}$ and any fixed $\delta \in (0, 1)$, the two equalities
\begin{align}
\lim_{n \to \infty} \frac{ 1 }{ 2^{n} } \Big| \Big\{ \bvec{s} \in \{ -, + \}^{n} \ \Big| \ I( W^{\bvec{s}} ) > 1 - \delta \Big\} \Big|
& =
I( W ) ,
\label{eq:strong1} \\
\lim_{n \to \infty} \frac{ 1 }{ 2^{n} } \Big| \Big\{ \bvec{s} \in \{ -, + \}^{n} \ \Big| \ I( W^{\bvec{s}} ) < \delta \Big\} \Big|
& =
1 - I( W )
\label{eq:strong2}
\end{align}
holds under the polar transforms \eqref{def:minus} and \eqref{def:plus} in which $(\mathcal{X}, \ast)$ forms a cyclic group $(\mathbb{Z}/q\mathbb{Z}, +)$.
The left-hand sides of \eqref{eq:strong1} and \eqref{eq:strong2} are the limiting proportions of almost noiseless and almost useless synthetic channels, respectively.
Moreover, Equations \eqref{eq:strong1} and \eqref{eq:strong2} imply that the limiting proportion of \emph{intermediate} synthetic channels is zero, i.e.,
\begin{align}
\lim_{n \to \infty} \frac{ 1 }{ 2^{n} } \Big| \Big\{ \bvec{s} \in \{ -, + \}^{n} \ \Big| \ \delta \le I( W^{\bvec{s}} ) \le 1 - \delta \Big\} \Big|
= 0
\label{eq:strong}
\end{align}
for every fixed $\delta > 0$.
In this paper, we call phenomena of \eqref{eq:strong} are the \emph{strong polarization}.
Moreover, for any $q \ge 2$ which is not only a prime number but also a composite number, \c{S}a\c{s}o\u{g}lu \cite{sasoglu} showed a sufficient condition of the strong polarization \eqref{eq:strong} for quasigroup operations%
\footnote{%
\c{S}a\c{s}o\u{g}lu said such a quasigroup operation to be \emph{polarizing}.%
}
$\ast$ used in the polar transforms \eqref{def:minus} and  \eqref{def:plus}.
Furthermore, Mori and Tanaka%
\footnote{%
Note that Mori and Tanaka showed the necessary and sufficient condition of the strong polarization \eqref{eq:strong} for more general polar transforms with $l \times l$ kernel, $l \ge 2$, over the finite field $\mathbb{F}_{q}$.
When $l = 2$, their condition can be reduced to that if an operation $\ast$ is defined by $a \ast b = a + \gamma \cdot b$ under the field operations with $\gamma \in \mathbb{F}_{q}^{\times}$, then the strong polarization \eqref{eq:strong} holds for every $q$-ary input DMC if and only if $\gamma$ is a primitive element of $\mathbb{F}_{q}$.
}
\cite{mori} considered the polar transforms \eqref{def:minus} and \eqref{def:plus} with quasigroup operation $\ast$ defined by field operations of $\mathbb{F}_{q}$, and they showed the necessary and sufficient condition of the strong polarization \eqref{eq:strong} under such an operation.
As shown in \eqref{eq:strong1} and \eqref{eq:strong2}, the asymptotic distributions of noiseless \eqref{eq:strong1} and useless channels \eqref{eq:strong2}, respectively, can be always and exactly characterized by only the symmetric capacity $I(W)$ for every DMC $W$%
\footnote{%
This fact comes from the conservation property $[I(W^{-}) + I(W^{+})]/2 = I(W)$ under an arbitrary quasigroup operation $\ast$ (cf. \cite{quasigroup}).
Note that in \cite{ergodic1, ergodic2}, allowing more weaker postulates of a closed binary operation $\ast$ than quasigroups, Nasser showed that the conservation property holds for every $q$-ary input DMC if and only if the map $(a, b) \mapsto (a \ast b, b)$ is bijective.
Such a postulate was said to be \emph{uniformly preserving}.
}.

\subsection{Multilevel Polarization}

Besides \sectref{sect:strong}, when the input alphabet size $q$ is a composite number, there are quasigroups $(\mathcal{X}, \ast)$ employed in the polar transforms \eqref{def:minus} and \eqref{def:plus} such that the strong polarization \eqref{eq:strong} does not hold in general (cf. \cite[Example~1]{sasoglu}).
That is, there is a $q$-ary input DMC $W$ such that the limiting proportion of intermediate synthetic channels $W^{\bvec{s}}$ is positive:
\begin{align}
\liminf_{n \to \infty} \frac{ 1 }{ 2^{n} } \Big| \Big\{ \bvec{s} \in \{ -, + \}^{n} \ \Big| \ \delta \le I( W^{\bvec{s}} ) \le 1 - \delta \Big\} \Big|
> 0
\label{eq:positive_intermediate}
\end{align}
for some $\delta > 0$.
In that cases, another type of polarization called \emph{multilevel polarization}%
\footnote{%
The strong polarization is a special case of the multilevel polarization; hence the former is said to be strong in this paper.%
}
has been examined by some researchers.
The notion of multilevel polarization is introduced later in this subsection.
When $q$ is a power of two, Park and Barg \cite{park} established the multilevel polarization theorem under the polar transforms with cyclic group $(\mathbb{Z}/q\mathbb{Z}, +)$.
Independent of \cite{park}, when $q$ is a prime power, Sahebi and Pradhan \cite{sahebi} examined the multilevel polarization theorem to arbitrary composite numbers $q$ under the polar transforms with arbitrary finite abelian group $(\mathcal{X}, +)$.
Nasser and Telatar \cite{quasigroup} established the multilevel polarization theorem under the polar transforms with arbitrary quasigroup $(\mathcal{X}, \ast)$.
Nasser clarified further the necessary and sufficient condition of multilevel polarization for algebraic structures $(\mathcal{X}, \ast)$ allowing more weaker postulates than quasigroups.

In the context of multilevel polarization, the limiting proportion of intermediate synthetic channels is allowed to be positive, as shown in \eqref{eq:positive_intermediate}.
Then, notions of \emph{partially noiseless} channels are required to achieve the symmetric capacity for arbitrary input DMCs.
Such notions are, however, independently introduced by several authors \cite{park, sahebi, ergodic1, ergodic2, quasigroup} as different types.
In particular, descriptions of multilevel polarization are slightly complicated if $(\mathcal{X}, \ast)$ is a quasigroup \cite{ergodic1, ergodic2, quasigroup}.

As a simple instance of them, following \cite[Section~VI]{quasigroup}, we now introduce a notion of multilevel polarization under the polar transforms with group $\mathcal{X}$ briefly as follows:
Let $N \trianglelefteq G$ be a shorthand for a normal subgroup $N$ of a group $G$.
For a DMC $W : \mathcal{X} \to \mathcal{Y}$ and a normal subgroup $\mathcal{N} \trianglelefteq \mathcal{X}$, the homomorphism channel $W[\mathcal{N}] : \mathcal{X}/\mathcal{N} \to \mathcal{Y}$ is defined by
\begin{align}
W[\mathcal{N}](y \mid a \mathcal{N})
\coloneqq
\frac{ 1 }{ |\mathcal{N}| } \sum_{x \in a \mathcal{N}} W(y \mid x) ,
\label{def:homomorphism}
\end{align}
where the quotient group of $G$ by $N \trianglelefteq G$ is denoted by $G/N$.
Then, Nasser and Telatar \cite[Theorem~6]{quasigroup} showed that%
\footnote{%
In \cite[Theorem~6]{quasigroup}, the rate of polarization for Bhattacharyya parameter is also shown; but we omit it in the paper for simplicity.%
}
\begin{align}
&
\sum_{\mathcal{N} \trianglelefteq \mathcal{X}} \lim_{n \to \infty} \frac{ 1 }{ 2^{n} } \Big| \Big\{ \bvec{s} \in \{ -, + \}^{n} \ \Big| \ \big| I(W^{\bvec{s}}) - \log [\mathcal{X} : \mathcal{N}] \big| < \delta
\ \mathrm{and} \
\big| I(W^{\bvec{s}}[\mathcal{N}]) - \log [\mathcal{X} : \mathcal{N}] \big| < \delta
\Big\} \Big| = 1
\label{eq:multilevel}
\end{align}
for every fixed $\delta > 0$, where $[G : N] = |G/N|$ denotes the index of a subgroup $N$ in a group $G$.

We now consider each term of the summation of \eqref{eq:multilevel}.
It is clear that the left-hand sides of \eqref{eq:strong1} and \eqref{eq:strong2} coincide with the terms of the summation with trivial normal subgroups $\mathcal{N} = \mathcal{X}$ and $\mathcal{N} = \{ e \}$, respectively, where $e \in \mathcal{X}$ is the identity element.
Thus, the strong polarization \eqref{eq:strong} is a special case of multilevel polarization \eqref{eq:multilevel}.
Moreover, other terms of the summation are the limiting proportions of partially noiseless synthetic channels $W^{\bvec{s}}$, because the condition
\begin{align}
\big| I(W^{\bvec{s}}[\mathcal{N}]) - \log [\mathcal{X} : \mathcal{N}] \big| < \delta
\label{eq:partially_noiseless}
\end{align}
implies an almost noiseless homomorphism channel $W^{\bvec{s}}[\mathcal{N}]$
for $\delta$ sufficiently small.
Together with \eqref{eq:partially_noiseless}, note that the condition
\begin{align}
\big| I(W^{\bvec{s}}) - \log [\mathcal{X} : \mathcal{N}] \big| < \delta
\end{align}
implies that the almost noiseless homomorphism channel $W^{\bvec{s}}[\mathcal{N}]$ has almost the same symmetric capacity as original one $W^{\bvec{s}}$;
this is a reason why polar codes can achieve the symmetric capacity with multilevel polarization.

Although the limiting proportions \eqref{eq:strong1} and \eqref{eq:strong2} are fully solved, each limiting proportion of partially noiseless synthetic channels, i.e., each term of the summation of \eqref{eq:multilevel}, is still unknown in general.
To refine the multilevel polarization theorem, this study investigates the limiting proportions of partially noiseless synthetic channels for a certain class of erasure channels.
The next subsection introduces such a class.

\subsection{A Generalized Erasure Channel}

For two integers $a, b \ge 1$, let $a|b$ be a shorthand for $a$ divides $b$.
We define a generalized erasure channel as follows:

\begin{definition}[{\cite[Definition~2]{itw2016}%
\footnote{%
Note that \defref{def:V} is slightly different to \cite[Definition~2]{itw2016}; and these are essentially equivalent under relabeling of the input alphabets.%
}}]
\label{def:V}
Let $\mathcal{X} = \mathbb{Z}/q\mathbb{Z}$ and%
\footnote{%
Note that $\bigcup_{d|q} (\mathbb{Z}/d\mathbb{Z})$ is defined to be a set but not an algebraic structure; and it is assumed that $(\mathbb{Z}/n_{1}\mathbb{Z}) \cap (\mathbb{Z}/n_{2}\mathbb{Z}) = \emptyset$ if $n_{1} \neq n_{2}$.
Namely, we consider $\bigcup_{d|q} (\mathbb{Z}/d\mathbb{Z})$ as a disjoint union or a direct sum.
}
$\mathcal{Y} = \bigcup_{d|q} (\mathbb{Z}/d\mathbb{Z})$.
For a given probability vector%
\footnote{%
A vector $(a_{i})_{i}$ is called a probability vector if $a_{i} \ge 0$ and $\sum_{i} a_{i} = 1$.%
}
$( \varepsilon_{d} )_{d|q}$, the erasure channel $V : \mathcal{X} \to \mathcal{Y}$ is defined by 
\begin{align}
V(y \mid x)
\coloneqq
\begin{dcases}
\varepsilon_{d}
& \mathrm{if} \ y = \varphi_{d}( x ) \ \mathrm{for} \ \mathrm{some} \ d|q ,
\\
0
& \mathrm{otherwise} ,
\end{dcases}
\end{align}
where $\varphi_{d} : \mathbb{Z}/q\mathbb{Z} \to \mathbb{Z}/d\mathbb{Z}$ is a surjective ring homomorphism given by $\varphi_{d} : x \mapsto (x + d\mathbb{Z})$ for each $d|q$.
\end{definition}

For a probability vector $( \varepsilon_{d} )_{d|q}$ and a sequence $\bvec{s} = s_{1}s_{2} \cdots s_{n} \in \{ -, + \}^{n}$, we define the vector $( \varepsilon_{d}^{\bvec{s}} )_{d|q}$ as
\begin{align}
\varepsilon_{d}^{\boldsymbol{s}}
\coloneqq
\begin{dcases}
\sum_{\substack{ d_{1}|q, d_{2}|q: \\ \gcd( d_{1}, d_{2} ) = d }} \varepsilon_{d_{1}}^{s_{1}s_{2} \cdots s_{n-1}} \varepsilon_{d_{2}}^{s_{1}s_{2} \cdots s_{n-1}}
& \mathrm{if} \ s_{n} = - ,
\\
\sum_{\substack{ d_{1}|q, d_{2}|q: \\ \lcm( d_{1}, d_{2} ) = d}} \varepsilon_{d_{1}}^{s_{1}s_{2} \cdots s_{n-1}} \varepsilon_{d_{2}}^{s_{1}s_{2} \cdots s_{n-1}}
& \mathrm{if} \ s_{n} = +
\end{dcases}
\label{def:eps_s}
\end{align}
for each $d|q$ recursively.
It can be verified by induction that the vector $( \varepsilon_{d}^{\bvec{s}} )_{d|q}$ is also a probability vector.
So far, the polar transforms of \eqref{def:minus} and \eqref{def:plus} with quasigroup $(\mathcal{X}, \ast)$ were discussed.
Henceforth, we consider the quasigroup operation $\ast$ defined by operations of the integer residue class ring $\mathcal{X} = \mathbb{Z}/q\mathbb{Z}$ as follows:
We define the quasigroup operation $a \ast b \coloneqq a + \gamma \cdot b$ for $a, b \in \mathcal{X}$ with a fixed unit $\gamma$ belonging to the ring $\mathcal{X}$.
Under polar transforms with this quasigroup operation, the following theorem holds.

\begin{theorem}[{\cite[Theorem~1]{itw2016}}]
\label{th:itw2016}
For an initial erasure channel $V : \mathcal{X} \to \mathcal{Y}$ with probability vector $( \varepsilon_{d} )_{d|q}$ and a sequence $\bvec{s} \in \{ -, + \}^{\ast}$, the synthetic channel $V^{\bvec{s}}$ is equivalent%
\footnote{%
The equivalence relation is introduced in \cite{itw2016}.
}
to the erasure channel with probability vector $( \varepsilon_{d}^{\bvec{s}} )_{d|q}$ given in \eqref{def:eps_s}.
\end{theorem}

Note that \cite[Theorem~1]{itw2016} is stated only for one-step polar transform; and it can be directly extended to $n$-step polar transforms as shown in \thref{th:itw2016} by induction.
Note that for any DMC $W$ and any sequence $\bvec{s} \in \{ -, + \}^{n}$ of length $n$, the output alphabet size of a synthetic channel $W^{\bvec{s}}$ grows double-exponentially as $n$ increases (cf. \sectref{sect:notations}).
This is a main issue of computational complexities for analyzing and constructing polar codes.
Fortunately, \thref{th:itw2016} enables us to analyze the synthetic channel $V^{\bvec{s}}$ by only calculating $( \varepsilon_{d}^{\boldsymbol{s}} )_{d|q}$ with the recursive formula \eqref{def:eps_s}.
Thus, the erasure channel $V$ is proposed in \cite{itw2016} to be a toy model of multilevel polarization.

In \cite[Theorem~2]{itw2016}, the asymptotic distribution of multilevel channel polarization of $V$ was characterized by its initial vector $( \varepsilon_{d} )_{d|q}$ when $q$ is a prime power, i.e., in the case of $q = p^{r}$ for some prime number $p$ and some integer $r \in \mathbb{N}$.
This implies that each term of the summation of \eqref{eq:multilevel} was solved for the erasure channels $V$ if $q = p^{r}$.
This study examine further the asymptotic distribution of multilevel channel polarization of $V$ for general composite numbers $q$.

\section{Main Results}

In this section, we consider erasure channels $V$ of \defref{def:V} with probability vector $( \varepsilon_{d} )_{d|q}$.
Henceforth, assume that the input alphabet size $q$ can be factorized by%
\footnote{Even if $q$ has only one prime factors $q = p_{1}^{r_{1}}$, in this study, we write $q = p_{1}^{r_{1}} p_{2}^{r_{2}} \cdots p_{m}^{r_{m}}$ for some $m \ge 2$ by setting $r_{2} = \cdots = r_{m} = 0$.}
$q = p_{1}^{r_{1}} p_{2}^{r_{2}} \cdots p_{m}^{r_{m}}$.
If a positive integer $d$ can be factorized by $d = p_{1}^{t_{1}} p_{2}^{t_{2}} \cdots p_{m}^{t_{m}}$, then we write it as $d = \langle \bvec{t} \rangle$ for short, where $\bvec{t} = (t_{1}, t_{2}, \dots, t_{m})$.
Namely, defining a partial order $\bvec{t} \le \bvec{u}$ between two $m$-tuples $\bvec{t}$ and $\bvec{u}$ by $t_{i} \le u_{i}$ for every $i = 1, 2, \dots, m$, we observe that $d$ divides $q$ if and only if $\bvec{0} \le \bvec{t} \le \bvec{r}$ for $d = \langle \bvec{t} \rangle$ and $q = \langle \bvec{r} \rangle$, where $\bvec{0} = (0, \dots, 0)$.
The key idea of our proofs is that for each integers $i$ and $j$ satisfying $1 \le i < j \le m$, we combine the probability masses $( \varepsilon_{\langle \bvec{t} \rangle}^{\bvec{s}} )_{\bvec{0} \le \bvec{t} \le \bvec{r}}$ into the following four masses:%
\begin{align}
\theta_{i, j}^{\bvec{s}}(a, b)
& \coloneqq
\sum_{\substack{ \bvec{t} : \bvec{0} \le \bvec{t} \le \bvec{r} , \\ t_{i} \ge a, t_{j} \ge b } } \varepsilon_{\langle \bvec{t} \rangle}^{\bvec{s}} ,
\label{def:theta} \\
\lambda_{i, j}^{\bvec{s}}(a, b)
& \coloneqq
\sum_{\substack{ \bvec{t} : \bvec{0} \le \bvec{t} \le \bvec{r} , \\ t_{i} \ge a, t_{j} < b } }  \varepsilon_{\langle \bvec{t} \rangle}^{\bvec{s}} ,
\label{def:lambda} \\
\rho_{i, j}^{\bvec{s}}(a, b)
& \coloneqq
\sum_{\substack{ \bvec{t} : \bvec{0} \le \bvec{t} \le \bvec{r} , \\ t_{i} < a, t_{j} \ge b } }  \varepsilon_{\langle \bvec{t} \rangle}^{\bvec{s}} ,
\label{def:rho} \\
\beta_{i, j}^{\bvec{s}}(a, b)
& \coloneqq
\sum_{\substack{ \bvec{t} : \bvec{0} \le \bvec{t} \le \bvec{r} , \\ t_{i} < a, t_{j} < b } }  \varepsilon_{\langle \bvec{t} \rangle}^{\bvec{s}}
\label{def:beta} 
\end{align}
for each integers $a, b \ge 1$, and each sequence $\bvec{s} \in \{ -, + \}^{\ast}$, where $( \varepsilon_{\langle \bvec{t} \rangle}^{\bvec{s}} )_{\bvec{0} \le \bvec{t} \le \bvec{r}} = ( \varepsilon_{d}^{\bvec{s}} )_{d|q}$ is recursively defined in \eqref{def:eps_s} with an initial probability vector $( \varepsilon_{d} )_{d|q}$.
If the sequence $\bvec{s}$ is empty, then we omit the superscripts $\bvec{s}$ as $\theta_{i, j}(a, b)$, $\lambda_{i, j}(a, b)$, $\rho_{i, j}(a, b)$, and $\beta_{i, j}(a, b)$.
Note that
\begin{align}
\theta_{i, j}^{\bvec{s}}(a, b) + \lambda_{i, j}^{\bvec{s}}(a, b) + \rho_{i, j}^{\bvec{s}}(a, b) + \beta_{i, j}^{\bvec{s}}(a, b)
=
\sum_{d|q} \varepsilon_{d}^{\bvec{s}}
=
1
\end{align}
for each $1 \le i < j \le m$, each $a, b \ge 1$, and each $\bvec{s} \in \{ -, + \}^{\ast}$.
We now give formulas for \eqref{def:theta}--\eqref{def:beta} under the recursive formula \eqref{def:eps_s} as follows:

\begin{lemma}
\label{lem:formulas}
For any $\bvec{s} \in \{ -, + \}^{\ast}$, $1 \le i < j \le m$, and $a, b \ge 1$, it holds that
\begin{align}
\theta_{i, j}^{\bvec{s}-}(a, b)
& =
\theta_{i, j}^{\bvec{s}}(a, b)^{2} ,
\\
\theta_{i, j}^{\bvec{s}+}(a, b)
& =
\theta_{i, j}^{\bvec{s}}(a, b) \, \big[ 2 - \theta_{i, j}^{\bvec{s}}(a, b) \big] + 2 \, \lambda_{i, j}^{\bvec{s}}(a, b) \, \rho_{i, j}^{\bvec{s}}(a, b) ,
\\
\lambda_{i, j}^{\bvec{s}-}(a, b)
& =
\lambda_{i, j}^{\bvec{s}}(a, b) \, \big[ \lambda_{i, j}^{\bvec{s}}(a, b) + 2 \, \theta_{i, j}^{\bvec{s}}(a, b) \big] ,
\\
\lambda_{i, j}^{\bvec{s}+}(a, b)
& =
\lambda_{i, j}^{\bvec{s}}(a, b) \, \big[ \lambda_{i, j}^{\bvec{s}}(a, b) + 2 \, \beta_{i, j}^{\bvec{s}}(a, b) \big] ,
\\
\rho_{i, j}^{\bvec{s}-}(a, b)
& =
\rho_{i, j}^{\bvec{s}}(a, b) \, \big[ \rho_{i, j}^{\bvec{s}}(a, b) + 2 \, \theta_{i, j}^{\bvec{s}}(a, b) \big] ,
\\
\rho_{i, j}^{\bvec{s}+}(a, b)
& =
\rho_{i, j}^{\bvec{s}}(a, b) \, \big[ \rho_{i, j}^{\bvec{s}}(a, b) + 2 \, \beta_{i, j}^{\bvec{s}}(a, b) \big] ,
\\
\beta_{i, j}^{\bvec{s}-}(a, b)
& =
\beta_{i, j}^{\bvec{s}}(a, b) \, \big[ 2 - \beta_{i, j}^{\bvec{s}}(a, b) \big] + 2 \, \lambda_{i, j}^{\bvec{s}}(a, b) \, \rho_{i, j}^{\bvec{s}}(a, b) ,
\\
\beta_{i, j}^{\bvec{s}+}(a, b)
& =
\beta_{i, j}^{\bvec{s}}(a, b)^{2} .
\end{align}
\end{lemma}

\begin{IEEEproof}[Proof of \lemref{lem:formulas}]
By symmetry, it suffices to prove only for the minus transforms.
Fix a sequence $\bvec{s} \in \{ -, + \}^{\ast}$, indices $1 \le i < j \le m$, and integers $a, b \ge 1$ arbitrarily.
A direct calculation shows
\begin{align}
\varepsilon_{\langle \bvec{t} \rangle}^{\bvec{s}-}
& \overset{\mathclap{\eqref{def:eps_s}}}{=}
\sum_{\substack{ d_{1}|q, d_{2}|q : \\ \gcd(d_{1}, d_{2}) = \langle \bvec{t} \rangle }} \varepsilon_{d_{1}}^{\bvec{s}} \, \varepsilon_{d_{2}}^{\bvec{s}}
\notag \\
& =
\sum_{\bvec{u} : \bvec{0} \le \bvec{u} \le \bvec{r}} \sum_{\bvec{v} : \bvec{0} \le \bvec{v} \le \bvec{r}} \varepsilon_{\langle \bvec{u} \rangle}^{\bvec{s}} \, \varepsilon_{\langle \bvec{v} \rangle}^{\bvec{s}} \prod_{k = 1}^{m} \mathbbm{1}[ t_{k} = \min\{ u_{k}, v_{k} \} ]
\notag \\
& =
\sum_{\bvec{u} : \bvec{0} \le \bvec{u} \le \bvec{r}} \sum_{\bvec{v} : \bvec{0} \le \bvec{v} \le \bvec{r}} \varepsilon_{\langle \bvec{u} \rangle}^{\bvec{s}} \, \varepsilon_{\langle \bvec{v} \rangle}^{\bvec{s}} \prod_{k = 1}^{m} \Big( \mathbbm{1}[ t_{k} = u_{k} \le v_{k} ] + \mathbbm{1}[ t_{k} = v_{k} < u_{k} ] \Big)
\notag \\
& =
\sum_{\bvec{b} \in \{ 0, 1 \}^{m}} \sum_{\bvec{u} : \bvec{0} \le \bvec{u} \le \bvec{r}} \sum_{\bvec{v} : \bvec{0} \le \bvec{v} \le \bvec{r}} \varepsilon_{\langle \bvec{u} \rangle}^{\bvec{s}} \, \varepsilon_{\langle \bvec{v} \rangle}^{\bvec{s}} \prod_{k = 1}^{m} \Big( \mathbbm{1}[ t_{k} = u_{k} \le v_{k} ] \mathbbm{1}[ b_{k} = 0 ] + \mathbbm{1}[ t_{k} = v_{k} < u_{k} ] \mathbbm{1} [ b_{k} = 1 ] \Big)
\notag \\
& =
\sum_{\bvec{b} \in \{ 0, 1 \}^{m}} \sum_{\bvec{u} : \bvec{0} \le \bvec{u} \le \bvec{r}}  \varepsilon_{\langle \bvec{w}^{(0)} \rangle}^{\bvec{s}} \, \varepsilon_{\langle \bvec{w}^{(1)} \rangle}^{\bvec{s}} \prod_{k = 1}^{m} \Big( \mathbbm{1}[ t_{k} \le u_{k} ] \mathbbm{1}[ b_{k} = 0 ] + \mathbbm{1}[ t_{k} < u_{k} ] \mathbbm{1} [ b_{k} = 1 ] \Big)
\label{eq:eps_minus_proof}
\end{align}
for every $\bvec{0} \le \bvec{t} \le \bvec{r}$, where $\bvec{u} = (u_{1}, \dots, u_{m})$, $\bvec{v} = (v_{1}, \dots, v_{m})$, and $\bvec{b} = (b_{1}, \dots, b_{m})$; the indicator function is denoted by
\begin{align}
\mathbbm{1}[ P ]
\coloneqq
\begin{cases}
1
& \text{if $P$ is true} ,
\\
0
& \text{if $P$ is false} ;
\end{cases}
\end{align}
and both $\bvec{w}^{(0)} = (w_{1}^{(0)}, \dots, w_{m}^{(0)})$ and $\bvec{w}^{(1)} = (w_{1}^{(1)}, \dots, w_{m}^{(1)})$ are defined as functions of $(\bvec{t}, \bvec{u}, \bvec{b})$ so that
\begin{align}
w_{k}^{(0)}
& =
\begin{cases}
t_{k}
& \mathrm{if} \ b_{k} = 0 ,
\\
u_{k}
& \mathrm{if} \ b_{k} = 1 ,
\end{cases}
\\
w_{k}^{(1)}
& =
\begin{cases}
u_{k}
& \mathrm{if} \ b_{k} = 0 ,
\\
t_{k}
& \mathrm{if} \ b_{k} = 1 ,
\end{cases}
\end{align}
respectively, for each $k = 1, 2, \dots, m$.
Letting an $m$-tuple $\bvec{c} = (c_{1}, \dots, c_{m})$ by
\begin{align}
c_{k}
=
\begin{cases}
a
& \mathrm{if} \ k = i ,
\\
b
& \mathrm{if} \ k = j ,
\\
0
& \mathrm{otherwise}
\end{cases}
\end{align}
for each $k = 1, 2, \dots, m$, we observe that
\begin{align}
\theta_{i, j}^{\bvec{s}-}(a, b)
& =
\sum_{\bvec{t} : \bvec{c} \le \bvec{t} \le \bvec{r}} \varepsilon_{\langle \bvec{t} \rangle}^{\bvec{s}-}
\notag \\
& \overset{\mathclap{\eqref{eq:eps_minus_proof}}}{=}
\sum_{\bvec{t} : \bvec{c} \le \bvec{t} \le \bvec{r}} \sum_{\bvec{b} \in \{ 0, 1 \}^{m}} \sum_{\bvec{u} : \bvec{0} \le \bvec{u} \le \bvec{r}}  \varepsilon_{\langle \bvec{w}^{(0)} \rangle}^{\bvec{s}} \, \varepsilon_{\langle \bvec{w}^{(1)} \rangle}^{\bvec{s}} \prod_{k = 1}^{m} \Big( \mathbbm{1}[ t_{k} \le u_{k} ] \mathbbm{1}[ b_{k} = 0 ] + \mathbbm{1}[ t_{k} < u_{k} ] \mathbbm{1} [ b_{k} = 1 ] \Big)
\notag \\
& =
\sum_{\bvec{t} : \bvec{c} \le \bvec{t} \le \bvec{r}} \sum_{\bvec{b} \in \{ 0, 1 \}^{m}} \sum_{\bvec{u} : \bvec{0} \le \bvec{u} \le \bvec{r}}  \varepsilon_{\langle \bvec{t} \rangle}^{\bvec{s}} \, \varepsilon_{\langle \bvec{u} \rangle}^{\bvec{s}} \prod_{k = 1}^{m} \Big( \mathbbm{1}[ t_{k} \le u_{k} ] \mathbbm{1}[ b_{k} = 0 ] + \mathbbm{1}[ c_{k} \le u_{k} < t_{k} ] \mathbbm{1} [ b_{k} = 1 ] \Big)
\notag \\
& =
\sum_{\bvec{t} : \bvec{c} \le \bvec{t} \le \bvec{r}} \sum_{\bvec{u} : \bvec{0} \le \bvec{u} \le \bvec{r}}  \varepsilon_{\langle \bvec{t} \rangle}^{\bvec{s}} \, \varepsilon_{\langle \bvec{u} \rangle}^{\bvec{s}} \prod_{k = 1}^{m} \Big( \mathbbm{1}[ t_{k} \le u_{k} ] + \mathbbm{1}[ c_{k} \le u_{k} < t_{k} ] \Big)
\notag \\
& =
\sum_{\bvec{t} : \bvec{c} \le \bvec{t} \le \bvec{r}} \sum_{\bvec{u} : \bvec{0} \le \bvec{u} \le \bvec{r}}  \varepsilon_{\langle \bvec{t} \rangle}^{\bvec{s}} \, \varepsilon_{\langle \bvec{u} \rangle}^{\bvec{s}} \prod_{k = 1}^{m} \mathbbm{1}[ c_{k} \le u_{k} ]
\notag \\
& =
\sum_{\bvec{t} : \bvec{c} \le \bvec{t} \le \bvec{r}} \sum_{\bvec{u} : \bvec{c} \le \bvec{u} \le \bvec{r}}  \varepsilon_{\langle \bvec{t} \rangle}^{\bvec{s}} \, \varepsilon_{\langle \bvec{u} \rangle}^{\bvec{s}}
\notag \\
& =
\bigg( \sum_{\bvec{t} : \bvec{c} \le \bvec{t} \le \bvec{r}} \varepsilon_{\langle \bvec{t} \rangle}^{\bvec{s}} \bigg)^{2}
\notag \\
& =
\theta_{i, j}^{\bvec{s}}(a, b)^{2} .
\label{eq:formulas1_proof}
\end{align}
Similarly, we have
\begin{align}
\lambda_{i, j}^{\bvec{s}-}(a, b)
& =
\sum_{\substack{ \bvec{t} : \bvec{0} \le \bvec{t} \le \bvec{r} , \\ t_{i} \ge a, t_{j} < b } } \varepsilon_{\langle \bvec{t} \rangle}^{\bvec{s}-}
\notag \\
& \overset{\mathclap{\eqref{eq:eps_minus_proof}}}{=}
\sum_{\substack{ \bvec{t} : \bvec{0} \le \bvec{t} \le \bvec{r} , \\ t_{i} \ge a, t_{j} < b } } \sum_{\bvec{b} \in \{ 0, 1 \}^{m}} \sum_{\bvec{u} : \bvec{0} \le \bvec{u} \le \bvec{r}}  \varepsilon_{\langle \bvec{w}^{(0)} \rangle}^{\bvec{s}} \, \varepsilon_{\langle \bvec{w}^{(1)} \rangle}^{\bvec{s}} \prod_{k = 1}^{m} \Big( \mathbbm{1}[ t_{k} \le u_{k} ] \mathbbm{1}[ b_{k} = 0 ] + \mathbbm{1}[ t_{k} < u_{k} ] \mathbbm{1} [ b_{k} = 1 ] \Big)
\notag \\
& =
\sum_{\substack{ \bvec{t} : \bvec{0} \le \bvec{t} \le \bvec{r} , \\ t_{i} \ge a, t_{j} < b } } \sum_{\bvec{b} \in \{ 0, 1 \}^{m}} \sum_{\substack{ \bvec{u} : \bvec{0} \le \bvec{u} \le \bvec{r} , \\ u_{j} < b}} \varepsilon_{\langle \bvec{w}^{(0)} \rangle}^{\bvec{s}} \, \varepsilon_{\langle \bvec{w}^{(1)} \rangle}^{\bvec{s}} \prod_{k = 1}^{m} \Big( \mathbbm{1}[ t_{k} \le u_{k} ] \mathbbm{1}[ b_{k} = 0 ] + \mathbbm{1}[ t_{k} < u_{k} ] \mathbbm{1} [ b_{k} = 1 ] \Big)
\notag \\
& \qquad
{} + \sum_{\substack{ \bvec{t} : \bvec{0} \le \bvec{t} \le \bvec{r} , \\ t_{i} \ge a, t_{j} < b } } \sum_{\bvec{b} \in \{ 0, 1 \}^{m}} \sum_{\substack{ \bvec{u} : \bvec{0} \le \bvec{u} \le \bvec{r} , \\ u_{j} \ge b}} \varepsilon_{\langle \bvec{w}^{(0)} \rangle}^{\bvec{s}} \, \varepsilon_{\langle \bvec{w}^{(1)} \rangle}^{\bvec{s}} \prod_{k = 1}^{m} \Big( \mathbbm{1}[ t_{k} \le u_{k} ] \mathbbm{1}[ b_{k} = 0 ] + \mathbbm{1}[ t_{k} < u_{k} ] \mathbbm{1} [ b_{k} = 1 ] \Big)
\notag \\
& =
\sum_{\substack{ \bvec{t} : \bvec{0} \le \bvec{t} \le \bvec{r} , \\ t_{i} \ge a, t_{j} < b } } \sum_{\bvec{b} \in \{ 0, 1 \}^{m}} \sum_{\substack{ \bvec{u} : \bvec{0} \le \bvec{u} \le \bvec{r} , \\ u_{j} < b}} \varepsilon_{\langle \bvec{t} \rangle}^{\bvec{s}} \, \varepsilon_{\langle \bvec{u} \rangle}^{\bvec{s}} \Big( \mathbbm{1}[ t_{i} \le u_{i} ] \mathbbm{1}[ b_{i} = 0 ] + \mathbbm{1}[ a \le u_{i} < t_{i} ] \mathbbm{1} [ b_{i} = 1 ] \Big)
\notag \\
& \qquad \qquad \qquad \qquad \qquad \qquad \qquad
{} \times \prod_{k = 1 : k \neq i}^{m} \Big( \mathbbm{1}[ t_{k} \le u_{k} ] \mathbbm{1}[ b_{k} = 0 ] + \mathbbm{1}[ 0 \le u_{k} < t_{k} ] \mathbbm{1} [ b_{k} = 1 ] \Big)
\notag \\
& \qquad
{} + \sum_{\substack{ \bvec{t} : \bvec{0} \le \bvec{t} \le \bvec{r} , \\ t_{i} \ge a, t_{j} < b } } \sum_{\bvec{b} \in \{ 0, 1 \}^{m}} \sum_{\substack{ \bvec{u} : \bvec{0} \le \bvec{u} \le \bvec{r} , \\ u_{j} \ge b}} \varepsilon_{\langle \bvec{t} \rangle}^{\bvec{s}} \, \varepsilon_{\langle \bvec{u} \rangle}^{\bvec{s}} \Big( \mathbbm{1}[ a \le u_{i} \le t_{i} ] \mathbbm{1}[ b_{i} = 0 ] + \mathbbm{1}[ t_{k} < u_{k} ] \mathbbm{1} [ b_{i} = 1 ] \Big)
\notag \\
& \qquad \qquad \qquad \qquad \qquad \qquad \qquad
{} \times \prod_{k = 1 : k \neq i}^{m} \Big( \mathbbm{1}[ 0 \le u_{k} \le t_{k} ] \mathbbm{1}[ b_{k} = 0 ] + \mathbbm{1}[ t_{k} < u_{k} ] \mathbbm{1} [ b_{k} = 1 ] \Big)
\notag \\
& \overset{\text{(a)}}{=}
\sum_{\substack{ \bvec{t} : \bvec{0} \le \bvec{t} \le \bvec{r} , \\ t_{i} \ge a, t_{j} < b } } \sum_{\substack{ \bvec{u} : \bvec{0} \le \bvec{u} \le \bvec{r} , \\ u_{j} < b}} \varepsilon_{\langle \bvec{t} \rangle}^{\bvec{s}} \, \varepsilon_{\langle \bvec{u} \rangle}^{\bvec{s}} \Big( \mathbbm{1}[ t_{i} \le u_{i} ] + \mathbbm{1}[ a \le u_{i} < t_{i} ] \Big) \prod_{k = 1 : k \neq i}^{m} \Big( \mathbbm{1}[ t_{k} \le u_{k} ] + \mathbbm{1}[ 0 \le u_{k} < t_{k} ] \Big)
\notag \\
& \qquad
{} + 2 \sum_{\substack{ \bvec{t} : \bvec{0} \le \bvec{t} \le \bvec{r} , \\ t_{i} \ge a, t_{j} < b } } \sum_{\substack{ \bvec{u} : \bvec{0} \le \bvec{u} \le \bvec{r} , \\ u_{j} \ge b}} \varepsilon_{\langle \bvec{t} \rangle}^{\bvec{s}} \, \varepsilon_{\langle \bvec{u} \rangle}^{\bvec{s}} \Big( \mathbbm{1}[ a \le u_{i} \le t_{i} ] + \mathbbm{1}[ t_{i} < u_{i} ] \Big) \prod_{k = 1 : k \neq i}^{m} \Big( \mathbbm{1}[ 0 \le u_{k} \le t_{k} ] + \mathbbm{1}[ t_{k} < u_{k} ] \Big)
\notag \\
& =
\sum_{\substack{ \bvec{t} : \bvec{0} \le \bvec{t} \le \bvec{r} , \\ t_{i} \ge a, t_{j} < b } } \sum_{\substack{ \bvec{u} : \bvec{0} \le \bvec{u} \le \bvec{r} , \\ u_{j} < b}} \varepsilon_{\langle \bvec{t} \rangle}^{\bvec{s}} \, \varepsilon_{\langle \bvec{u} \rangle}^{\bvec{s}} \, \mathbbm{1}[ a \le u_{i} ] + 2 \sum_{\substack{ \bvec{t} : \bvec{0} \le \bvec{t} \le \bvec{r} , \\ t_{i} \ge a, t_{j} < b } } \sum_{\substack{ \bvec{u} : \bvec{0} \le \bvec{u} \le \bvec{r} , \\ u_{j} \ge b}} \varepsilon_{\langle \bvec{t} \rangle}^{\bvec{s}} \, \varepsilon_{\langle \bvec{u} \rangle}^{\bvec{s}} \, \mathbbm{1}[ a \le u_{i} ]
\notag \\
& =
\left( \sum_{\substack{ \bvec{t} : \bvec{0} \le \bvec{t} \le \bvec{r} , \\ t_{i} \ge a, t_{j} < b } } \varepsilon_{\langle \bvec{t} \rangle}^{\bvec{s}} \right)^{2} + 2 \left( \sum_{\substack{ \bvec{t} : \bvec{0} \le \bvec{t} \le \bvec{r} , \\ t_{i} \ge a, t_{j} < b } } \varepsilon_{\langle \bvec{t} \rangle}^{\bvec{s}} \right) \left( \sum_{\substack{ \bvec{u} : \bvec{0} \le \bvec{u} \le \bvec{r} , \\ u_{i} \ge a, u_{j} \ge b}} \varepsilon_{\langle \bvec{u} \rangle}^{\bvec{s}} \right)
\notag \\
& =
\lambda_{i, j}^{\bvec{s}}(a, b)^{2} + 2 \, \lambda_{i, j}^{\bvec{s}}(a, b) \, \theta_{i, j}^{\bvec{s}}(a, b) ,
\label{eq:formulas2_proof}
\end{align}
where the factor $2$ in (a) comes from the fact that $t_{j} < b$ and $u_{j} \ge b$ imply $\mathbbm{1}[ t_{j} < u_{j} ] = 1$.
Since $\lambda_{i, j}^{\bvec{s}}(a, b) = \rho_{j, i}^{\bvec{s}}(b, a)$, we readily see from \eqref{eq:formulas2_proof} that
\begin{align}
\rho_{i, j}^{\bvec{s}-}(a, b)
=
\rho_{i, j}^{\bvec{s}}(a, b)^{2} + 2 \, \rho_{i, j}^{\bvec{s}}(a, b) \, \theta_{i, j}^{\bvec{s}}(a, b) .
\label{eq:formulas3_proof}
\end{align}
Finally, as $\theta_{i, j}^{\bvec{s}}(a, b) + \lambda_{i, j}^{\bvec{s}}(a, b) + \rho_{i, j}^{\bvec{s}}(a, b) + \beta_{i, j}^{\bvec{s}}(a, b) = 1$, it follows from \eqref{eq:formulas1_proof}--\eqref{eq:formulas3_proof} that
\begin{align}
\beta_{i, j}^{\bvec{s}-}(a, b)
& =
1 - \theta_{i, j}^{\bvec{s}-}(a, b) - \lambda_{i, j}^{\bvec{s}-}(a, b) - \rho_{i, j}^{\bvec{s}-}(a, b)
\notag \\
& =
1 - \theta_{i, j}^{\bvec{s}}(a, b)^{2} - \big[ \lambda_{i, j}^{\bvec{s}}(a, b)^{2} + 2 \, \lambda_{i, j}^{\bvec{s}}(a, b) \, \theta_{i, j}^{\bvec{s}}(a, b) \big] - \big[ \rho_{i, j}^{\bvec{s}}(a, b)^{2} + 2 \, \rho_{i, j}^{\bvec{s}}(a, b) \, \theta_{i, j}^{\bvec{s}}(a, b) \big]
\notag \\
& =
1 - \theta_{i, j}^{\bvec{s}}(a, b)^{2} - \big[ \lambda_{i, j}^{\bvec{s}}(a, b) + \rho_{i, j}^{\bvec{s}}(a, b) \big]^{2} - 2 \, \theta_{i, j}^{\bvec{s}}(a, b) \, \big[ \lambda_{i, j}^{\bvec{s}}(a, b) + \rho_{i, j}^{\bvec{s}}(a, b) \big] + 2 \, \lambda_{i, j}^{\bvec{s}}(a, b) \, \rho_{i, j}^{\bvec{s}}(a, b)
\notag \\
& =
1 - \big[ \theta_{i, j}^{\bvec{s}}(a, b) + \lambda_{i, j}^{\bvec{s}}(a, b) + \rho_{i, j}^{\bvec{s}}(a, b) \big]^{2} + 2 \, \lambda_{i, j}^{\bvec{s}}(a, b) \, \rho_{i, j}^{\bvec{s}}(a, b)
\notag \\
& =
\big[ \theta_{i, j}^{\bvec{s}}(a, b) + \lambda_{i, j}^{\bvec{s}}(a, b) + \rho_{i, j}^{\bvec{s}}(a, b) \big] \big[ 1 - \theta_{i, j}^{\bvec{s}}(a, b) - \lambda_{i, j}^{\bvec{s}}(a, b) - \rho_{i, j}^{\bvec{s}}(a, b) \big] + \beta_{i, j}^{\bvec{s}}(a, b) + 2 \, \lambda_{i, j}^{\bvec{s}}(a, b) \, \rho_{i, j}^{\bvec{s}}(a, b)
\notag \\
& =
\beta_{i, j}^{\bvec{s}}(a, b) \big[ 1 + \theta_{i, j}^{\bvec{s}}(a, b) + \lambda_{i, j}^{\bvec{s}}(a, b) + \rho_{i, j}^{\bvec{s}}(a, b) \big] + 2 \, \lambda_{i, j}^{\bvec{s}}(a, b) \, \rho_{i, j}^{\bvec{s}}(a, b)
\notag \\
& =
\beta_{i, j}^{\bvec{s}}(a, b) \big[ 2 - \beta_{i, j}^{\bvec{s}}(a, b) \big] + 2 \, \lambda_{i, j}^{\bvec{s}}(a, b) \, \rho_{i, j}^{\bvec{s}}(a, b) .
\end{align}
This completes the proof of \lemref{lem:formulas}.
\end{IEEEproof}

\lemref{lem:formulas} can characterize the average value of \eqref{def:theta}--\eqref{def:beta} over one-step polar transform, as shown in the following lemma.

\begin{lemma}
\label{lem:quasi-conservation}
For any $\bvec{s} \in \{ -, + \}^{\ast}$, $1 \le i < j \le m$, and $a, b \ge 1$, it holds that
\begin{align}
\frac{ 1 }{ 2 } \Big[ \theta_{i, j}^{\bvec{s}-}(a, b) + \theta_{i, j}^{\bvec{s}+}(a, b) \Big]
& =
\theta_{i, j}^{\bvec{s}}(a, b) + \lambda_{i, j}^{\bvec{s}}(a, b) \, \rho_{i, j}^{\bvec{s}}(a, b) ,
\\
\frac{ 1 }{ 2 } \Big[ \lambda_{i, j}^{\bvec{s}-}(a, b) + \lambda_{i, j}^{\bvec{s}+}(a, b) \Big]
& =
\lambda_{i, j}^{\bvec{s}}(a, b) \, \big[ 1 - \rho_{i, j}^{\bvec{s}}(a, b) \big] ,
\\
\frac{ 1 }{ 2 } \Big[ \rho_{i, j}^{\bvec{s}-}(a, b) + \rho_{i, j}^{\bvec{s}+}(a, b) \Big]
& =
\rho_{i, j}^{\bvec{s}}(a, b) \, \big[ 1 - \lambda_{i, j}^{\bvec{s}}(a, b) \big] ,
\\
\frac{ 1 }{ 2 } \Big[ \beta_{i, j}^{\bvec{s}-}(a, b) + \beta_{i, j}^{\bvec{s}+}(a, b) \Big]
& =
\beta_{i, j}^{\bvec{s}}(a, b) + \lambda_{i, j}^{\bvec{s}}(a, b) \, \rho_{i, j}^{\bvec{s}}(a, b) .
\end{align}
\end{lemma}

\begin{IEEEproof}[Proof of \lemref{lem:quasi-conservation}]
\lemref{lem:quasi-conservation} straightforwardly follows from \lemref{lem:formulas}.
\end{IEEEproof}

The idea of \lemref{lem:quasi-conservation} comes from the conservation property $[I(W^{-}) + I(W^{+})]/2 = I(W)$; and note that in general, these quantities are not conserved on the polar transform.
In fact, \lemref{lem:quasi-conservation} looks like sub or super-martingales with inequalities
\begin{align}
\frac{ 1 }{ 2 } \Big[ \theta_{i, j}^{\bvec{s}-}(a, b) + \theta_{i, j}^{\bvec{s}+}(a, b) \Big]
& \ge
\theta_{i, j}^{\bvec{s}}(a, b) ,
\label{ineq:sub_theta} \\
\frac{ 1 }{ 2 } \Big[ \lambda_{i, j}^{\bvec{s}-}(a, b) + \lambda_{i, j}^{\bvec{s}+}(a, b) \Big]
& \le
\lambda_{i, j}^{\bvec{s}}(a, b) ,
\label{ineq:super_lambda} \\
\frac{ 1 }{ 2 } \Big[ \rho_{i, j}^{\bvec{s}-}(a, b) + \rho_{i, j}^{\bvec{s}+}(a, b) \Big]
& \le
\rho_{i, j}^{\bvec{s}}(a, b) ,
\label{ineq:super_rho} \\
\frac{ 1 }{ 2 } \Big[ \beta_{i, j}^{\bvec{s}-}(a, b) + \beta_{i, j}^{\bvec{s}+}(a, b) \Big]
& \ge
\beta_{i, j}^{\bvec{s}}(a, b)
\label{ineq:sub_beta}
\end{align}
when the sequence $\bvec{s} \in \{ -, + \}^{\ast}$ seems a uniformly distributed Bernoulli process, i.e., when $V^{\bvec{s}}$ is discussed as a polarization process.

The following lemma is a nice property between $\lambda_{i, j}^{\bvec{s}}(a, b)$ and $\rho_{i, j}^{\bvec{s}}(a, b)$; it shows that the inequality between $\lambda_{i, j}^{\bvec{s}}(a, b)$ and $\rho_{i, j}^{\bvec{s}}(a, b)$ is invariant under any polar transforms $\bvec{s} \in \{ -, + \}^{\ast}$.

\begin{lemma}
\label{lem:ineq}
For each $1 \le i < j \le m$ and $a, b \ge 1$, it holds that $\lambda_{i, j}^{\bvec{s}}(a, b) \le \rho_{i, j}^{\bvec{s}}(a, b)$ for every  $\bvec{s} \in \{ -, + \}^{\ast}$ if and only if $\lambda_{i, j}(a, b) \le \rho_{i, j}(a, b)$.
\end{lemma}

\begin{IEEEproof}[Proof of \lemref{lem:ineq}]
Let $1 \le i < j \le m$ and $a, b \ge 1$ be given.
By the symmetry $\lambda_{i, j}^{\bvec{s}}(a, b) = \rho_{j, i}^{\bvec{s}}(b, a)$, it suffices to prove the ``if part''.
We prove the lemma by induction.
If the sequence $\bvec{s}$ is empty, then the lemma is obvious.
Hence, it suffices to show that if $\lambda_{i, j}^{\bvec{s}}(a, b) \le \rho_{i, j}^{\bvec{s}}(a, b)$, then both $\lambda_{i, j}^{\bvec{s}-}(a, b) \le \rho_{i, j}^{\bvec{s}-}(a, b)$ and $\lambda_{i, j}^{\bvec{s}+}(a, b) \le \rho_{i, j}^{\bvec{s}+}(a, b)$ hold.
It follows from \lemref{lem:formulas} that
\begin{align}
\lambda_{i, j}^{\bvec{s}-}(a, b)
& =
\lambda_{i, j}^{\bvec{s}}(a, b) \big[ \lambda_{i, j}^{\bvec{s}}(a, b) + 2 \, \theta_{i, j}^{\bvec{s}}(a, b) \big]
\notag \\
& \overset{\mathclap{\text{(a)}}}{\le}
\rho_{i, j}^{\bvec{s}}(a, b) \big[ \rho_{i, j}^{\bvec{s}}(a, b) + 2 \, \theta_{i, j}^{\bvec{s}}(a, b) \big]
\notag \\
& =
\rho_{i, j}^{\bvec{s}-}(a, b) ,
\label{eq:minus_ineq}
\end{align}
where (a) follows by the hypothesis $\lambda_{i, j}^{\bvec{s}}(a, b) \le \rho_{i, j}^{\bvec{s}}(a, b)$.
Similar to \eqref{eq:minus_ineq}, we also have
\begin{align}
\lambda_{i, j}^{\bvec{s}+}(a, b)
& =
\lambda_{i, j}^{\bvec{s}}(a, b) \big[ \lambda_{i, j}^{\bvec{s}}(a, b) + 2 \, \beta_{i, j}^{\bvec{s}}(a, b) \big]
\notag \\
& \overset{\mathclap{\text{(b)}}}{\le}
\rho_{i, j}^{\bvec{s}}(a, b) \big[ \rho_{i, j}^{\bvec{s}}(a, b) + 2 \, \beta_{i, j}^{\bvec{s}}(a, b) \big]
\notag \\
& =
\rho_{i, j}^{\bvec{s}+}(a, b) ,
\label{eq:plus_ineq}
\end{align}
$\lambda_{i, j}^{\bvec{s}+}(a, b) \le \rho_{i, j}^{\bvec{s}+}(a, b)$.
This completes the proof of \lemref{lem:ineq}.
\end{IEEEproof}

\lemref{lem:ineq} also follows from \lemref{lem:formulas} by induction.
We now define the average value of \eqref{def:theta}--\eqref{def:beta} as follows:
\begin{align}
\mu_{i, j}^{(n)}[\theta](a, b)
& \coloneqq
\frac{ 1 }{ 2^{n} } \sum_{\bvec{s} \in \{ -, + \}^{n}} \theta_{i, j}^{\bvec{s}}(a, b) ,
\label{def:mu_theta}
\\
\mu_{i, j}^{(n)}[\lambda](a, b)
& \coloneqq
\frac{ 1 }{ 2^{n} } \sum_{\bvec{s} \in \{ -, + \}^{n}} \lambda_{i, j}^{\bvec{s}}(a, b) ,
\label{def:mu_lambda}
\\
\mu_{i, j}^{(n)}[\rho](a, b)
& \coloneqq
\frac{ 1 }{ 2^{n} } \sum_{\bvec{s} \in \{ -, + \}^{n}} \rho_{i, j}^{\bvec{s}}(a, b) ,
\label{def:mu_rho}
\\
\mu_{i, j}^{(n)}[\beta](a, b)
& \coloneqq
\frac{ 1 }{ 2^{n} } \sum_{\bvec{s} \in \{ -, + \}^{n}} \beta_{i, j}^{\bvec{s}}(a, b) .
\label{def:mu_beta}
\end{align}
For convenience, when $n = 0$, we write $\mu_{i, j}^{(0)}[\theta](a, b) \coloneqq \theta_{i, j}(a, b)$, $\mu_{i, j}^{(0)}[\lambda](a, b) \coloneqq \lambda_{i, j}(a, b)$, $\mu_{i, j}^{(0)}[\rho](a, b) \coloneqq \rho_{i, j}(a, b)$, and $\mu_{i, j}^{(0)}[\beta](a, b) \coloneqq \beta_{i, j}(a, b)$.
Then, the following lemma holds.

\begin{lemma}
\label{lem:martingale}
For any $n \ge 0$, $1 \le i < j \le m$, and $a, b \ge 1$, it holds that
\begin{align}
\mu_{i, j}^{(n)}[\lambda](a, b) - \mu_{i, j}^{(n)}[\rho](a, b)
& =
\lambda_{i, j}(a, b) - \rho_{i, j}(a, b) ,
\label{eq:lambda_minus_rho} \\
\mu_{i, j}^{(n)}[\theta](a, b) + \mu_{i, j}^{(n)}[\lambda](a, b)
& =
\theta_{i, j}(a, b)+ \lambda_{i, j}(a, b) ,
\label{eq:theta_plus_lambda} \\
\mu_{i, j}^{(n)}[\theta](a, b) + \mu_{i, j}^{(n)}[\rho](a, b)
& =
\theta_{i, j}(a, b) + \rho_{i, j}(a, b) ,
\label{eq:theta_plus_rho} \\
\mu_{i, j}^{(n)}[\beta](a, b) + \mu_{i, j}^{(n)}[\lambda](a, b)
& =
\beta_{i, j}(a, b) + \lambda_{i, j}(a, b) ,
\label{eq:beta_plus_lambda} \\
\mu_{i, j}^{(n)}[\beta](a, b) + \mu_{i, j}^{(n)}[\rho](a, b)
& =
\beta_{i, j}(a, b) + \rho_{i, j}(a, b) .
\label{eq:beta_plus_rho} 
\end{align}
\end{lemma}

\begin{IEEEproof}[Proof of \lemref{lem:martingale}]
Let $1 \le i < j \le m$ and $a, b \ge 1$ be given.
For each $n \in \mathbb{N}_{0}$, we have
\begin{align}
\mu_{i, j}^{(n+1)}[\lambda](a, b) - \mu_{i, j}^{(n+1)}[\rho](a, b)
& =
\frac{ 1 }{ 2^{n+1} } \sum_{\bvec{s} \in \{ -, + \}^{n}} \Big( \lambda_{i, j}^{\bvec{s}-}(a, b) + \lambda_{i, j}^{\bvec{s}+}(a, b) \Big) - \frac{ 1 }{ 2^{n+1} } \sum_{\bvec{s} \in \{ -, + \}^{n}} \Big( \rho_{i, j}^{\bvec{s}-}(a, b) + \rho_{i, j}^{\bvec{s}+}(a, b) \Big)
\notag \\
& \overset{\mathclap{\text{(a)}}}{=}
\frac{ 1 }{ 2^{n} } \sum_{\bvec{s} \in \{ -, + \}^{n}} \lambda_{i, j}^{\bvec{s}}(a, b) \big[ 1 - \rho_{i, j}^{\bvec{s}}(a, b) \big] - \frac{ 1 }{ 2^{n} } \sum_{\bvec{s} \in \{ -, + \}^{n}} \rho_{i, j}^{\bvec{s}}(a, b) \big[ 1 - \lambda_{i, j}^{\bvec{s}}(a, b) \big]
\notag \\
& =
\frac{ 1 }{ 2^{n} } \sum_{\bvec{s} \in \{ -, + \}^{n}} \lambda_{i, j}^{\bvec{s}}(a, b) - \frac{ 1 }{ 2^{n} } \sum_{\bvec{s} \in \{ -, + \}^{n}} \rho_{i, j}^{\bvec{s}}(a, b)
\notag \\
& =
\mu_{i, j}^{(n)}[\lambda](a, b) - \mu_{i, j}^{(n)}[\rho](a, b) ,
\label{eq:proof_martingale}
\end{align}
where (a) follows by \lemref{lem:quasi-conservation}.
This proves \eqref{eq:lambda_minus_rho} by induction.
The rest of equalities \eqref{eq:theta_plus_lambda}--\eqref{eq:beta_plus_rho} can be similarly proved by \lemref{lem:quasi-conservation}, as in \eqref{eq:proof_martingale}.
This completes the proof of \lemref{lem:martingale}.
\end{IEEEproof}

\lemref{lem:martingale} implies that the left-hand sides of \eqref{eq:lambda_minus_rho}--\eqref{eq:beta_plus_rho} has martingale-like properties with respect to a polarization process $V^{\bvec{s}}$ when $\bvec{s}$ seems a uniformly distributed Bernoulli process.
On the other hand, the following lemma follows from \eqref{ineq:sub_theta}--\eqref{ineq:sub_beta}.

\begin{lemma}
\label{lem:convergent}
The four sequences $( \mu_{i, j}^{(n)}[\theta](a, b) )_{n=1}^{\infty}$, $( \mu_{i, j}^{(n)}[\lambda](a, b) )_{n=1}^{\infty}$, $( \mu_{i, j}^{(n)}[\rho](a, b) )_{n=1}^{\infty}$, and $( \mu_{i, j}^{(n)}[\beta](a, b) )_{n=1}^{\infty}$ are convergent for each $1 \le i < j \le m$ and $a, b \ge 1$.
\end{lemma}

\begin{IEEEproof}[Proof of \lemref{lem:convergent}]
Let $1 \le i < j \le m$ and $a, b \ge 1$ be given.
It follows from \eqref{ineq:sub_theta}--\eqref{ineq:sub_beta} that
\begin{itemize}
\item
the number $\mu_{i, j}^{(n)}[\theta](a, b)$ is nondecreasing as $n$ increases;
\item
the number $\mu_{i, j}^{(n)}[\lambda](a, b)$ is nonincreasing as $n$ increases;
\item
the number $\mu_{i, j}^{(n)}[\rho](a, b)$ is nonincreasing as $n$ increases; and
\item
the number $\mu_{i, j}^{(n)}[\beta](a, b)$ is nondecreasing as $n$ increases.
\end{itemize}
Therefore, since these numbers are bounded as
\begin{align}
0
& \le
\mu_{i, j}^{(n)}[\theta](a, b)
\le
1 ,
\\
0
& \le
\mu_{i, j}^{(n)}[\lambda](a, b)
\le
1 ,
\\
0
& \le
\mu_{i, j}^{(n)}[\rho](a, b)
\le
1 ,
\\
0
& \le
\mu_{i, j}^{(n)}[\beta](a, b)
\le
1
\end{align}
for every $n \in \mathbb{N}_{0}$, we obtain the claim of \lemref{lem:convergent}.
\end{IEEEproof}

By \lemref{lem:convergent}, we can define the following limits:
\begin{align}
\mu_{i, j}^{(\infty)}[\theta](a, b)
& \coloneqq
\lim_{n \to \infty} \mu_{i, j}^{(n)}[\theta](a, b) ,
\\
\mu_{i, j}^{(\infty)}[\lambda](a, b)
& \coloneqq
\lim_{n \to \infty} \mu_{i, j}^{(n)}[\lambda](a, b) ,
\\
\mu_{i, j}^{(\infty)}[\rho](a, b)
& \coloneqq
\lim_{n \to \infty} \mu_{i, j}^{(n)}[\rho](a, b) ,
\\
\mu_{i, j}^{(\infty)}[\beta](a, b)
& \coloneqq
\lim_{n \to \infty} \mu_{i, j}^{(n)}[\beta](a, b) ,
\end{align}
The following theorem shows that these limits can be solved by the initial probability vector $( \varepsilon_{d} )_{d|q}$.

\begin{lemma}
\label{lem:fujisaki}
For any $1 \le i < j \le m$ and $a, b \ge 1$, it holds that
\begin{align}
\mu_{i, j}^{(\infty)}[\theta](a, b)
& =
\theta_{i, j}(a, b) + \min\{ \lambda_{i, j}(a, b), \rho_{i, j}(a, b) \} ,
\label{eq:theta_inf} \\
\mu_{i, j}^{(\infty)}[\lambda](a, b)
& =
\big| \lambda_{i, j}(a, b) - \rho_{i, j}(a, b) \big|^{+} ,
\label{eq:lambda_inf} \\
\mu_{i, j}^{(\infty)}[\rho](a, b)
& =
\big| \rho_{i, j}(a, b) - \lambda_{i, j}(a, b) \big|^{+} ,
\label{eq:rho_inf} \\
\mu_{i, j}^{(\infty)}[\beta](a, b)
& =
\beta_{i, j}(a, b) + \min\{ \lambda_{i, j}(a, b), \rho_{i, j}(a, b) \} ,
\label{eq:beta_inf} 
\end{align}
where $| c |^{+} \coloneqq \max\{ 0, c \}$ for $c \in \mathbb{R}$.
\end{lemma}

\begin{IEEEproof}[Proof of \lemref{lem:fujisaki}]
Let $1 \le i < j \le m$ and $a, b \ge 1$ be given.
Since $\lambda_{i, j}^{\bvec{s}}(a, b) = \rho_{j, i}^{\bvec{s}}(b, a)$, we may assume without loss of generality that $\lambda_{i, j}(a, b) \le \rho_{i, j}(a, b)$.
A simple calculation yields
\begin{align}
\mu_{i, j}^{(n+1)}[\lambda](a, b)
& =
\frac{ 1 }{ 2^{n} } \sum_{\bvec{s} \in \{ -, + \}^{n}} \frac{ 1 }{ 2 } \Big( \lambda_{i, j}^{\bvec{s}-}(a, b) + \lambda_{i, j}^{\bvec{s}+}(a, b) \Big)
\notag \\
& \overset{\mathclap{\text{(a)}}}{=}
\frac{ 1 }{ 2^{n} } \sum_{\bvec{s} \in \{ -, + \}^{n}} \lambda_{i, j}^{\bvec{s}}(a, b) \big[ 1 - \rho_{i, j}^{\bvec{s}}(a, b) \big]
\notag \\
& \overset{\mathclap{\text{(b)}}}{\le}
\frac{ 1 }{ 2^{n} } \sum_{\bvec{s} \in \{ -, + \}^{n}} \lambda_{i, j}^{\bvec{s}}(a, b) \big[1 - \lambda_{i, j}^{\bvec{s}}(a, b) \big]
\notag \\
& \overset{\mathclap{\text{(c)}}}{=}
\mu_{i, j}^{(n)}[\lambda](a, b) - \nu_{i, j}^{(n)}[\lambda](a, b) ,
\label{eq:lambda_mu_nu}
\end{align}
where (a) follows by \lemref{lem:quasi-conservation}, (b) follows by \lemref{lem:ineq}, and (c) follows by the definition of the second moment:
\begin{align}
\nu_{i, j}^{(n)}[\lambda](a, b)
\coloneqq
\frac{ 1 }{ 2^{n} } \sum_{\bvec{s} \in \{ -, + \}^{n}} \lambda_{i, j}^{\bvec{s}}(a, b)^{2} .
\end{align}
It follows from \eqref{eq:lambda_mu_nu} that
\begin{align}
0
\le
\nu_{i, j}^{(n)}[\lambda](a, b)
\le
\mu_{i, j}^{(n)}[\lambda](a, b) - \mu_{i, j}^{(n+1)}[\lambda](a, b) ,
\end{align}
and the squeeze theorem shows that $\nu_{i, j}^{(n)}[\lambda](a, b) \to 0$ as $n \to \infty$ (cf. \lemref{lem:convergent}).
On the other hand, we observe that
\begin{align}
\mu_{i, j}^{(n)}[\lambda](a, b)^{2}
& =
\Bigg[ \frac{ 1 }{ 2^{n} } \sum_{\bvec{s} \in \{ -, + \}^{n}} \lambda_{i, j}^{\bvec{s}}(a, b) \Bigg]^{2}
\notag \\
& =
\frac{ 1 }{ 2^{2n} } \sum_{\bvec{s}_{1} \in \{ -, + \}^{n}} \Bigg[ \lambda_{i, j}^{\bvec{s}_{1}}(a, b)^{2} + \sum_{\substack{ \bvec{s}_{2} \in \{ -, + \}^{n} : \\ \bvec{s}_{2} \neq \bvec{s}_{1} }}\lambda_{i, j}^{\bvec{s}_{1}}(a, b) \, \lambda_{i, j}^{\bvec{s}_{2}}(a, b) \Bigg]
\notag \\
& \le
\frac{ 1 }{ 2^{2n} } \sum_{\bvec{s}_{1} \in \{ -, + \}^{n}} \Bigg[ \lambda_{i, j}^{\bvec{s}_{1}}(a, b)^{2} + \sum_{\substack{ \bvec{s}_{2} \in \{ -, + \}^{n} : \\ \lambda_{i, j}^{\bvec{s}_{2}}(a, b) \ge \lambda_{i, j}^{\bvec{s}_{1}}(a, b) }} \lambda_{i, j}^{\bvec{s}_{2}}(a, b)^{2} + \sum_{\substack{ \bvec{s}_{3} \in \{ -, + \}^{n} : \\ \lambda_{i, j}^{\bvec{s}_{3}}(a, b) < \lambda_{i, j}^{\bvec{s}_{1}}(a, b) }} \lambda_{i, j}^{\bvec{s}_{1}}(a, b)^{2} \Bigg]
\notag \\
& \le
\frac{ 1 }{ 2^{2n} } \sum_{\bvec{s}_{1} \in \{ -, + \}^{n}} \Bigg[ \lambda_{i, j}^{\bvec{s}_{1}}(a, b)^{2} + \sum_{\bvec{s}_{2} \in \{ -, + \}^{n}} \lambda_{i, j}^{\bvec{s}_{2}}(a, b)^{2} + (2^{n}-1) \, \lambda_{i, j}^{\bvec{s}_{1}}(a, b)^{2} \Bigg]
\notag \\
& =
2 \, \nu_{i, j}^{(n)}[\lambda](a, b) ,
\end{align}
which implies that
\begin{align}
0
\le
\mu_{i, j}^{(n)}[\lambda](a, b)
\le
\sqrt{ 2 \, \nu_{i, j}^{(n)}[\lambda](a, b) } .
\label{ineq:Holder}
\end{align}
Note that the second inequality of \eqref{ineq:Holder} can be seen as a version of H\"{o}lder's inequality.
Then, it also follows by the squeeze theorem that $\mu_{i, j}^{(\infty)}[\lambda](a, b) = 0$, because $\nu_{i, j}^{(n)}[\lambda](a, b) \to 0$ as $n \to \infty$.
Hence, we have
\begin{align}
\mu_{i, j}^{(\infty)}[\rho](a, b)
& =
\mu_{i, j}^{(\infty)}[\rho](a, b) - \mu_{i, j}^{(\infty)}[\lambda](a, b)
\notag \\
& =
\lim_{n \to \infty} \Big( \mu_{i, j}^{(n)}[\rho](a, b) - \mu_{i, j}^{(n)}[\lambda](a, b) \Big)
\notag \\
& \overset{\mathclap{\text{(a)}}}{=}
\rho_{i, j}(a, b) - \lambda_{i, j}(a, b) ,
\label{eq:deriving_mu-rho-infty}
\\
\mu_{i, j}^{(\infty)}[\theta](a, b)
& =
\mu_{i, j}^{(\infty)}[\theta](a, b) + \mu_{i, j}^{(\infty)}[\lambda](a, b)
\notag \\
& =
\lim_{n \to \infty} \Big( \mu_{i, j}^{(n)}[\theta](a, b) + \mu_{i, j}^{(n)}[\lambda](a, b) \Big)
\notag \\
& \overset{\mathclap{\text{(b)}}}{=}
\theta_{i, j}(a, b) + \lambda_{i, j}(a, b) ,
\\
\mu_{i, j}^{(\infty)}[\beta](a, b)
& =
\mu_{i, j}^{(\infty)}[\beta](a, b) + \mu_{i, j}^{(\infty)}[\lambda](a, b)
\notag \\
& =
\lim_{n \to \infty} \Big( \mu_{i, j}^{(\infty)}[\beta](a, b) + \mu_{i, j}^{(n)}[\lambda](a, b) \Big)
\notag \\
& \overset{\mathclap{\text{(c)}}}{=}
\beta_{i, j}(a, b) + \lambda_{i, j}(a, b) ,
\end{align}
where (a)--(c) follow by \lemref{lem:martingale}.
Considering the counterpart hypothesis $\lambda_{i, j}(a, b) \ge \rho_{i, j}(a, b)$, we have \eqref{eq:theta_inf}--\eqref{eq:beta_inf}.
This completes the proof of \lemref{lem:fujisaki}.
\end{IEEEproof}

\begin{algorithm}[t]
\DontPrintSemicolon
\KwData{An initial probability vector $( \varepsilon_{d} )_{d|q}$}
\KwResult{The asymptotic distribution $( \mu_{d}^{(\infty)} )_{d|q}$}
	Initialize $( \mu_{d}^{(\infty)} )_{d|q}$ by the zero vector $(0, \dots, 0)$\;
	$\alpha \longleftarrow 0$\;
	$\bvec{t} = (t_{1}, \dots, t_{m}) \longleftarrow \bvec{0} = (0, \dots, 0)$\;
	\While{$0 \le \alpha < 1$}{
		$(i, j) \longleftarrow (1, 2)$\;
		\While{$j \le m$}{
			\uIf{$\lambda_{i, j}(t_{i}+1, t_{j}+1) \le \rho_{i, j}(t_{i}+1, t_{j}+1)$}{
			$k \longleftarrow j$ and $l \longleftarrow i$\;
			$i \longleftarrow i+1$ and $j \longleftarrow j + 1$\;
			}
			\Else{
			$k \longleftarrow i$ and $l \longleftarrow i$\;
			$j \longleftarrow j + 1$\;
			}
		}
		$\mu_{\langle \bvec{t} \rangle}^{(\infty)} \longleftarrow \beta_{l, m}(t_{l}+1, t_{m}+1) + \min\{ \lambda_{l, m}(t_{l}+1, t_{m}+1), \rho_{l, m}(t_{l}+1, t_{m}+1) \} - \alpha$\;
		$\alpha \longleftarrow \alpha + \mu_{\langle \bvec{t} \rangle}^{(\infty)}$\;
		$t_{k} \longleftarrow t_{k} + 1$\;
	}
\caption{Solving asymptotic distribution}
\label{alg:main}
\end{algorithm}

As in \eqref{def:mu_theta}--\eqref{def:mu_beta}, we define the average value of the recursive formula \eqref{def:eps_s} over all sequences $\bvec{s} \in \{ -, + \}^{n}$ of length $n$ as
\begin{align}
\mu_{d}^{(n)}
\coloneqq
\frac{ 1 }{ 2^{n} } \sum_{\bvec{s} \in \{ -, + \}^{n}} \varepsilon_{d}^{\bvec{s}}
\label{def:mu_d}
\end{align}
for each $d|q$ and $n \in \mathbb{N}$.
Similar to \eqref{def:mu_theta}--\eqref{def:mu_beta}, the value $\mu_{d}^{(n)}$ always has a limit%
\footnote{The existence of the limit directly follows from the proof of \thref{th:mu_d}.}
$\mu_{d}^{(\infty)} \coloneqq \lim_{n \to \infty} \mu_{d}^{(n)}$ for each $d|q$, and the asymptotic distribution $( \mu_{d}^{(\infty)} )_{d|q}$ can be algorithmically calculated as shown in the following theorem.

\begin{theorem}
\label{th:mu_d}
The probability vector $( \mu_{d}^{(\infty)} )_{d|q}$ can be calculated by Algorithm~\ref{alg:main} running in%
\footnote{Note that $\mathrm{O}( \cdot )$ is the Big-O notation, but $\omega( \cdot )$ and $\Omega( \cdot )$ are number theoretic notations, i.e., these are not the little-omega and Big-Omega notations, respectively, in this paper.}
$\mathrm{O}( \omega(q) \, \Omega(q) \, \tau( q ) )$, where $\omega( q ) \le m$ denotes the number of distinct prime factors of $q$; $\Omega(q) \coloneqq \sum_{i = 1}^{m} r_{i}$ denotes the number of prime factors of $q$ with multiplicity; and $\tau( q ) \coloneqq \prod_{i = 1}^{m} (r_{i}+1)$ denotes the number of positive divisors of $q$.
\end{theorem}

\begin{IEEEproof}[Proof of \thref{th:mu_d}]
Note that even if $\omega( q ) = 1$, Algorithm~\ref{alg:main} still works well by setting $m = 2$ and $r_{2} = 0$, i.e., the input alphabet is denoted by $q = p_{1}^{r_{1}} = p_{1}^{r_{1}} p_{2}^{0} = p_{1}^{r_{1}} p_{2}^{r_{2}}$.
In this case, note that $\lambda_{1, 2}(a, b) \ge 0$ but $\rho_{1, 2}(a, b) = 0$ for every $a, b \ge 1$.
If $\omega( q ) \ge 2$, then $m = \omega(q)$ is sufficient.

First, suppose that $\bvec{t}^{(0)} = (t_{1}^{(0)}, \dots, t_{m}^{(0)}) = (0, \dots, 0) = \bvec{0}$ as in Line 3 of Algorithm~\ref{alg:main}.
That is, consider the first step of the while loop in Lines 4--15 of Algorithm~\ref{alg:main}.
If $\lambda_{i, j}(t_{i}^{(0)}+1, t_{j}^{(0)}+1) = \lambda_{i, j}(1, 1) \le \rho_{i, j}(1, 1) = \rho_{i, j}(t_{i}^{(0)}+1, t_{j}^{(0)}+1)$ as in Line~7 of Algorithm~\ref{alg:main}, then it follows from \lemref{lem:fujisaki} that
\begin{align}
\mu_{i, j}^{(\infty)}[\lambda](t_{i}^{(0)}+1, t_{j}^{(0)}+1)
=
\mu_{i, j}^{(\infty)}[\lambda](1, 1)
=
0 .
\label{eq:alg1_lambda0}
\end{align}
Given that
\begin{align}
\mu_{i, j}^{(n)}[\lambda](a, b)
=
\sum_{\bvec{t} : \bvec{0} \le \bvec{t} \le \bvec{r}, t_{i} \ge a, t_{j} < b} \mu_{\langle \bvec{t} \rangle}^{(\infty)} ,
\end{align}
Equation \eqref{eq:alg1_lambda0} implies that
$
\mu_{\langle \bvec{t} \rangle}^{(\infty)}
=
0
$
for every $\bvec{0} \le \bvec{t} \le \bvec{r}$ satisfying $0 = t_{i}^{(0)} < t_{i} \le r_{i}$ and $t_{j} \le t_{j}^{(0)} = 0$.
Similarly, we observe that if $\lambda_{i, j}(t_{i}^{(0)}+1, t_{j}^{(0)}+1) = \lambda_{i, j}(1, 1) < \rho_{i, j}(1, 1) = \rho_{i, j}(t_{i}^{(0)}+1, t_{j}^{(0)}+1)$ as in Line~10 of Algorithm~\ref{alg:main}, then
$
\mu_{\langle \bvec{t} \rangle}^{(\infty)}
=
0
$
for every $\bvec{0} \le \bvec{t} \le \bvec{r}$ satisfying $t_{i} \le t_{i}^{(0)} = 0$ and $0 = t_{j}^{(0)} < t_{j} \le r_{j}$.
Therefore, by the while loop in Lines 5--12 of Algorithm~\ref{alg:main}, one can get the number $k$ such that for each $1 \le k^{\prime} \le m$ satisfying $k^{\prime} \neq k$, it holds that
$
\mu_{\langle \bvec{t} \rangle}^{(\infty)}
=
0
$
for every $\bvec{0} \le \bvec{t} \le \bvec{r}$ satisfying $t_{k} \le t_{k}^{(0)} = 0$ and $0 = t_{k^{\prime}}^{(0)} < t_{k^{\prime}} \le r_{k^{\prime}}$.
Note that $l = k$ if $k < m$; and $l = m$ if $k < m$.
Given that
\begin{align}
\mu_{i, j}^{(n)}[\beta](a, b)
=
\sum_{\bvec{t} : \bvec{0} \le \bvec{t} \le \bvec{r}, t_{i} < a, t_{j} < b} \mu_{\langle \bvec{t} \rangle}^{(n)} ,
\label{eq:mu_beta_eps}
\end{align}
we have
\begin{align}
\mu_{\langle \bvec{t}^{(0)} \rangle}^{(\infty)}
& =
\mu_{l, m}^{(n)}[\beta](t_{l}^{(0)}+1, t_{l}^{(0)}+1)
=
\mu_{l, m}^{(n)}[\beta](1, 1) ,
\label{eq:sol_alg_first1} \\
\mu_{\langle \bvec{t} \rangle}^{(\infty)}
& =
0
\qquad \mathrm{for} \ \mathrm{every} \ \bvec{0} \le \bvec{t} \le \bvec{r} \ \mathrm{satisfying} \ t_{k} = 0 \ \mathrm{and} \ 1 \le t_{k^{\prime}} \le r_{k^{\prime}} \ \mathrm{for} \ \mathrm{each} \ k^{\prime} \neq k .
\label{eq:sol_alg_first2} 
\end{align}
Note that it follows from \lemref{lem:fujisaki} that
\begin{align}
\mu_{l, m}^{(n)}[\beta](t_{l}^{(0)}+1, t_{l}^{(0)}+1)
=
\beta_{l, m}(t_{l}^{(0)}+1, t_{l}^{(0)}+1) + \min\{ \lambda_{l, m}(t_{l}^{(0)}+1, t_{l}^{(0)}+1), \rho_{i, j}(t_{l}^{(0)}+1, t_{l}^{(0)}+1) \} .
\end{align}
Therefore, by the first step $\bvec{t}^{(0)} = \bvec{0}$ of the while loop in Lines 4--15 of Algorithm~\ref{alg:main}, one can obtain $\mu_{\langle \bvec{t} \rangle}^{(\infty)}$ for every $\bvec{0} \le \bvec{t} \le \bvec{r}$ satisfying $t_{k} \le t_{k}^{(0)} = 0$, as in \eqref{eq:sol_alg_first1} and \eqref{eq:sol_alg_first2}.
To continue the while loop, after $\bvec{t}^{(1)} = (t_{1}^{(1)}, \dots, t_{m}^{(1)})$ is created in Line~15 of Algorithm~\ref{alg:main} as
\begin{align}
t_{k^{\prime}}^{(1)}
=
\begin{cases}
t_{k^{\prime}}^{(0)}
& \mathrm{if} \ k^{\prime} \neq k ,
\\
t_{k^{\prime}}^{(0)} + 1
& \mathrm{if} \ k^{\prime} = k ,
\end{cases}
\end{align}
for each $1 \le k^{\prime} \le m$, we go back to Line~4 of Algorithm~\ref{alg:main} whenever $0 \le \alpha < 1$.
The case $\alpha = 1$ occurs if and only if $\beta_{l, m}(1, 1) = 1$.
In this case, we have $\mu_{\langle \bvec{t}^{(0)} \rangle}^{(\infty)} = 1$ and $\mu_{\langle \bvec{t} \rangle}^{(\infty)} = 0$ for every $\bvec{t} \neq \bvec{t}^{(0)}$; and we just finish the algorithm.

Second, suppose that for some $\bvec{0} \le \bvec{t}^{(h)} = (t_{1}^{(h)}, \dots, t_{m}^{(h)}) \le \bvec{r}$ with $0 \le h \le \Omega( q )$, the value $\mu_{\langle \bvec{t} \rangle}^{(\infty)}$ has been already solved for every $\bvec{0} \le \bvec{t} \le \bvec{r}$ satisfying $t_{\bar{k}} < t_{\bar{k}}^{(h)}$ for some $1 \le \bar{k} \le m$.
That is, consider the $h$th-step of the while loop in Lines 4--15 of Algorithm~\ref{alg:main}.
By Lines 2 and 14 of Algorithm~\ref{alg:main}, it follows that
\begin{align}
\alpha
& =
\sum_{\substack{ \bvec{t} : \bvec{0} \le \bvec{t} \le \bvec{r} , \\ t_{\bar{k}} < t_{\bar{k}}^{(h)} \, \text{for some} \, 1 \le \bar{k} \le m }} \mu_{\langle \bvec{t} \rangle}^{(\infty)}
\notag \\
& =
\sum_{\substack{ \bvec{t} : \bvec{0} \le \bvec{t} \le \bvec{r} , \\ t_{\bar{k}} < t_{\bar{k}}^{(h)} \, \text{for some} \, 1 \le \bar{k} \le m }} \varepsilon_{\langle \bvec{t} \rangle} .
\end{align}
Similar to the previous paragraph, by the while loop in Line 6--12 of Algorithm~\ref{alg:main}, one can obtain the integer $k$ such that for each $1 \le k^{\prime} \le m$ satisfying $k^{\prime} \neq k$, it holds that
$
\mu_{\langle \bvec{t} \rangle}^{(\infty)}
=
0
$
for every $\bvec{0} \le \bvec{t} \le \bvec{r}$ satisfying $t_{k} \le t_{k}^{(h)}$ and $t_{k^{\prime}}^{(h)} < t_{k^{\prime}} \le r_{k^{\prime}}$.
Note that $l = k$ if $k < m$; and $l = m$ if $k < m$.
Therefore, given that \eqref{eq:mu_beta_eps}, we have
\begin{align}
\mu_{\langle \bvec{t}^{(h)} \rangle}^{(\infty)}
& =
\mu_{l, m}^{(n)}[\beta](t_{l}^{(h)}+1, t_{m}^{(h)}+1) - \alpha ,
\\
\mu_{\langle \bvec{t} \rangle}^{(\infty)}
& =
0
\qquad \mathrm{for} \ \mathrm{every} \ \bvec{0} \le \bvec{t} \le \bvec{r} \ \mathrm{satisfying} \ t_{k} \le t_{k}^{(h)} \ \mathrm{and} \ t_{k^{\prime}}^{(h)} < u_{k^{\prime}} \le r_{k^{\prime}} \ \mathrm{for} \ \mathrm{each} \ k^{\prime} \neq k .
\end{align}
Note that it follows from \lemref{lem:fujisaki} that
\begin{align}
\mu_{l, m}^{(n)}[\beta](t_{l}^{(h)}+1, t_{l}^{(h)}+1)
=
\beta_{l, m}(t_{l}^{(h)}+1, t_{l}^{(h)}+1) + \min\{ \lambda_{l, m}(t_{l}^{(h)}+1, t_{l}^{(h)}+1), \rho_{i, j}(t_{l}^{(h)}+1, t_{l}^{(h)}+1) \} .
\end{align}
Then, by setting $\bvec{0} \le \bvec{t}^{(h+1)} \le \bvec{r}$ as
\begin{align}
t_{k^{\prime}}^{(h+1)}
=
\begin{cases}
t_{k^{\prime}}^{(h)}
& \mathrm{if} \ k^{\prime} \neq k ,
\\
t_{k^{\prime}}^{(h)} + 1
& \mathrm{if} \ k^{\prime} = k
\end{cases}
\label{eq:t_next}
\end{align}
for each $1 \le k^{\prime} \le m$, we observe that $\mu_{\langle \bvec{t} \rangle}^{(\infty)}$ has been solved for every $\bvec{0} \le \bvec{t} \le \bvec{r}$ satisfying $t_{\bar{k}} < t_{\bar{k}}^{(h+1)}$ for some $1 \le \bar{k} \le m$.
Note that \eqref{eq:t_next} is done in Line~15 of Algorithm~\ref{alg:main}.
If $0 \le \alpha < 1$, then
\begin{align}
0
\le
\sum_{\substack{ \bvec{t} : \bvec{0} \le \bvec{t} \le \bvec{r} , \\ t_{\bar{k}} < t_{\bar{k}}^{(h+1)} \, \text{for some} \, 1 \le \bar{k} \le m }} \mu_{\langle \bvec{t} \rangle}^{(\infty)}
<
1 ,
\end{align}
and we go back to Line~4 of Algorithm~\ref{alg:main}.
Note that $\bvec{0} \le \bvec{t}^{(h+1)} \le \bvec{r}$, provided that $\alpha < 1$.
On the other hand, if $\alpha = 1$, then
\begin{align}
\sum_{\substack{ \bvec{t} : \bvec{0} \le \bvec{t} \le \bvec{r} , \\ t_{\bar{k}} < t_{\bar{k}}^{(h+1)} \, \text{for some} \, 1 \le \bar{k} \le m }} \mu_{\langle \bvec{t} \rangle}^{(\infty)}
=
1 
\end{align}
which implies that the asymptotic distribution $( \mu_{d}^{(\infty)} )_{d|q} = ( \mu_{\langle \bvec{t} \rangle}^{(\infty)} )_{\bvec{0} \le \bvec{t} \le \bvec{r}}$ is solved.
Note that if $h = \Omega( q )$, i.e., if $\bvec{t}^{(h)} = \bvec{r}$, then $\alpha = 1$ always holds after Line~14 of Algorithm~\ref{alg:main}.

Finally, we verify the computational complexity of Algorithm~\ref{alg:main}.
By Line~15 of Algorithm~\ref{alg:main}, the while loop in Lines~4--15 of Algorithm~\ref{alg:main} is repeated at most $\Omega( q ) = r_{1} + r_{2} + \dots r_{m} + 1$ times.
The while loop in Lines~6--12 of Algorithm~\ref{alg:main} is repeated at $m-1$ times.
In Line~7 of Algorithm~\ref{alg:main}, both $\lambda_{i, j}(t_{i} + 1, t_{j} + 1)$ and $\lambda_{i, j}(t_{i} + 1, t_{j} + 1)$ can be calculated by a given initial probability vector $( \varepsilon_{d} )_{d|q}$ at most $\tau( q ) = (r_{1}+1) (r_{2}+1) \cdots (r_{m}+1)$ times addition.
Similarly, in Line~13 of Algorithm~\ref{alg:main}, the values $\beta_{l, m}(t_{l}+1, t_{m}+1)$, $\lambda_{l, m}(t_{l}+1, t_{m}+1)$, and $\lambda_{l, m}(t_{l}+1, t_{m}+1)$ can also be calculated by a given initial probability vector $( \varepsilon_{d} )_{d|q}$ at most $\tau( q ) = (r_{1}+1) (r_{2}+1) \cdots (r_{m}+1)$ times addition.
Therefore, we conclude that Algorithm~\ref{alg:main} runs in $\mathrm{O}( \omega( q ) \, \Omega( q ) \, \tau( q ) )$.
Note that calculations in Algorithm~\ref{alg:main} are only addition and subtraction, i.e., there is nether multiplication nor division.
This complete the proof of \thref{th:mu_d}.
\end{IEEEproof}

By \thref{th:mu_d}, we can immediately observe the following corollary.

\begin{corollary}
\label{cor:mu_d}
The asymptotic distribution $( \mu_{d}^{(\infty)} )_{d|q}$ has at most $\Omega( q ) + 1$ positive probability masses.
\end{corollary}

An example of Algorithm~\ref{alg:main} is as follows:

\begin{example}
\label{ex:mu_d}
Consider an erasure channel $V$ defined in \defref{def:V} with an initial probability vector $(\varepsilon_{d})_{d|q}$ as follows:
The input alphabet size is $q = 4500 = 2^{2} \cdot 3^{2} \cdot 5^{3}$, where note that the set of positive divisors $d$ of $q$ is $\{ 1, \allowbreak 2, \allowbreak 3, \allowbreak 4, \allowbreak 5, \allowbreak 6, \allowbreak 9, \allowbreak 10, \allowbreak 12, \allowbreak 15, \allowbreak 18, \allowbreak 20, \allowbreak 25, \allowbreak 30, \allowbreak 36, \allowbreak 45, \allowbreak 50, \allowbreak 60, \allowbreak 75, \allowbreak 90, \allowbreak 100, \allowbreak 125, \allowbreak 150, \allowbreak 180, \allowbreak 225, \allowbreak 250, \allowbreak 300, \allowbreak 375, \allowbreak 450, \allowbreak 500, \allowbreak 750, \allowbreak 900, \allowbreak 1125, \allowbreak 1500, \allowbreak 2250, \allowbreak 4500 \}$.
The initial probability vector $(\varepsilon_{d})_{d|q}$ is given by%
\footnote{The elements $\varepsilon_{d}$ of $(\varepsilon_{d})_{d|q}$ are sorted in increasing order of indices $d$.}
$(\varepsilon_{d})_{d|q} = (1/150) \times (0, \allowbreak 1, \allowbreak 2, \allowbreak 3, \allowbreak 4, \allowbreak 5, \allowbreak 6, \allowbreak 7, \allowbreak 8, \allowbreak 9, \allowbreak 0, \allowbreak 1, \allowbreak 2, \allowbreak 3, \allowbreak 4, \allowbreak 5, \allowbreak 6, \allowbreak 7, \allowbreak 8, \allowbreak 9, \allowbreak 0, \allowbreak 1, \allowbreak 2, \allowbreak 3, \allowbreak 4, \allowbreak 5, \allowbreak 6, \allowbreak 7, \allowbreak 8, \allowbreak 9, \allowbreak 0, \allowbreak 1, \allowbreak 2, \allowbreak 3, \allowbreak 4, \allowbreak 5)$.
Then, Algorithm~\ref{alg:main} solves the asymptotic distribution $(\mu_{d}^{(\infty)})_{d|q} = (29/150, \allowbreak 0, \allowbreak 0, \allowbreak 0, \allowbreak 1/15, \allowbreak 0, \allowbreak 0, \allowbreak 0, \allowbreak 0, \allowbreak 11/150, \allowbreak 0, \allowbreak 0, \allowbreak 0, \allowbreak 9/50, \allowbreak 0, \allowbreak 0, \allowbreak 0, \allowbreak 0, \allowbreak 0, \allowbreak 0, \allowbreak 0, \allowbreak 0, \allowbreak 11/75, \allowbreak 0, \allowbreak 0, \allowbreak 0, \allowbreak 0, \allowbreak 0, \allowbreak 1/150, \allowbreak 0, \allowbreak 0, \allowbreak 7/75, \allowbreak 0, \allowbreak 0, \allowbreak 0, \allowbreak 6/25)$.
We summarize this result in \tabref{table:mu_d}.
The calculation process of Algorithm~\ref{alg:main} is shown in Appendix.
\end{example}

\begin{table}[!t]
\caption{Example of Algorithm~\ref{alg:main} with the setting of \exref{ex:mu_d}. The input alphabet size is $q = 4500 = 2^{2} \cdot 3^{2} \cdot 5^{3}$.
An initial probability vector $( \varepsilon_{d} )_{d|q}$ and its resultant asymptotic distribution $( \mu_{d}^{(\infty)} )_{d|q}$ are summarized in the table.}
\label{table:mu_d}
\hspace{-5em}
\begin{tabular}{rrrrrrrrrrrrrrrrrrr}
\toprule
divisor $d$ & $1$ & $2$ & $3$ & $4$ & $5$ & $6$ & $9$ & $10$ & $12$ & $15$ & $18$ & $20$ & $25$ & $30$ & $36$ & $45$ & $50$ & $60$ \\
\midrule
$( \varepsilon_{d} )_{d|q}$ & $0$ & $\mathlarger{\mathlarger{\sfrac{1}{150}}}$ & $\mathlarger{\mathlarger{\sfrac{2}{150}}}$ & $\mathlarger{\mathlarger{\sfrac{3}{150}}}$ & $\mathlarger{\mathlarger{\sfrac{4}{150}}}$ & $\mathlarger{\mathlarger{\sfrac{5}{150}}}$ & $\mathlarger{\mathlarger{\sfrac{6}{150}}}$ & $\mathlarger{\mathlarger{\sfrac{7}{150}}}$ & $\mathlarger{\mathlarger{\sfrac{8}{150}}}$ & $\mathlarger{\mathlarger{\sfrac{9}{150}}}$ & $0$ & $\mathlarger{\mathlarger{\sfrac{1}{150}}}$ & $\mathlarger{\mathlarger{\sfrac{2}{150}}}$ & $\mathlarger{\mathlarger{\sfrac{3}{150}}}$ & $\mathlarger{\mathlarger{\sfrac{4}{150}}}$ & $\mathlarger{\mathlarger{\sfrac{5}{150}}}$ & $\mathlarger{\mathlarger{\sfrac{6}{150}}}$ & $\mathlarger{\mathlarger{\sfrac{7}{150}}}$ \\
$( \mu_{d}^{(\infty)} )_{d|q}$ & $\mathlarger{\mathlarger{\sfrac{29}{150}}}$ & $0$ & $0$ & $0$ & $\mathlarger{\mathlarger{\sfrac{1}{15}}}$ & $0$ & $0$ & $0$ & $0$ & $\mathlarger{\mathlarger{\sfrac{11}{150}}}$ & $0$ & $0$ & $0$ & $\mathlarger{\mathlarger{\sfrac{9}{50}}}$ & $0$ & $0$ & $0$ & $0$ \\
\bottomrule
\toprule
divisor $d$ & $75$ & $90$ & $100$ & $125$ & $150$ & $180$ & $225$ & $250$ & $300$ & $375$ & $450$ & $500$ & $750$ & $900$ & $1125$ & $1500$ & $2250$ & $4500$ \\
\midrule
$( \varepsilon_{d} )_{d|q}$ & $\mathlarger{\mathlarger{\sfrac{8}{150}}}$ & $\mathlarger{\mathlarger{\sfrac{9}{150}}}$ & $0$ & $\mathlarger{\mathlarger{\sfrac{1}{150}}}$ & $\mathlarger{\mathlarger{\sfrac{2}{150}}}$ & $\mathlarger{\mathlarger{\sfrac{3}{150}}}$ & $\mathlarger{\mathlarger{\sfrac{4}{150}}}$ & $\mathlarger{\mathlarger{\sfrac{5}{150}}}$ & $\mathlarger{\mathlarger{\sfrac{6}{150}}}$ & $\mathlarger{\mathlarger{\sfrac{7}{150}}}$ & $\mathlarger{\mathlarger{\sfrac{8}{150}}}$ & $\mathlarger{\mathlarger{\sfrac{9}{150}}}$ & 0 & $\mathlarger{\mathlarger{\sfrac{1}{150}}}$ & $\mathlarger{\mathlarger{\sfrac{2}{150}}}$ & $\mathlarger{\mathlarger{\sfrac{3}{150}}}$ & $\mathlarger{\mathlarger{\sfrac{4}{150}}}$ & $\mathlarger{\mathlarger{\sfrac{5}{150}}}$ \\
$( \mu_{d}^{(\infty)} )_{d|q}$ & $0$ & $0$ & $0$ & $0$ & $\mathlarger{\mathlarger{\sfrac{11}{75}}}$ & $0$ & $0$ & $0$ & $0$ & $0$ & $\mathlarger{\mathlarger{\sfrac{1}{150}}}$ & $0$ & $0$ & $\mathlarger{\mathlarger{\sfrac{7}{75}}}$ & $0$ & $0$ & $0$ & $\mathlarger{\mathlarger{\sfrac{6}{25}}}$ \\
\bottomrule
\end{tabular}
\end{table}

The following theorem shows that $( \varepsilon_{d}^{\bvec{s}} )_{d|q}$ tends to a unit vector $(0, \dots, 1, \dots, 0)$ for almost all polarization process $\bvec{s} \in \{ s_{1}, s_{1}s_{2}, s_{1}s_{2}s_{3}, \dots \}$, and limiting proportions of them are exactly characterized by the asymptotic distribution $( \mu_{d}^{(\infty)} )_{d|q}$.

\begin{theorem}
\label{th:polarization}
For any fixed $\delta \in (0, 1)$, it holds that
\begin{align}
\lim_{n \to \infty} \frac{ 1 }{ 2^{n} } \Big| \Big\{ \bvec{s} \in \{ -, + \}^{n} \ \Big| \ \delta \le \varepsilon_{d}^{\bvec{s}} \le 1 - \delta \Big\} \Big|
& =
0 ,
\label{eq:proportion0} \\
\lim_{n \to \infty} \frac{ 1 }{ 2^{n} } \Big| \Big\{ \bvec{s} \in \{ -, + \}^{n} \ \Big| \ \varepsilon_{d}^{\bvec{s}} > 1 - \delta \Big\} \Big|
& =
\mu_{d}^{(\infty)}
\label{eq:proportion1}
\end{align}
for every $d|q$, where $(\mu_{d}^{(\infty)})_{d|q}$ can be calculated by Algorithm~\ref{alg:main} (cf. \thref{th:mu_d}).
\end{theorem}

To prove \thref{th:polarization}, we give the following simple and useful lemma.

\begin{lemma}
\label{lem:additive}
For each $n \in \mathbb{N}$, let a nonempty collection $\mathcal{F}_{n}$ of subsets of a set be a field%
\footnote{%
Note that this field $\mathcal{F}_{n}$ is a measure theoretic notion satisfying $A^{\complement} \in \mathcal{F}_{n}$ if $A \in \mathcal{F}_{n}$; and $A \cup B \in \mathcal{F}_{n}$ if $A, B \in \mathcal{F}_{n}$, where $A^{\complement}$ denotes the complement of a set $A$.
},
and let $f_{n} : \mathcal{F}_{n} \to [0, 1]$ be an additive set function.
For each $i \in \mathbb{N}$, let $( S_{i, n} )_{n}$ be a sequence of sets such that $S_{i, n} \in \mathcal{F}_{n}$ for every $n \in \mathbb{N}$ and $f_{n}( S_{i, n} ) \to 1$ as $n \to \infty$.
Then, it holds that
\begin{align}
\lim_{n \to \infty} f_{n} \bigg( \bigcap_{i=1}^{k} S_{i, n} \bigg)
=
1
\quad
\mathrm{for} \ k \in \mathbb{N} .
\end{align}
\end{lemma}

\begin{IEEEproof}[Proof of \lemref{lem:additive}]
We prove \lemref{lem:additive} by induction.
Define
\begin{align}
S_{n}^{(k)}
\coloneqq
\bigcap_{i = 1}^{k} S_{i, n}
\end{align}
for each $k, n \in \mathbb{N}$.
By hypothesis, it is clear that
\begin{align}
\lim_{n \to \infty} f_{n}\big( S_{n}^{(1)} \big)
=
\lim_{n \to \infty} f_{n}\big( S_{1, n} \big)
=
1 .
\end{align}
Suppose that
\begin{align}
\lim_{n \to \infty} f_{n}\big( S_{n}^{(k-1)} \big)
=
1 .
\end{align}
for a fixed integer $k \in \mathbb{N}$.
Then, we have
\begin{align}
1
& =
\lim_{n \to \infty} f_{n}\big( S_{n}^{(k-1)} \big)
\notag \\
& \ge
\liminf_{n \to \infty} f_{n}\big( S_{n}^{(k)} \big)
\notag \\
& =
\liminf_{n \to \infty} \Big( f_{n}\big( S_{n}^{(k-1)} \big) + f_{n}\big( S_{k, n} \big) - f_{n}\big( S_{n}^{(k-1)} \cup S_{k, n} \big) \Big)
\notag \\
& \ge
\liminf_{n \to \infty} f_{n}\big( S_{n}^{(k-1)} \big) + \liminf_{n \to \infty} f_{n}\big( S_{k, n} \big) - \limsup_{n \to \infty} f_{n}\big( S_{n}^{(k-1)} \cup S_{k, n} \big)
\notag \\
& \ge
1 + 1 - 1
\notag \\
& =
1 ,
\end{align}
which implies that
\begin{align}
\lim_{n \to \infty} f_{n}\big( S_{n}^{(k)} \big)
=
1 .
\end{align}
This completes the proof of \lemref{lem:additive}.
\end{IEEEproof}

\begin{IEEEproof}[Proof of \thref{th:polarization}]
This proof is inspired by Alsan and Telatar's simple proof of polarization \cite[Theorem~1]{alsan3}.
Let $1 \le i < j \le m$ and $a, b \ge 1$ be given.
Define
\begin{align}
\nu_{i, j}^{(n)}[\theta](a, b)
\coloneqq
\frac{ 1 }{ 2^{n} } \sum_{\bvec{s} \in \{ -, + \}^{n}} \theta_{i, j}^{\bvec{s}}(a, b)^{2}
\end{align}
for each $n \in \mathbb{N}$.
Then, we have that for a fixed $\delta \in (0, 1)$,
\begin{align}
\nu_{i, j}^{(n+1)}[\theta](a, b)
& =
\frac{ 1 }{ 2^{n} } \sum_{\bvec{s} \in \{ -, + \}^{n}} \frac{ 1 }{ 2 } \Big[ \theta_{i, j}^{\bvec{s}-}(a, b)^{2} + \theta_{i, j}^{\bvec{s}+}(a, b)^{2} \Big]
\notag \\
& \overset{\mathclap{\text{(a)}}}{=}
\frac{ 1 }{ 2^{n} } \sum_{\bvec{s} \in \{ -, + \}^{n}} \Bigg[ \bigg( \frac{ 1 }{ 2 } \Big( \theta_{i, j}^{\bvec{s}-}(a, b) + \theta_{i, j}^{\bvec{s}+}(a, b) \Big) \bigg)^{2} + \bigg( \frac{ 1 }{ 2 } \Big(\theta_{i, j}^{\bvec{s}-}(a, b) - \theta_{i, j}^{\bvec{s}+}(a, b) \Big) \bigg)^{2} \Bigg]
\notag \\
& \overset{\mathclap{\text{(b)}}}{=}
\frac{ 1 }{ 2^{n} } \sum_{\bvec{s} \in \{ -, + \}^{n}} \Bigg[ \Big( \theta_{i, j}^{\bvec{s}}(a, b) + \lambda_{i, j}^{\bvec{s}}(a, b) \, \rho_{i, j}^{\bvec{s}}(a, b) \Big)^{2} + \Big( \theta_{i, j}^{\bvec{s}}(a, b) \, \big[ 1 - \theta_{i, j}^{\bvec{s}}(a, b) \big] + \lambda_{i, j}^{\bvec{s}}(a, b) \, \rho_{i, j}^{\bvec{s}}(a, b) \Big)^{2} \Bigg]
\notag \\
& \ge
\frac{ 1 }{ 2^{n} } \sum_{\bvec{s} \in \{ -, + \}^{n}} \Big[ \theta_{i, j}^{\bvec{s}}(a, b)^{2} + \theta_{i, j}^{\bvec{s}}(a, b)^{2} \big[1 - \theta_{i, j}^{\bvec{s}}(a, b) \big]^{2} \Big]
\notag \\
& \ge
\nu_{i, j}^{(n)}[\theta](a, b) + \frac{ 1 }{ 2^{n} } \sum_{\substack{ \bvec{s} \in \{ -, + \}^{n} : \\ \delta \le \theta_{i, j}^{\bvec{s}}(a, b) \le 1 - \delta }} \theta_{i, j}^{\bvec{s}}(a, b)^{2} \big[1 - \theta_{i, j}^{\bvec{s}}(a, b) \big]^{2}
\notag \\
& \ge
\nu_{i, j}^{(n)}[\theta](a, b) + \frac{ 1 }{ 2^{n} } \sum_{\substack{ \bvec{s} \in \{ -, + \}^{n} : \\ \delta \le \theta_{i, j}^{\bvec{s}}(a, b) \le 1 - \delta }} \delta^{2} (1 - \delta)^{2} ,
\label{eq:nu_theta}
\end{align}
where (a) follows from the identity
\begin{align}
\frac{ x^{2} + y^{2} }{ 2 } = \Big( \frac{ x + y }{ 2 } \Big)^{2} + \Big( \frac{ x - y }{ 2 } \Big)^{2} ,
\end{align}
and (b) follows by \lemref{lem:formulas}.
This implies that the sequence $\big( \nu_{i, j}^{(n)}[\theta](a, b) \big)_{n=1}^{\infty}$ is nondecreasing.
As $\nu_{i, j}^{(n)}[\theta](a, b) \le 1$ for every $n \in \mathbb{N}$, the sequence $\big( \nu_{i, j}^{(n)}[\theta](a, b) \big)_{n=1}^{\infty}$ is convergent; thus, it holds that $\nu_{i, j}^{(n+1)}[\theta](a, b) - \nu_{i, j}^{(n)}[\theta](a, b) \to 0$ as $n \to \infty$.
We get from \eqref{eq:nu_theta} that
\begin{align}
0
\le
\frac{ 1 }{ 2^{n} } \Big| \Big\{ \bvec{s} \in \{ -, + \}^{n} \ \Big| \ \delta \le \theta_{i, j}^{\bvec{s}}(a, b) \le 1 - \delta \Big\} \Big|
\le
\frac{ \nu_{i, j}^{(n+1)}[\theta](a, b) - \nu_{i, j}^{(n)}[\theta](a, b) }{ \delta^{2} (1 - \delta)^{2} } .
\end{align}
As $\delta \in (0, 1)$ is a fixed number that does not depend on $n \in \mathbb{N}$, this implies that
\begin{align}
\lim_{n \to \infty} \frac{ 1 }{ 2^{n} } \Big| \Big\{ \bvec{s} \in \{ -, + \}^{n} \ \Big| \ \delta \le \theta_{i, j}^{\bvec{s}}(a, b) \le 1 - \delta \Big\} \Big|
=
0 .
\label{eq:theta_limit}
\end{align}

We now prove \eqref{eq:proportion0} by induction.
By \thref{th:mu_d}, it follows from Algorithm~\ref{alg:main} that there exists a sequence $( \bvec{t}^{(h)} = (t_{1}^{(h)}, \dots, t_{m}^{(h)}) )_{h = 0}^{\Omega( q )}$ satisfying the following:
(i) $\bvec{t}^{(0)} = \bvec{0} = (0, \dots, 0)$ and $\bvec{t}^{(\Omega( q ))} = \bvec{r} = (r_{1}, \dots, r_{m})$;
(ii) for any $0 \le h < \Omega( q )$, there exists $1 \le k \le m$ such that
\begin{align}
t_{k^{\prime}}^{(h+1)}
=
\begin{cases}
t_{k^{\prime}}^{(h)}
& \mathrm{if} \ k^{\prime} \neq k ,
\\
t_{k^{\prime}}^{(h)} + 1
& \mathrm{if} \ k^{\prime} = k
\end{cases}
\label{eq:kprime_plus1}
\end{align}
for each $1 \le k^{\prime} \le m$;
and (iii) $\mu_{\langle \bvec{t} \rangle}^{(\infty)} = 0$ if $\bvec{t} \neq \bvec{t}^{(h)}$ for every $0 \le h \le \Omega( q )$ (see also \corref{cor:mu_d}).
If $\mu_{d}^{(\infty)} = 0$, then we observe that for a fixed $\delta \in (0, 1)$,
\begin{align}
0
& =
\mu_{d}^{(\infty)}
\notag \\
& =
\lim_{n \to \infty} \frac{ 1 }{ 2^{n} } \sum_{\bvec{s} \in \{-, +\}^{n}} \varepsilon_{d}^{\bvec{s}}
\notag \\
& \ge
\limsup_{n \to \infty} \frac{ 1 }{ 2^{n} } \sum_{\bvec{s} \in \{-, +\}^{n} : \varepsilon_{d} \ge \delta} \varepsilon_{d}^{\bvec{s}}
\notag \\
& \ge
\limsup_{n \to \infty} \frac{ 1 }{ 2^{n} } \sum_{\bvec{s} \in \{-, +\}^{n} : \varepsilon_{d} \ge \delta} \delta
\notag \\
& =
\delta \limsup_{n \to \infty} \frac{ 1 }{ 2^{n} } \Big| \Big\{ \bvec{s} \in \{ -, + \}^{n} \ \Big| \ \varepsilon_{d}^{\bvec{s}} \ge \delta \Big\} \Big| ,
\end{align}
which implies that
\begin{align}
\mu_{d}^{(\infty)} = 0
\quad \Longrightarrow \quad
\lim_{n \to \infty} \frac{ 1 }{ 2^{n} } \Big| \Big\{ \bvec{s} \in \{ -, + \}^{n} \ \Big| \ \varepsilon_{d}^{\bvec{s}} < \delta \Big\} \Big|
=
1 .
\label{eq:eps_vanish}
\end{align}
It follows from \eqref{eq:kprime_plus1} that there exists a pair $(i, j)$ such that $\mu_{\langle \bvec{t} \rangle}^{(\infty)} = 0$ for every $\bvec{0} \le \bvec{t} \le \bvec{r}$ satisfying $t_{i} = r_{i}$ and $t_{j} = r_{j}$, except for $\bvec{t} = \bvec{r}$.
For such a pair $(i, j)$, we have
\begin{align}
0
& \overset{\mathclap{\text{(a)}}}{=}
\lim_{n \to \infty} \frac{ 1 }{ 2^{n} } \Big| \Big\{ \bvec{s} \in \{-, +\}^{n} \ \Big| \ \delta \le \theta_{i, j}^{\bvec{s}}(r_{i}, r_{j}) \le 1 - \delta \Big\} \Big|
\notag \\
& \ge
\limsup_{n \to \infty} \frac{ 1 }{ 2^{n} } \Bigg| \Big\{ \bvec{s} \in \{-, +\}^{n} \ \Big| \ \delta \le \theta_{i, j}^{\bvec{s}}(r_{i}, r_{j}) \le 1 - \delta \Big\} \cap \bigg( \bigcap_{\bvec{t} : \bvec{0} \le \bvec{t} \le \bvec{r}, t_{i} = r_{i}, t_{j} = r_{j}, \bvec{t} \neq \bvec{r}} \Big\{ \bvec{s} \in \{ -, + \}^{n} \ \Big| \ \varepsilon_{\langle \bvec{t} \rangle}^{\bvec{s}} < \delta \Big\} \bigg) \Bigg|
\notag \\
& \overset{\mathclap{\text{(b)}}}{\ge}
\limsup_{n \to \infty} \frac{ 1 }{ 2^{n} } \Bigg| \Big\{ \bvec{s} \in \{-, +\}^{n} \ \Big| \ \delta \le \varepsilon_{q}^{\bvec{s}} \le 1 - \delta^{\prime} \Big\} \cap \bigg( \bigcap_{\bvec{t} : \bvec{0} \le \bvec{t} \le \bvec{r}, t_{i} = r_{i}, t_{j} = r_{j}, \bvec{t} \neq \bvec{r}} \Big\{ \bvec{s} \in \{ -, + \}^{n} \ \Big| \ \varepsilon_{\langle \bvec{t} \rangle}^{\bvec{s}} < \delta \Big\} \bigg) \Bigg|
\notag \\
& \overset{\mathclap{\text{(c)}}}{\ge}
\limsup_{n \to \infty} \frac{ 1 }{ 2^{n} } \Bigg| \Big\{ \bvec{s} \in \{-, +\}^{n} \ \Big| \ \delta^{\prime} \le \varepsilon_{q}^{\bvec{s}} \le 1 - \delta^{\prime} \Big\} \cap \bigg( \bigcap_{\bvec{t} : \bvec{0} \le \bvec{t} \le \bvec{r}, t_{i} = r_{i}, t_{j} = r_{j}, \bvec{t} \neq \bvec{r}} \Big\{ \bvec{s} \in \{ -, + \}^{n} \ \Big| \ \varepsilon_{\langle \bvec{t} \rangle}^{\bvec{s}} < \delta \Big\} \bigg) \Bigg|
\notag \\
& \overset{\mathclap{\text{(d)}}}{=}
\limsup_{n \to \infty} \frac{ 1 }{ 2^{n} } \Bigg( \Big| \Big\{ \bvec{s} \in \{-, +\}^{n} \ \Big| \ \delta^{\prime} \le \varepsilon_{q}^{\bvec{s}} \le 1 - \delta^{\prime} \Big\} \Big| + \Bigg| \bigcap_{\bvec{t} : \bvec{0} \le \bvec{t} \le \bvec{r}, t_{i} = r_{i}, t_{j} = r_{j}, \bvec{t} \neq \bvec{r}} \Big\{ \bvec{s} \in \{ -, + \}^{n} \ \Big| \ \varepsilon_{\langle \bvec{t} \rangle}^{\bvec{s}} < \delta \Big\} \Bigg|
\notag \\
& \qquad \qquad \qquad
{} - \Bigg| \Big\{ \bvec{s} \in \{-, +\}^{n} \ \Big| \ \delta^{\prime} \le \varepsilon_{q}^{\bvec{s}} \le 1 - \delta^{\prime} \Big\} \cup \Bigg( \bigcap_{\bvec{t} : \bvec{0} \le \bvec{t} \le \bvec{r}, t_{i} = r_{i}, t_{j} = r_{j}, \bvec{t} \neq \bvec{r}} \Big\{ \bvec{s} \in \{ -, + \}^{n} \ \Big| \ \varepsilon_{\langle \bvec{t} \rangle}^{\bvec{s}} < \delta \Big\} \Bigg) \Bigg| \Bigg)
\notag \\
& \overset{\mathclap{\text{(e)}}}{\ge}
\limsup_{n \to \infty} \frac{ 1 }{ 2^{n} } \Bigg( \Big| \Big\{ \bvec{s} \in \{-, +\}^{n} \ \Big| \ \delta^{\prime} \le \varepsilon_{q}^{\bvec{s}} \le 1 - \delta^{\prime} \Big\} \Big| + \Bigg| \bigcap_{\bvec{t} : \bvec{0} \le \bvec{t} \le \bvec{r}, t_{i} = r_{i}, t_{j} = r_{j}, \bvec{t} \neq \bvec{r}} \Big\{ \bvec{s} \in \{ -, + \}^{n} \ \Big| \ \varepsilon_{\langle \bvec{t} \rangle}^{\bvec{s}} < \delta \Big\} \Bigg| - 2^{n} \Bigg)
\notag \\
& \overset{\mathclap{\text{(f)}}}{\ge}
\limsup_{n \to \infty} \frac{ 1 }{ 2^{n} } \Big| \Big\{ \bvec{s} \in \{-, +\}^{n} \ \Big| \ \delta^{\prime} \le \varepsilon_{q}^{\bvec{s}} \le 1 - \delta^{\prime} \Big\} \Big| + \liminf_{n \to \infty} \frac{ 1 }{ 2^{n} } \Bigg| \bigcap_{\bvec{t} : \bvec{0} \le \bvec{t} \le \bvec{r}, t_{i} = r_{i}, t_{j} = r_{j}, \bvec{t} \neq \bvec{r}} \Big\{ \bvec{s} \in \{ -, + \}^{n} \ \Big| \ \varepsilon_{\langle \bvec{t} \rangle}^{\bvec{s}} < \delta \Big\} \Bigg| - 1
\notag \\
& \overset{\mathclap{\text{(g)}}}{=}
\limsup_{n \to \infty} \frac{ 1 }{ 2^{n} } \Big| \Big\{ \bvec{s} \in \{-, +\}^{n} \ \Big| \ \delta^{\prime} \le \varepsilon_{q}^{\bvec{s}} \le 1 - \delta^{\prime} \Big\} \Big| ,
\label{eq:eps_q_vanish}
\end{align}
where (a) follows from \eqref{eq:theta_limit};
(b) follows from the fact that
\begin{align}
\theta_{i, j}^{\bvec{s}}(r_{i}, r_{j})
& =
\sum_{\bvec{t} : \bvec{0} \le \bvec{t} \le \bvec{r}, t_{i} \ge r_{i}, t_{j} \ge r_{j}} \varepsilon_{\langle \bvec{t} \rangle}^{\bvec{s}}
=
\varepsilon_{q}^{\bvec{s}} + \sum_{\bvec{t} : \bvec{0} \le \bvec{t} \le \bvec{r}, t_{i} \ge r_{i}, t_{j} \ge r_{j}, \bvec{t} \neq \bvec{r}} \varepsilon_{\langle \bvec{t} \rangle}^{\bvec{s}}
\end{align}
and the definition of $\delta^{\prime} = \delta^{\prime}(\delta, q) \coloneqq \delta \, \tau( q )$;
(c) follows from the fact that $\delta^{\prime} \ge \delta$;
(d) follows by the inclusion-exclusion principle;
(e) follows from the fact that
\begin{align}
\Bigg| \Big\{ \bvec{s} \in \{-, +\}^{n} \ \Big| \ \delta^{\prime} \le \varepsilon_{q}^{\bvec{s}} \le 1 - \delta^{\prime} \Big\} \cup \Bigg( \bigcap_{\bvec{t} : \bvec{0} \le \bvec{t} \le \bvec{r}, t_{i} = r_{i}, t_{j} = r_{j}, \bvec{t} \neq \bvec{r}} \Big\{ \bvec{s} \in \{ -, + \}^{n} \ \Big| \ \varepsilon_{\langle \bvec{t} \rangle}^{\bvec{s}} < \delta \Big\} \Bigg) \Bigg|
\le
2^{n} ;
\end{align}
(f) follows from the fact that
\begin{align}
\limsup_{n \to \infty} (a_{n} + b_{n})
\ge
\limsup_{n \to \infty} a_{n} + \liminf_{n \to \infty} b_{n}
\label{eq:limsup_liminf}
\end{align}
for two sequences $(a_{n})_{n}$ and $(b_{n})_{n}$; and
(g) follows from \lemref{lem:additive} and \eqref{eq:eps_vanish}.
Since $\delta \in (0, 1)$ is an arbitrary fixed number and $q = \langle \bvec{r} \rangle = \langle \bvec{t}^{(\Omega(q))} \rangle$, it follows from \eqref{eq:eps_q_vanish} that
\begin{align}
\lim_{n \to \infty} \frac{ 1 }{ 2^{n} } \Big| \Big\{ \bvec{s} \in \{-, +\}^{n} \ \Big| \ \delta \le \varepsilon_{\langle \bvec{t}^{(\Omega( q ))} \rangle}^{\bvec{s}} \le 1 - \delta \Big\} \Big|
=
0 .
\end{align}
We now suppose that for some $0 \le h \le \Omega( q )$, it holds that
\begin{align}
\lim_{n \to \infty} \frac{ 1 }{ 2^{n} } \Big| \Big\{ \bvec{s} \in \{-, +\}^{n} \ \Big| \ \delta \le \varepsilon_{\langle \bvec{t}^{(h^{\prime})} \rangle}^{\bvec{s}} \le 1 - \delta \Big\} \Big|
=
0
\qquad
\mathrm{for} \ \mathrm{all} \ h < h^{\prime} \le \Omega( q ) .
\label{eq:hypo_eps_h}
\end{align}
It follows from \eqref{eq:kprime_plus1} that there exists a pair $(i, j)$ such that $\mu_{\langle \bvec{t} \rangle}^{(\infty)} = 0$ for every $\bvec{0} \le \bvec{t} \le \bvec{r}$ satisfying $t_{i} \ge t_{i}^{(h)}$ and $t_{j} \ge t_{j}^{(h)}$, except for $\bvec{t} = \bvec{t}^{(h^{\prime})}$ for every $h \le h^{\prime} \le \Omega( q )$.
For such a pair $(i, j)$, similar to \eqref{eq:eps_q_vanish}, we have
\begin{align}
0
& \overset{\mathclap{\text{(a)}}}{=}
\lim_{n \to \infty} \frac{ 1 }{ 2^{n} } \Big| \Big\{ \bvec{s} \in \{-, +\}^{n} \ \Big| \ \delta \le \theta_{i, j}^{\bvec{s}}(t_{i}^{(h)}, t_{j}^{(h)}) \le 1 - \delta \Big\} \Big|
\notag \\
& \ge
\limsup_{n \to \infty} \frac{ 1 }{ 2^{n} } \left| \Big\{ \bvec{s} \in \{-, +\}^{n} \ \Big| \ \delta \le \theta_{i, j}^{\bvec{s}}(t_{i}^{(h)}, t_{j}^{(h)}) \le 1 - \delta \Big\} \cap \left( \bigcap_{\substack{ \bvec{t} : \bvec{0} \le \bvec{t} \le \bvec{r}, \\ t_{i} \ge t_{i}^{(h)}, t_{j} \ge t_{j}^{(h)}, \\ \bvec{t} \neq \bvec{t}^{(h^{\prime})} \, \forall h^{\prime} \ge h }} \Big\{ \bvec{s} \in \{ -, + \}^{n} \ \Big| \ \varepsilon_{\langle \bvec{t} \rangle}^{\bvec{s}} < \delta \Big\} \right) \right|
\notag \\
& \overset{\mathclap{\text{(b)}}}{\ge}
\limsup_{n \to \infty} \frac{ 1 }{ 2^{n} } \left| \Bigg\{ \bvec{s} \in \{-, +\}^{n} \ \Bigg| \ \delta \le \sum_{h^{\prime} = h}^{\Omega( q )} \varepsilon_{\langle \bvec{t}^{(h^{\prime})} \rangle}^{\bvec{s}} \le 1 - \delta^{\prime} \Bigg\} \cap \left( \bigcap_{\substack{ \bvec{t} : \bvec{0} \le \bvec{t} \le \bvec{r}, \\ t_{i} \ge t_{i}^{(h)}, t_{j} \ge t_{j}^{(h)}, \\ \bvec{t} \neq \bvec{t}^{(h^{\prime})} \, \forall h^{\prime} \ge h }} \Big\{ \bvec{s} \in \{ -, + \}^{n} \ \Big| \ \varepsilon_{\langle \bvec{t} \rangle}^{\bvec{s}} < \delta \Big\} \right) \right|
\notag \\
& \overset{\mathclap{\text{(c)}}}{\ge}
\limsup_{n \to \infty} \frac{ 1 }{ 2^{n} } \left| \Bigg\{ \bvec{s} \in \{-, +\}^{n} \ \Bigg| \ \delta^{\prime} \le \sum_{h^{\prime} = h}^{\Omega( q )} \varepsilon_{\langle \bvec{t}^{(h^{\prime})} \rangle}^{\bvec{s}} \le 1 - \delta^{\prime} \Bigg\} \cap \left( \bigcap_{\substack{ \bvec{t} : \bvec{0} \le \bvec{t} \le \bvec{r}, \\ t_{i} \ge t_{i}^{(h)}, t_{j} \ge t_{j}^{(h)}, \\ \bvec{t} \neq \bvec{t}^{(h^{\prime})} \, \forall h^{\prime} \ge h }} \Big\{ \bvec{s} \in \{ -, + \}^{n} \ \Big| \ \varepsilon_{\langle \bvec{t} \rangle}^{\bvec{s}} < \delta \Big\} \right) \right|
\notag \\
& \overset{\mathclap{\text{(d)}}}{\ge}
\limsup_{n \to \infty} \frac{ 1 }{ 2^{n} } \left( \Bigg| \Bigg\{ \bvec{s} \in \{-, +\}^{n} \ \Bigg| \ \delta^{\prime} \le \sum_{h^{\prime} = h}^{\Omega( q )} \varepsilon_{\langle \bvec{t}^{(h^{\prime})} \rangle}^{\bvec{s}} \le 1 - \delta^{\prime} \Bigg\} \Bigg| + \left| \bigcap_{\substack{ \bvec{t} : \bvec{0} \le \bvec{t} \le \bvec{r}, \\ t_{i} \ge t_{i}^{(h)}, t_{j} \ge t_{j}^{(h)}, \\ \bvec{t} \neq \bvec{t}^{(h^{\prime})} \, \forall h^{\prime} \ge h }} \Big\{ \bvec{s} \in \{ -, + \}^{n} \ \Big| \ \varepsilon_{\langle \bvec{t} \rangle}^{\bvec{s}} < \delta \Big\} \right| - 2^{n} \right)
\notag \\
& \overset{\mathclap{\text{(e)}}}{\ge}
\limsup_{n \to \infty} \frac{ 1 }{ 2^{n} } \Bigg| \Bigg\{ \bvec{s} \in \{-, +\}^{n} \ \Bigg| \ \delta^{\prime} \le \sum_{h^{\prime} = h}^{\Omega( q )} \varepsilon_{\langle \bvec{t}^{(h^{\prime})} \rangle}^{\bvec{s}} \le 1 - \delta^{\prime} \Bigg\} \Bigg| + \liminf_{n \to \infty} \frac{ 1 }{ 2^{n} } \left| \bigcap_{\substack{ \bvec{t} : \bvec{0} \le \bvec{t} \le \bvec{r}, \\ t_{i} \ge t_{i}^{(h)}, t_{j} \ge t_{j}^{(h)}, \\ \bvec{t} \neq \bvec{t}^{(h^{\prime})} \, \forall h^{\prime} \ge h }} \Big\{ \bvec{s} \in \{ -, + \}^{n} \ \Big| \ \varepsilon_{\langle \bvec{t} \rangle}^{\bvec{s}} < \delta \Big\} \right| - 1
\notag \\
& \overset{\mathclap{\text{(f)}}}{=}
\limsup_{n \to \infty} \frac{ 1 }{ 2^{n} } \Bigg| \Bigg\{ \bvec{s} \in \{-, +\}^{n} \ \Bigg| \ \delta^{\prime} \le \sum_{h^{\prime} = h}^{\Omega( q )} \varepsilon_{\langle \bvec{t}^{(h^{\prime})} \rangle}^{\bvec{s}} \le 1 - \delta^{\prime} \Bigg\} \Bigg| ,
\label{eq:eps_sum_vanish}
\end{align}
where (a) follows from \eqref{eq:theta_limit};
(b) follows from the fact that
\begin{align}
\theta_{i, j}^{\bvec{s}}(t_{i}^{(h)}, t_{j}^{(h)})
& =
\sum_{\substack{ \bvec{t} : \bvec{0} \le \bvec{t} \le \bvec{r}, \\ t_{i} \ge t_{i}^{(h)}, t_{j} \ge t_{j}^{(h)} }} \varepsilon_{\langle \bvec{t} \rangle}^{\bvec{s}}
=
\left( \sum_{h^{\prime} = h}^{\Omega( q )} \varepsilon_{\langle \bvec{t}^{(h^{\prime})} \rangle} \right) + \left( \sum_{\substack{ \bvec{t} : \bvec{0} \le \bvec{t} \le \bvec{r}, \\ t_{i} \ge t_{i}^{(h)}, t_{j} \ge t_{j}^{(h)}, \\ \bvec{t} \neq \bvec{t}^{(h^{\prime})} \, \forall h^{\prime} \ge h }} \varepsilon_{\langle \bvec{t} \rangle}^{\bvec{s}} \right)
\end{align}
and the definition of $\delta^{\prime} = \delta^{\prime}(\delta, q) \coloneqq \delta \, \tau( q )$;
(c) follows from the fact that $\delta^{\prime} \ge \delta$;
(d) follows by the inclusion-exclusion principle and the fact that
\begin{align}
\left| \Bigg\{ \bvec{s} \in \{-, +\}^{n} \ \Bigg| \ \delta^{\prime} \le \sum_{h^{\prime} = h}^{\Omega( q )} \varepsilon_{\langle \bvec{t}^{(h^{\prime})} \rangle}^{\bvec{s}} \le 1 - \delta^{\prime} \Bigg\} \cup \left( \bigcap_{\substack{ \bvec{t} : \bvec{0} \le \bvec{t} \le \bvec{r}, \\ t_{i} \ge t_{i}^{(h)}, t_{j} \ge t_{j}^{(h)}, \\ \bvec{t} \neq \bvec{t}^{(h^{\prime})} \, \forall h^{\prime} \ge h }} \Big\{ \bvec{s} \in \{ -, + \}^{n} \ \Big| \ \varepsilon_{\langle \bvec{t} \rangle}^{\bvec{s}} < \delta \Big\} \right) \right|
\le
2^{n} ;
\end{align}
(e) follows from \eqref{eq:limsup_liminf}; and
(f) follows from \lemref{lem:additive} and \eqref{eq:eps_vanish}.
Since $\delta \in (0, 1)$ is an arbitrary fixed number, it follows from \eqref{eq:eps_sum_vanish} that
\begin{align}
\lim_{n \to \infty} \frac{ 1 }{ 2^{n} } \Bigg| \Bigg\{ \bvec{s} \in \{-, +\}^{n} \ \Bigg| \ \delta \le \sum_{h^{\prime} = h}^{\Omega( q )} \varepsilon_{\langle \bvec{t}^{(h^{\prime})} \rangle}^{\bvec{s}} \le 1 - \delta \Bigg\} \Bigg|
& =
0 .
\label{eq:eps_sum_vanish2}
\end{align}
Furthermore, we observe that
\begin{align}
0
& \overset{\mathclap{\text{(a)}}}{=}
\lim_{n \to \infty} \frac{ 1 }{ 2^{n} } \Bigg| \Bigg\{ \bvec{s} \in \{-, +\}^{n} \ \Bigg| \ \delta \le \sum_{h^{\prime} = h}^{\Omega( q )} \varepsilon_{\langle \bvec{t}^{(h^{\prime})} \rangle}^{\bvec{s}} \le 1 - \delta \Bigg\} \Bigg|
\notag \\
& \ge
\limsup_{n \to \infty} \frac{ 1 }{ 2^{n} } \Bigg| \Bigg\{ \bvec{s} \in \{-, +\}^{n} \ \Bigg| \ \delta \le \sum_{h^{\prime} = h}^{\Omega( q )} \varepsilon_{\langle \bvec{t}^{(h^{\prime})} \rangle}^{\bvec{s}} \le 1 - \delta \Bigg\} \cap \Bigg( \bigcap_{h^{\prime} = h + 1}^{\Omega( q )} \Big\{ \bvec{s} \in \{-, +\}^{n} \ \Big| \ \delta \le \varepsilon_{\langle \bvec{t}^{(h)} \rangle}^{\bvec{s}} \le 1 - \delta \Big\}^{\complement} \Bigg) \Bigg|
\notag \\
& \overset{\mathclap{\text{(b)}}}{\ge}
\limsup_{n \to \infty} \frac{ 1 }{ 2^{n} } \Bigg| \Big\{ \bvec{s} \in \{-, +\}^{n} \ \Big| \ \delta \le \varepsilon_{\langle \bvec{t}^{(h)} \rangle}^{\bvec{s}} \le 1 - \delta^{\prime} \Big\} \cap \Bigg( \bigcap_{h^{\prime} = h + 1}^{\Omega( q )} \Big\{ \bvec{s} \in \{-, +\}^{n} \ \Big| \ \delta \le \varepsilon_{\langle \bvec{t}^{(h)} \rangle}^{\bvec{s}} \le 1 - \delta \Big\}^{\complement} \Bigg) \Bigg|
\notag \\
& \overset{\mathclap{\text{(c)}}}{\ge}
\limsup_{n \to \infty} \frac{ 1 }{ 2^{n} } \Bigg| \Big\{ \bvec{s} \in \{-, +\}^{n} \ \Big| \ \delta^{\prime} \le \varepsilon_{\langle \bvec{t}^{(h)} \rangle}^{\bvec{s}} \le 1 - \delta^{\prime} \Big\} \cap \Bigg( \bigcap_{h^{\prime} = h + 1}^{\Omega( q )} \Big\{ \bvec{s} \in \{-, +\}^{n} \ \Big| \ \delta \le \varepsilon_{\langle \bvec{t}^{(h)} \rangle}^{\bvec{s}} \le 1 - \delta \Big\}^{\complement} \Bigg) \Bigg|
\notag \\
& \overset{\mathclap{\text{(d)}}}{\ge}
\limsup_{n \to \infty} \frac{ 1 }{ 2^{n} } \Bigg( \Big| \Big\{ \bvec{s} \in \{-, +\}^{n} \ \Big| \ \delta^{\prime} \le \varepsilon_{\langle \bvec{t}^{(h)} \rangle}^{\bvec{s}} \le 1 - \delta^{\prime} \Big\} \Big| + \Bigg| \bigcap_{h^{\prime} = h + 1}^{\Omega( q )} \Big\{ \bvec{s} \in \{-, +\}^{n} \ \Big| \ \delta \le \varepsilon_{\langle \bvec{t}^{(h)} \rangle}^{\bvec{s}} \le 1 - \delta \Big\}^{\complement} \Bigg| - 2^{n} \Bigg)
\notag \\
& \overset{\mathclap{\text{(e)}}}{\ge}
\limsup_{n \to \infty} \frac{ 1 }{ 2^{n} } \Big| \Big\{ \bvec{s} \in \{-, +\}^{n} \ \Big| \ \delta^{\prime} \le \varepsilon_{\langle \bvec{t}^{(h)} \rangle}^{\bvec{s}} \le 1 - \delta^{\prime} \Big\} \Big| + \liminf_{n \to \infty} \frac{ 1 }{ 2^{n} } \Bigg| \bigcap_{h^{\prime} = h + 1}^{\Omega( q )} \Big\{ \bvec{s} \in \{-, +\}^{n} \ \Big| \ \delta \le \varepsilon_{\langle \bvec{t}^{(h)} \rangle}^{\bvec{s}} \le 1 - \delta \Big\}^{\complement} \Bigg| - 1
\notag \\
& \overset{\mathclap{\text{(f)}}}{=}
\limsup_{n \to \infty} \frac{ 1 }{ 2^{n} } \Big| \Big\{ \bvec{s} \in \{-, +\}^{n} \ \Big| \ \delta^{\prime} \le \varepsilon_{\langle \bvec{t}^{(h)} \rangle}^{\bvec{s}} \le 1 - \delta^{\prime} \Big\} \Big| ,
\label{eq:eps_h_vanish}
\end{align}
where (a) follows from \eqref{eq:eps_sum_vanish2};
(b) follows by the definition of $\delta^{\prime} = \delta^{\prime}(\delta, q) \coloneqq \delta \, \tau( q )$;
(c) follows from the fact that $\delta^{\prime} \ge \delta$;
(d) follows by the inclusion-exclusion principle and the fact that
\begin{align}
\Bigg| \Big\{ \bvec{s} \in \{-, +\}^{n} \ \Big| \ \delta^{\prime} \le \varepsilon_{\langle \bvec{t}^{(h)} \rangle}^{\bvec{s}} \le 1 - \delta^{\prime} \Big\} \cup \Bigg( \bigcap_{h^{\prime} = h + 1}^{\Omega( q )} \Big\{ \bvec{s} \in \{-, +\}^{n} \ \Big| \ \delta \le \varepsilon_{\langle \bvec{t}^{(h)} \rangle}^{\bvec{s}} \le 1 - \delta \Big\}^{\complement} \Bigg) \Bigg|
\le
2^{n} ;
\end{align}
(e) follows from \eqref{eq:limsup_liminf}; and
(f) follows from \lemref{lem:additive} and the hypothesis \eqref{eq:hypo_eps_h}.
Since $\delta \in (0, 1)$ is an arbitrary fixed number, it follows from \eqref{eq:eps_h_vanish} that
\begin{align}
\lim_{n \to \infty} \frac{ 1 }{ 2^{n} } \Big| \Big\{ \bvec{s} \in \{-, +\}^{n} \ \Big| \ \delta \le \varepsilon_{\langle \bvec{t}^{(h)} \rangle}^{\bvec{s}} \le 1 - \delta \Big\} \Big|
& =
0 ,
\label{eq:eps_h_vanish2}
\end{align}
which implies by induction together with \eqref{eq:eps_vanish} that \eqref{eq:proportion0} of \thref{th:polarization} holds, i.e.,
\begin{align}
\lim_{n \to \infty} \frac{ 1 }{ 2^{n} } \Big| \Big\{ \bvec{s} \in \{-, +\}^{n} \ \Big| \ \delta \le \varepsilon_{d}^{\bvec{s}} \le 1 - \delta \Big\} \Big|
& =
0
\end{align}
for every $\delta \in (0, 1)$ and every $d|q$.

Finally, we prove \eqref{eq:proportion1} of \thref{th:polarization}.
It follows by the definition \eqref{def:mu_d} that
\begin{align}
\mu_{d}^{(n)}
& =
\frac{ 1 }{ 2^{n} } \sum_{\bvec{s} \in \{ -, + \}^{n}} \varepsilon_{d}^{\bvec{s}}
\notag \\
& \le
\frac{ 1 }{ 2^{n} } \sum_{\substack{ \bvec{s} \in \{ -, + \}^{n} : \\ \varepsilon_{d} < \delta }} \delta + \frac{ 1 }{ 2^{n} } \sum_{\substack{ \bvec{s} \in \{ -, + \}^{n} : \\ \delta \le \varepsilon_{d} \le 1 - \delta }} (1 - \delta) + \frac{ 1 }{ 2^{n} } \sum_{\substack{ \bvec{s} \in \{ -, + \}^{n} : \\ \varepsilon_{d} > 1 - \delta }} 1
\notag \\
& =
\delta + \frac{ 1 }{ 2^{n} } \sum_{\substack{ \bvec{s} \in \{ -, + \}^{n} : \\ \delta \le \varepsilon_{d} \le 1 - \delta }} (1 - 2 \delta) + \frac{ 1 }{ 2^{n} } \sum_{\substack{ \bvec{s} \in \{ -, + \}^{n} : \\ \varepsilon_{d} > 1 - \delta }} (1 - \delta) ,
\notag
\end{align}
which implies together with \eqref{eq:proportion0} that
\begin{align}
\mu_{d}^{(\infty)}
\le
\delta + (1 - \delta) \liminf_{n \to \infty} \frac{ 1 }{ 2^{n} } \Big| \Big\{ \bvec{s} \in \{ -, + \}^{n} \ \Big| \ \varepsilon_{d}^{(n)} > 1 - \delta \Big\} \Big|
\label{eq:proportion1_1}
\end{align}
In addition, we also get
\begin{align}
\mu_{d}^{(n)}
& =
\frac{ 1 }{ 2^{n} } \sum_{\bvec{s} \in \{ -, + \}^{n}} \varepsilon_{d}^{\bvec{s}}
\notag \\
& \ge
\frac{ 1 }{ 2^{n} } \sum_{\substack{ \bvec{s} \in \{ -, + \}^{n} : \\ \delta \le \varepsilon_{d} \le 1 - \delta }} \delta + \frac{ 1 }{ 2^{n} } \sum_{\substack{ \bvec{s} \in \{ -, + \}^{n} : \\ \varepsilon_{d} > 1 - \delta }} (1-\delta) ,
\end{align}
which also implies together with \eqref{eq:proportion0} that
\begin{align}
(1-\delta) \limsup_{n \to \infty} \frac{ 1 }{ 2^{n} } \Big| \Big\{ \bvec{s} \in \{ -, + \}^{n} \ \Big| \ \varepsilon_{d}^{(n)} > 1 - \delta \Big\} \Big|
\le
\mu_{d}^{(\infty)}
\label{eq:proportion1_2}
\end{align}
As $\delta > 0$ can be chosen arbitrarily small, it follows from \eqref{eq:proportion1_1} and \eqref{eq:proportion1_2} that \eqref{eq:proportion1}.
This completes the proof of \thref{th:polarization}.
\end{IEEEproof}

Considering the input alphabet $\mathcal{X} = \mathbb{Z}/q\mathbb{Z}$ as an abelian group, as in \eqref{eq:multilevel}, we can conclude a multilevel polarization theorem of erasure channels $V : \mathcal{X} \to \mathcal{Y}$ as follows:

\begin{corollary}
\label{cor:multilevel}
Let $V$ be an erasure channel with initial probability vector $(\varepsilon_{d})_{d|q}$.
For any $d|q$, it holds that
\begin{align}
\lim_{n \to \infty} \frac{ 1 }{ 2^{n} } \Big| \Big\{ \bvec{s} \in \{ -, + \}^{n} \ \Big| \ \big| I(V^{\bvec{s}}) - \log d \big| < \delta \ \mathrm{and} \ \big| I(V^{\bvec{s}}[\ker \varphi_{d}]) - \log d \big| < \delta \Big\} \Big|
=
\mu_{d}^{(\infty)}
\end{align}
for every fixed $\delta > 0$,
where $V^{\bvec{s}}[\ker \varphi_{d}]$ is a homomorphism channel of $V^{\bvec{s}}$ defined in \eqref{def:homomorphism}; the function $\varphi_{d} : x \mapsto (x + d\mathbb{Z})$ is a group homomorphism; and $\ker \varphi_{d} \coloneqq \{ x \in \mathbb{Z}/q\mathbb{Z} \mid \phi_{d}( x ) = d \mathbb{Z} \}$ denotes the kernel of $\varphi_{d}$.
\end{corollary}

\begin{IEEEproof}[Proof of \corref{cor:multilevel}]
It follows from \eqref{def:homomorphism} that for each $d|q$, the channel $V[\ker \varphi_{d}] : \mathcal{X} / \ker \varphi_{d} \to \mathcal{Y}$ is given by
\begin{align}
V[\ker \varphi_{d}](y \mid x)
& =
\begin{dcases}
\bar{\varepsilon}_{d_{1}}[d]
& \mathrm{if} \ y = \psi_{d_{1}}( x ) \ \mathrm{for} \ \mathrm{some} \ d_{1}|d ,
\\
0
& \mathrm{otherwise} ,
\end{dcases}
\end{align}
where $\psi_{d_{1}} : x \mapsto (x + d_{1} \mathbb{Z})$ is a group homomorphism for each $d_{1}|d$; and the probability vector $( \bar{\varepsilon}_{d_{1}}[d] )_{d_{1}|d}$ is given by
\begin{align}
\bar{\varepsilon}_{d_{1}}[d]
\coloneqq
\varepsilon_{d_{1}} + \sum_{\substack{ d_{2}|q : \\ d_{2} \neq d_{1}, d_{2} \nmid d, d_{1}|d_{2} }} \varepsilon_{d_{2}} .
\end{align}
That is, the channel $V[\ker \varphi_{d}]$ is also an erasure channel of \defref{def:V}.
Since $I(V) = \sum_{d|q} \varepsilon_{d} \, (\log d)$ (cf. \cite[Equation~(28)]{itw2016}), we have
\begin{align}
I( V[\ker \varphi_{d}] )
=
\sum_{d_{1}|d} \Bigg( \varepsilon_{d_{1}} + \sum_{\substack{ d_{2}|q : \\ d_{2} \neq d_{1}, d_{2} \nmid d, d_{1}|d_{2} }} \varepsilon_{d_{2}} \Bigg) \log d_{1} .
\end{align}
Therefore, \thref{th:polarization} directly provides \corref{cor:multilevel}.
\end{IEEEproof}

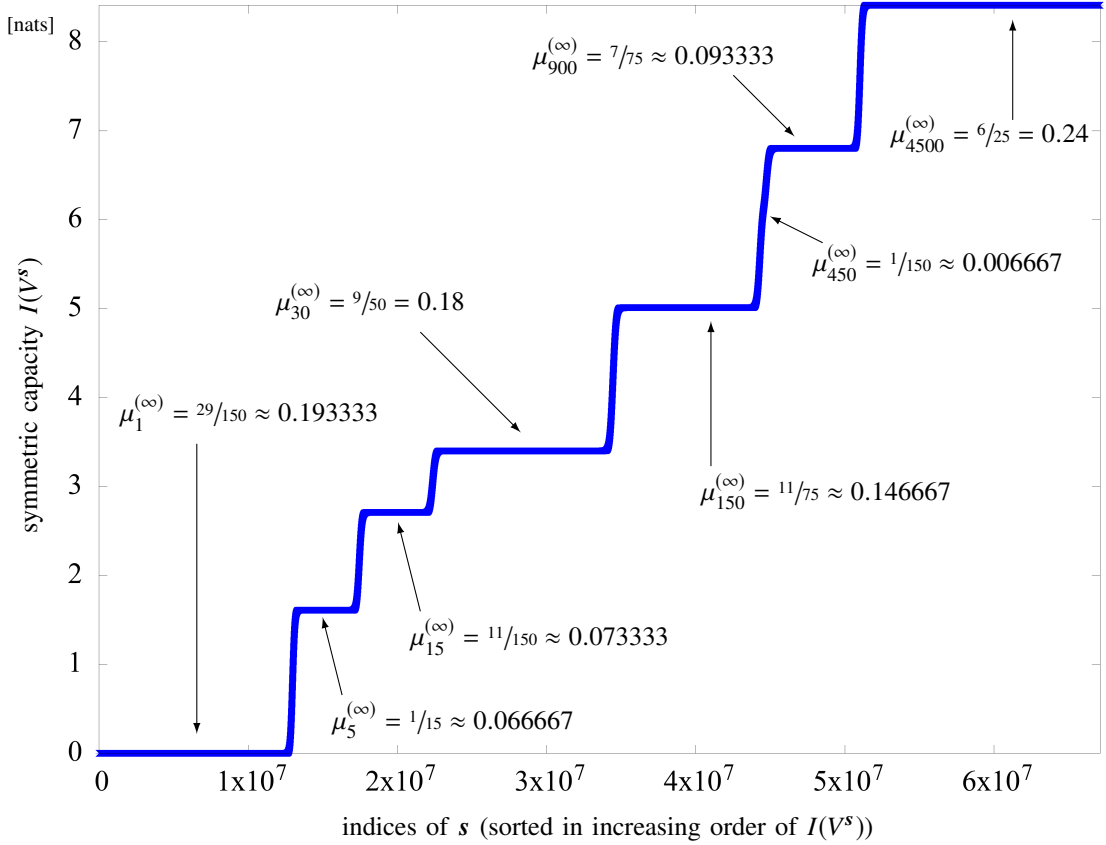
\begin{figure}[!t]
\centering
\begin{overpic}[width = 0.9\hsize, clip]{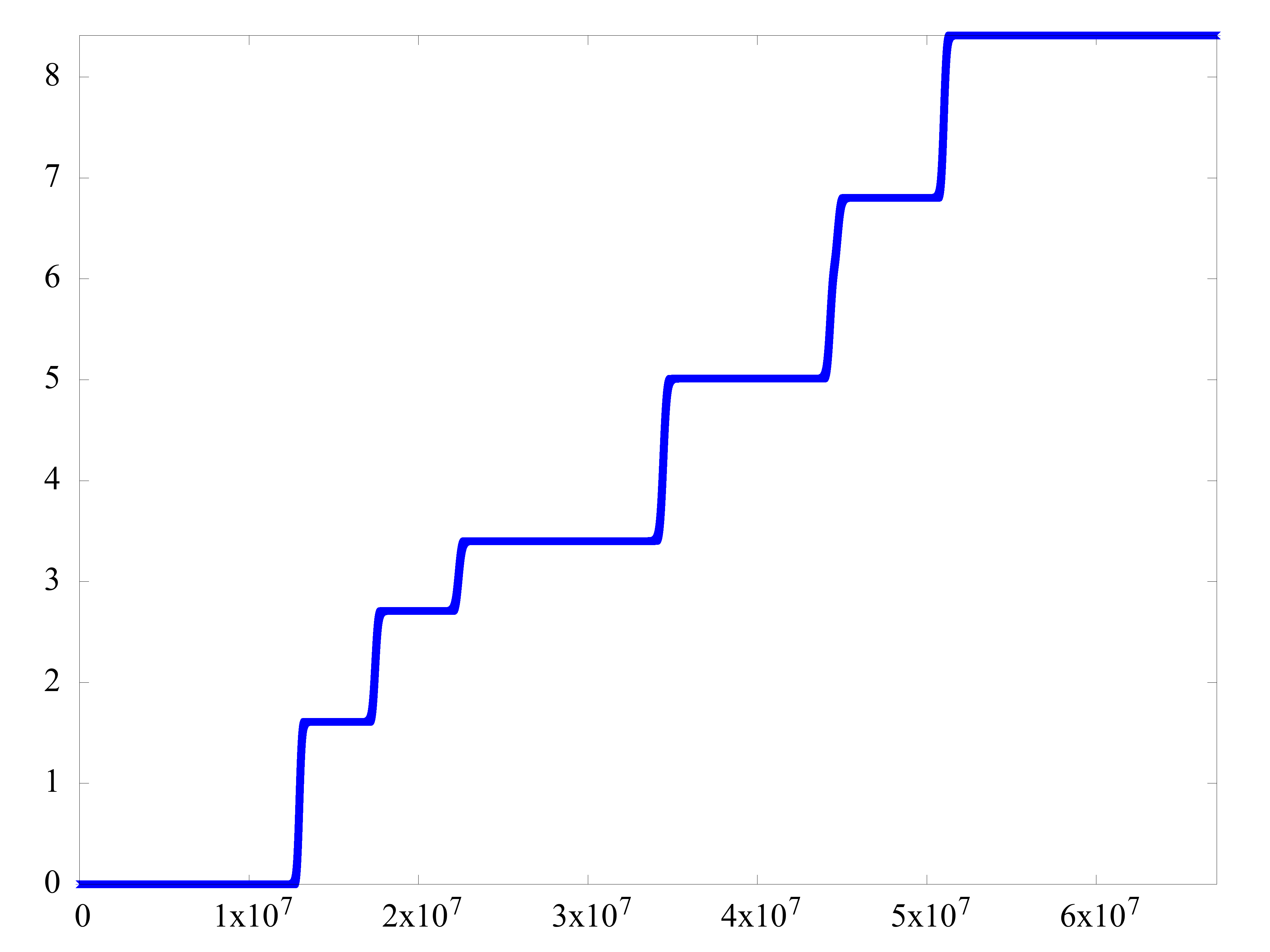}
\put(-2, 70){\footnotesize [nats]}
\put(-1, 25){\rotatebox{90}{symmetric capacity $I( V^{\bvec{s}} )$}}
\put(28, -2){indices of $\bvec{s}$ (sorted in increasing order of $I( V^{\bvec{s}} )$)}
\put(8, 35){$\mu_{1}^{(\infty)} = \sfrac{29}{150} \approx 0.193333$}
\put(15, 33){\vector(0, -1){26}}
\put(27, 7.5){$\mu_{5}^{(\infty)} = \sfrac{1}{15} \approx 0.066667$}
\put(28, 10.5){\vector(-1, 4){1.75}}
\put(34, 15){$\mu_{15}^{(\infty)} = \sfrac{11}{150} \approx 0.073333$}
\put(35, 18){\vector(-1, 4){2}}
\put(22, 45){$\mu_{30}^{(\infty)} = \sfrac{9}{50} = 0.18$}
\put(35, 43){\vector(1, -1){9}}
\put(60, 28){$\mu_{150}^{(\infty)} = \sfrac{11}{75} \approx 0.146667$}
\put(61, 31){\vector(0, 1){13}}
\put(70, 48.5){$\mu_{450}^{(\infty)} = \sfrac{1}{150} \approx 0.006667$}
\put(71, 51){\vector(-2, 1){4.75}}
\put(45, 67){$\mu_{900}^{(\infty)} = \sfrac{7}{75} \approx 0.093333$}
\put(63, 66){\vector(1, -1){5.5}}
\put(77, 60){$\mu_{4500}^{(\infty)} = \sfrac{6}{25} = 0.24$}
\put(88, 62){\vector(0, 1){9}}
\end{overpic}
\caption{Plots of the symmetric capacities $I(V^{\bvec{s}})$ of polar transforms $\bvec{s} \in \{ -, + \}^{n}$ with $n = 26$-step polar transforms. The initial channel $V$ is given in \exref{ex:mu_d} (see also \tabref{table:mu_d}).
Note that the logarithm is measured in nats.
The proportion of polar transforms $\bvec{s} \in \{ -, + \}^{n}$ satisfying $I(V^{\bvec{s}}) \approx \log d$ and $I(V^{\bvec{s}}[ \ker \varphi_{d} ]) \approx \log d$ is approximately equal to $\mu_{d}^{(\infty)}$ for each $d|q$ (cf. \exref{ex:mu_d} and \corref{cor:multilevel}).
For example, the proportion of $\bvec{s} \in \{ -, + \}^{n}$ satisfying $I(V^{\bvec{s}}) \approx \log 30$ and $I(V^{\bvec{s}}[ \ker \varphi_{30} ]) \approx \log 30$ is approximately equal to $\mu_{30}^{(\infty)} = 0.18$.}
\label{fig:mu_d}
\end{figure}

By \corref{cor:multilevel}, it holds that
\begin{align}
\sum_{d|q} \lim_{n \to \infty} \frac{ 1 }{ 2^{n} } \Big| \Big\{ \bvec{s} \in \{ -, + \}^{n} \ \Big| \ \big| I(V^{\bvec{s}}) - \log d \big| < \delta \ \mathrm{and} \ \big| I(V^{\bvec{s}}[\ker \varphi_{d}]) - \log d \big| < \delta \Big\} \Big|
=
1 ,
\label{eq:multilevel_sum_V}
\end{align}
which is a version of \eqref{eq:multilevel}.
Therefore, \corref{cor:multilevel} shows the limiting proportion of each term of the sum of \eqref{eq:multilevel} for every generalized erasure channel $V$ defined in \defref{def:V}.
Figure~\ref{fig:mu_d} shows an example of multilevel polarization for the channel $V$ given in \exref{ex:mu_d} (see also \tabref{table:mu_d}), where \figref{fig:mu_d} is plotted by the recursive formula \eqref{def:eps_s} (see also \thref{th:itw2016}).

\section{Conclusion}

We have examined asymptotic distributions of multilevel polarization for erasure channels $V$ defined in \defref{def:V}.
\thref{th:mu_d} showed how to calculate the asymptotic distributions $( \mu_{d}^{(\infty)} )_{d|q}$ defined in \eqref{def:mu_d}, and its example was given in \exref{ex:mu_d} (see also \tabref{table:mu_d}).
\thref{th:polarization} showed that the asymptotic distributions of multilevel polarization of $V$ with an initial probability vector $( \varepsilon_{d} )_{d|q}$ can be fully characterized by $( \mu_{d}^{(\infty)} )_{d|q}$ (see also \corref{cor:multilevel}).
As future works, asymptotic distributions of multilevel polarization for general DMCs are of interest to refine polarization theorem.

\appendix[Example of Algorithm~\ref{alg:main}]

We show an example of the calculation process of Algorithm~\ref{alg:main} in the setting of \exref{ex:mu_d} as follows:
\begin{itemize}
\item
$m = 3$;
\item
$(p_{1}, p_{2}, p_{3}) = (2, 3, 5)$;
\item
$\bvec{r} = (r_{1}, r_{2}, r_{3}) = (2, 2, 3)$;
\item
the input alphabet size $q = p_{1}^{r_{1}} p_{2}^{r_{2}} p_{3}^{r_{3}} = 2^{2} \cdot 3^{2} \cdot 5^{3} = 4500$;
\item
the initial probability vector $( \varepsilon_{d} )_{d|q} = (\varepsilon_{1}, \varepsilon_{2}, \dots, \varepsilon_{4500})$ is given as \exref{ex:mu_d} (see also \tabref{table:mu_d}).
\end{itemize}
Note that $d = p_{1}^{t_{1}} p_{2}^{t_{2}} p_{3}^{t_{3}} = \langle t_{1}, t_{2}, t_{3} \rangle = \langle \bvec{t} \rangle$.
In Lines 1--3 of Algorithm~\ref{alg:main}, we first initialize as follows:
\begin{itemize}
\item
$( \mu_{d}^{(\infty)} )_{d|q} = ( \mu_{1}^{(\infty)}, \mu_{2}^{(\infty)}, \dots, \mu_{4500}^{(\infty)} ) = (0, 0, \dots, 0)$;
\item
$\alpha = 0$; and
\item
$\bvec{t} = (t_{1}, t_{2}, t_{3}) = (0, 0, 0)$.
\end{itemize}
It is clear that the condition $0 \le \alpha = 0 < 1$ of Line~4 holds.

Consider the first step of the while loop in Lines 4--15 of Algorithm~\ref{alg:main} with the following parameters: $\alpha = 0$ and $\bvec{t} = (t_{1}, t_{2}, t_{3}) = (0, 0, 0)$.
Set $(i, j) = (1, 2)$ as in Line~5, and go to the while loop in Lines~6--12 of Algorithm~\ref{alg:main}.
It can be verified that
\begin{align*}
\lambda_{1, 2}(1, 1)
& =
\sum_{u_{1} = 1}^{2} \sum_{u_{2} = 0}^{0} \sum_{u_{3} = 0}^{3} \varepsilon_{\langle u_{1}, u_{2}, u_{3} \rangle}
=
\varepsilon_{2} + \varepsilon_{4} + \varepsilon_{10} + \varepsilon_{20} + \varepsilon_{50} + \varepsilon_{100} + \varepsilon_{250} + \varepsilon_{500}
=
\frac{ 16 }{ 75 } ,
\\
\rho_{1, 2}(1, 1)
& =
\sum_{u_{1} = 0}^{0} \sum_{u_{2} = 1}^{2} \sum_{u_{3} = 0}^{3} \varepsilon_{\langle u_{1}, u_{2}, u_{3} \rangle}
=
\varepsilon_{3} + \varepsilon_{9} + \varepsilon_{15} + \varepsilon_{45} + \varepsilon_{75} + \varepsilon_{225} + \varepsilon_{375} + \varepsilon_{1125}
=
\frac{ 43 }{ 150 } .
\end{align*}
Since $\lambda_{1, 2}(1, 1) < \rho_{1, 2}(1, 1)$, store $(k, l) = (2, 1)$ as in Line~8; reset $(i, j) = (2, 3)$ as in Line~9; and go back to Line~6.
It can be verified that
\begin{align*}
\lambda_{2, 3}(1, 1)
& =
\sum_{u_{1} = 0}^{2} \sum_{u_{2} = 1}^{2} \sum_{u_{3} = 0}^{0} \varepsilon_{\langle u_{1}, u_{2}, u_{3} \rangle}
=
\varepsilon_{3} + \varepsilon_{6} + \varepsilon_{9} + \varepsilon_{12} + \varepsilon_{18} + \varepsilon_{36}
=
\frac{ 1 }{ 6 } ,
\\
\rho_{2, 3}(1, 1)
& =
\sum_{u_{1} = 0}^{2} \sum_{u_{2} = 0}^{0} \sum_{u_{3} = 1}^{3} \varepsilon_{\langle u_{1}, u_{2}, u_{3} \rangle}
=
\varepsilon_{5} + \varepsilon_{10} + \varepsilon_{20} + \varepsilon_{25} + \varepsilon_{50} + \varepsilon_{100} + \varepsilon_{125} + \varepsilon_{250} + \varepsilon_{500}
=
\frac{ 7 }{ 30 } .
\end{align*}
Since $\lambda_{2, 3}(1, 1) < \rho_{2, 3}(1, 1)$, store $(k, l) = (3, 2)$ as in Line~8; reset $(i, j) = (3, 4)$ as in Line~9; and go back to Line~6.
As $j = 4 > 3 = m$, the while loop in Lines~6--12 of Algorithm~\ref{alg:main} is finished and we go to Line~13.
It can be verified that
\begin{align*}
\beta_{2, 3}(1, 1)
=
\sum_{u_{1} = 0}^{2} \sum_{u_{2} = 0}^{0} \sum_{u_{3} = 0}^{0} \varepsilon_{\langle u_{1}, u_{2}, u_{3} \rangle}
=
\varepsilon_{1} + \varepsilon_{2} + \varepsilon_{4}
=
\frac{ 2 }{ 75 } .
\end{align*}
Since $\lambda_{2, 3}(1, 1) < \rho_{2, 3}(1, 1)$, we get in Line~13 that
\begin{align}
\mu_{\langle 0, 0, 0 \rangle}^{(\infty)}
=
\mu_{1}^{(\infty)}
=
\beta_{2, 3}(1, 1) + \lambda_{2, 3}(1, 1) - \alpha
=
\frac{ 29 }{ 150 } .
\end{align}
Resetting $\alpha = 29/150$ and $t_{3} = 1$ as in Lines~14 and~15, respectively, i.e., $\bvec{t} = (t_{1}, t_{2}, t_{3}) = (0, 0, 1)$, we go back to Line~4.
As $0 \le \alpha = 29/150 < 1$, we continue the while loop in Lines~4--15 of Algorithm~\ref{alg:main}.

Consider the second step of the while loop in Lines~4--15 of Algorithm~\ref{alg:main} with the following parameters: $\alpha = 29/150$ and $\bvec{t} = (t_{1}, t_{2}, t_{3}) = (0, 0, 1)$.
Set $(i, j) = (1, 2)$ as in Line~5, and go to the while loop in Lines~6--12 of Algorithm~\ref{alg:main}.
Since $\lambda_{2, 3}(1, 1) < \rho_{2, 3}(1, 1)$, store $(k, l) = (3, 2)$ as in Line~8; reset $(i, j) = (3, 4)$ as in Line~9; and go back to Line~6.
It can be verified that
\begin{align*}
\lambda_{2, 3}(1, 2)
& =
\sum_{u_{1} = 0}^{2} \sum_{u_{2} = 1}^{2} \sum_{u_{3} = 0}^{1} \varepsilon_{\langle u_{1}, u_{2}, u_{3} \rangle}
=
\varepsilon_{3} + \varepsilon_{6} + \varepsilon_{9} + \varepsilon_{12} + \varepsilon_{15} + \varepsilon_{18} + \varepsilon_{30} + \varepsilon_{36} + \varepsilon_{45} + \varepsilon_{60} + \varepsilon_{90} + \varepsilon_{180}
=
\frac{ 61 }{ 150 } ,
\\
\rho_{2, 3}(1, 2)
& =
\sum_{u_{1} = 0}^{2} \sum_{u_{2} = 0}^{0} \sum_{u_{3} = 2}^{3} \varepsilon_{\langle u_{1}, u_{2}, u_{3} \rangle}
=
\varepsilon_{25} + \varepsilon_{50} + \varepsilon_{100} + \varepsilon_{125} + \varepsilon_{250} + \varepsilon_{500}
=
\frac{ 23 }{ 150 } .
\end{align*}
Since $\lambda_{2, 3}(1, 2) > \rho_{2, 3}(1, 2)$, store $(k, l) = (2, 2)$ as in Line~11; reset $(i, j) = (2, 4)$ as in Line~12; and go back to Line~6.
As $j = 4 > 3 = m$, the while loop in Lines~6--12 of Algorithm~\ref{alg:main} is finished and we go to Line~13.
It can be verified that
\begin{align*}
\beta_{2, 3}(1, 2)
=
\sum_{u_{1} = 0}^{2} \sum_{u_{2} = 0}^{0} \sum_{u_{3} = 0}^{1} \varepsilon_{\langle u_{1}, u_{2}, u_{3} \rangle}
=
\varepsilon_{1} + \varepsilon_{2} + \varepsilon_{4} + \varepsilon_{5} + \varepsilon_{10} + \varepsilon_{20}
=
\frac{ 8 }{ 75 } .
\end{align*}
Since $\lambda_{2, 3}(1, 2) > \rho_{2, 3}(1, 2)$, we get in Line~13 that
\begin{align}
\mu_{\langle 0, 0, 1 \rangle}^{(\infty)}
=
\mu_{5}^{(\infty)}
=
\beta_{2, 3}(1, 2) + \rho_{2, 3}(1, 2) - \alpha
=
\frac{ 1 }{ 15 } .
\end{align}
Resetting $\alpha = (29/150) + (1/15) = 59/150$ and $t_{2} = 1$ as in Lines~14 and~15, respectively, i.e., $\bvec{t} = (t_{1}, t_{2}, t_{3}) = (0, 1, 1)$, we go back to Line~4.
As $0 \le \alpha = 59/150 < 1$, we continue the while loop in Lines~4--15 of Algorithm~\ref{alg:main}.

Consider the third step of the while loop in Lines~4--15 of Algorithm~\ref{alg:main} with the following parameters: $\alpha = 59/150$ and $\bvec{t} = (t_{1}, t_{2}, t_{3}) = (0, 1, 1)$.
Set $(i, j) = (1, 2)$ as in Line~5, and go to the while loop in Lines~6--12 of Algorithm~\ref{alg:main}.
It can be verified that
\begin{align*}
\lambda_{1, 2}(1, 2)
& =
\sum_{u_{1} = 1}^{2} \sum_{u_{2} = 0}^{1} \sum_{u_{3} = 0}^{3} \varepsilon_{\langle u_{1}, u_{2}, u_{3} \rangle}
\notag \\
& =
\varepsilon_{2} + \varepsilon_{4} + \varepsilon_{6} + \varepsilon_{10} + \varepsilon_{12} + \varepsilon_{20} + \varepsilon_{30} + \varepsilon_{50} + \varepsilon_{60} + \varepsilon_{100} + \varepsilon_{150} + \varepsilon_{250} + \varepsilon_{300} + \varepsilon_{500} + \varepsilon_{750} + \varepsilon_{1500}
=
\frac{ 11 }{ 25 } ,
\\
\rho_{1, 2}(1, 2)
& =
\sum_{u_{1} = 0}^{0} \sum_{u_{2} = 2}^{2} \sum_{u_{3} = 0}^{3} \varepsilon_{\langle u_{1}, u_{2}, u_{3} \rangle}
=
\varepsilon_{9} + \varepsilon_{45} + \varepsilon_{225} + \varepsilon_{1125}
=
\frac{ 17 }{ 150 } .
\end{align*}
Since $\lambda_{1, 2}(1, 2) > \rho_{1, 2}(1, 2)$, store $(k, l) = (1, 1)$ as in Line~11; reset $(i, j) = (1, 3)$ as in Line~12; and go back to Line~6.
It can be verified that
\begin{align*}
\lambda_{1, 3}(1, 2)
& =
\sum_{u_{1} = 1}^{2} \sum_{u_{2} = 0}^{2} \sum_{u_{3} = 0}^{1} \varepsilon_{\langle u_{1}, u_{2}, u_{3} \rangle}
=
\varepsilon_{2} + \varepsilon_{4} + \varepsilon_{6} + \varepsilon_{10} + \varepsilon_{12} + \varepsilon_{18} + \varepsilon_{20} + \varepsilon_{30} + \varepsilon_{36} + \varepsilon_{60} + \varepsilon_{90} + \varepsilon_{180}
=
\frac{ 17 }{ 50 } ,
\\
\rho_{1, 3}(1, 2)
& =
\sum_{u_{1} = 0}^{0} \sum_{u_{2} = 0}^{2} \sum_{u_{3} = 2}^{3} \varepsilon_{\langle u_{1}, u_{2}, u_{3} \rangle}
=
\varepsilon_{25} + \varepsilon_{75} + \varepsilon_{125} + \varepsilon_{225} + \varepsilon_{375} + \varepsilon_{1125}
=
\frac{ 4 }{ 25 } .
\end{align*}
Since $\lambda_{1, 3}(1, 2) > \rho_{1, 3}(1, 2)$, store $(k, l) = (1, 1)$ as in Line~11; reset $(i, j) = (1, 4)$ as in Line~12; and go back to Line~6.
As $j = 4 > 3 = m$, the while loop in Lines~6--12 of Algorithm~\ref{alg:main} is finished and we go to Line~13.
It can be verified that
\begin{align*}
\beta_{1, 3}(1, 2)
=
\sum_{u_{1} = 0}^{0} \sum_{u_{2} = 0}^{0} \sum_{u_{3} = 0}^{1} \varepsilon_{\langle u_{1}, u_{2}, u_{3} \rangle}
=
\varepsilon_{1} + \varepsilon_{3} + \varepsilon_{5} + \varepsilon_{9} + \varepsilon_{15} + \varepsilon_{45}
=
\frac{ 13 }{ 75 } .
\end{align*}
Since $\lambda_{1, 2}(1, 2) > \rho_{1, 2}(1, 2)$, we get in Line~13 that
\begin{align}
\mu_{\langle 0, 1, 1 \rangle}^{(\infty)}
=
\mu_{15}^{(\infty)}
=
\beta_{1, 3}(1, 2) + \rho_{1, 3}(1, 2) - \alpha
=
\frac{ 11 }{ 150 } .
\end{align}
Resetting $\alpha = (59/150) + (11/150) = 1/3$ and $t_{1} = 1$ as in Lines~14 and~15, respectively, i.e., $\bvec{t} = (t_{1}, t_{2}, t_{3}) = (1, 1, 1)$, we go back to Line~4.
As $0 \le \alpha = 1/3 < 1$, we continue the while loop in Lines~4--15 of Algorithm~\ref{alg:main}.

Consider the fourth step of the while loop in Lines~4--15 of Algorithm~\ref{alg:main} with the following parameters: $\alpha = 1/3$ and $\bvec{t} = (t_{1}, t_{2}, t_{3}) = (1, 1, 1)$.
Set $(i, j) = (1, 2)$ as in Line~5, and go to the while loop in Lines~6--12 of Algorithm~\ref{alg:main}.
It can be verified that
\begin{align*}
\lambda_{1, 2}(2, 2)
& =
\sum_{u_{1} = 2}^{2} \sum_{u_{2} = 0}^{1} \sum_{u_{3} = 0}^{3} \varepsilon_{\langle u_{1}, u_{2}, u_{3} \rangle}
=
\varepsilon_{4} + \varepsilon_{12} + \varepsilon_{20} + \varepsilon_{60} + \varepsilon_{100} + \varepsilon_{300} + \varepsilon_{500} + \varepsilon_{1500}
=
\frac{ 37 }{ 150 } ,
\\
\rho_{1, 2}(2, 2)
& =
\sum_{u_{1} = 0}^{1} \sum_{u_{2} = 2}^{2} \sum_{u_{3} = 0}^{3} \varepsilon_{\langle u_{1}, u_{2}, u_{3} \rangle}
=
\varepsilon_{9} + \varepsilon_{18} + \varepsilon_{45} + \varepsilon_{90} + \varepsilon_{225} + \varepsilon_{450} + \varepsilon_{1125} + \varepsilon_{2250}
=
\frac{ 19 }{ 75 } .
\end{align*}
Since $\lambda_{1, 2}(2, 2) < \rho_{1, 2}(2, 2)$, store $(k, l) = (2, 1)$ as in Line~8; reset $(i, j) = (2, 3)$ as in Line~9; and go back to Line~6.
It can be verified that
\begin{align*}
\lambda_{2, 3}(2, 2)
& =
\sum_{u_{1} = 0}^{2} \sum_{u_{2} = 2}^{2} \sum_{u_{3} = 0}^{1} \varepsilon_{\langle u_{1}, u_{2}, u_{3} \rangle}
=
\varepsilon_{9} + \varepsilon_{18} + \varepsilon_{36} + \varepsilon_{45} + \varepsilon_{90} + \varepsilon_{180} + \varepsilon_{500} + \varepsilon_{1500}
=
\frac{ 37 }{ 150 } ,
\\
\rho_{2, 3}(2, 2)
& =
\sum_{u_{1} = 0}^{2} \sum_{u_{2} = 0}^{1} \sum_{u_{3} = 2}^{3} \varepsilon_{\langle u_{1}, u_{2}, u_{3} \rangle}
\notag \\
& =
\varepsilon_{25} + \varepsilon_{50} + \varepsilon_{75} + \varepsilon_{100} + \varepsilon_{125} + \varepsilon_{150} + \varepsilon_{250} + \varepsilon_{300} + \varepsilon_{375} + \varepsilon_{500} + \varepsilon_{750} + \varepsilon_{1500}
=
\frac{ 49 }{ 150 } .
\end{align*}
Since $\lambda_{2, 3}(2, 2) < \rho_{2, 3}(2, 2)$, store $(k, l) = (3, 2)$ as in Line~8; reset $(i, j) = (3, 4)$ as in Line~9; and go back to Line~6.
As $j = 4 > 3 = m$, the while loop in Lines~6--12 of Algorithm~\ref{alg:main} is finished and we go to Line~13.
It can be verified that
\begin{align*}
\beta_{2, 3}(2, 2)
=
\sum_{u_{1} = 0}^{2} \sum_{u_{2} = 0}^{1} \sum_{u_{3} = 0}^{1} \varepsilon_{\langle u_{1}, u_{2}, u_{3} \rangle}
=
\varepsilon_{1} + \varepsilon_{2} + \varepsilon_{3} + \varepsilon_{4} + \varepsilon_{5} + \varepsilon_{6} + \varepsilon_{10} + \varepsilon_{12} + \varepsilon_{15} + \varepsilon_{20} + \varepsilon_{30} + \varepsilon_{60}
=
\frac{ 1 }{ 3 } .
\end{align*}
Since $\lambda_{2, 3}(2, 2) < \rho_{2, 3}(2, 2)$, we get in Line~13 that
\begin{align}
\mu_{\langle 1, 1, 1 \rangle}^{(\infty)}
=
\mu_{30}^{(\infty)}
=
\beta_{2, 3}(2, 2) + \lambda_{2, 3}(2, 2) - \alpha
=
\frac{ 9 }{ 50 } .
\end{align}
Resetting $\alpha = (1/3) + (9/50) = 77/150$ and $t_{3} = 2$ as in Lines~14 and~15, respectively, i.e., $\bvec{t} = (t_{1}, t_{2}, t_{3}) = (1, 1, 2)$, we go back to Line~4.
As $0 \le \alpha = 77/150 < 1$, we continue the while loop in Lines~4--15 of Algorithm~\ref{alg:main}.

Consider the fifth step of the while loop in Lines~4--15 of Algorithm~\ref{alg:main} with the following parameters: $\alpha = 77/150$ and $\bvec{t} = (t_{1}, t_{2}, t_{3}) = (1, 1, 2)$.
Set $(i, j) = (1, 2)$ as in Line~5, and go to the while loop in Lines~6--12 of Algorithm~\ref{alg:main}.
Since $\lambda_{1, 2}(2, 2) < \rho_{1, 2}(2, 2)$, store $(k, l) = (2, 1)$ as in Line~8; reset $(i, j) = (2, 3)$ as in Line~9; and go back to Line~6.
It can be verified that
\begin{align*}
\lambda_{2, 3}(2, 3)
& =
\sum_{u_{1} = 0}^{2} \sum_{u_{2} = 2}^{2} \sum_{u_{3} = 0}^{2} \varepsilon_{\langle u_{1}, u_{2}, u_{3} \rangle}
=
\varepsilon_{9} + \varepsilon_{18} + \varepsilon_{36} + \varepsilon_{45} + \varepsilon_{90} + \varepsilon_{180} + \varepsilon_{225} + \varepsilon_{450} + \varepsilon_{900}
=
\frac{ 4 }{ 15 } ,
\\
\rho_{2, 3}(2, 3)
& =
\sum_{u_{1} = 0}^{2} \sum_{u_{2} = 0}^{1} \sum_{u_{3} = 3}^{3} \varepsilon_{\langle u_{1}, u_{2}, u_{3} \rangle}
=
\varepsilon_{125} + \varepsilon_{250} + \varepsilon_{375} + \varepsilon_{500} + \varepsilon_{750} + \varepsilon_{1500}
=
\frac{ 1 }{ 6 } .
\end{align*}
Since $\lambda_{2, 3}(2, 2) > \rho_{2, 3}(2, 2)$, store $(k, l) = (2, 2)$ as in Line~11; reset $(i, j) = (2, 4)$ as in Line~12; and go back to Line~6.
As $j = 4 > 3 = m$, the while loop in Lines~6--12 of Algorithm~\ref{alg:main} is finished and we go to Line~13.
It can be verified that
\begin{align*}
\beta_{2, 3}(2, 3)
& =
\sum_{u_{1} = 0}^{2} \sum_{u_{2} = 0}^{1} \sum_{u_{3} = 0}^{2} \varepsilon_{\langle u_{1}, u_{2}, u_{3} \rangle}
\notag \\
& =
\varepsilon_{1} + \varepsilon_{2} + \varepsilon_{3} + \varepsilon_{4} + \varepsilon_{5} + \varepsilon_{6} + \varepsilon_{10} + \varepsilon_{12} + \varepsilon_{15} + \varepsilon_{20} + \varepsilon_{25} + \varepsilon_{30} + \varepsilon_{50} + \varepsilon_{60} + \varepsilon_{75} + \varepsilon_{100} + \varepsilon_{150} + \varepsilon_{300}
=
\frac{ 37 }{ 75 } .
\end{align*}
Since $\lambda_{2, 3}(2, 3) > \rho_{2, 3}(2, 3)$, we get in Line~13 that
\begin{align}
\mu_{\langle 1, 1, 2 \rangle}^{(\infty)}
=
\mu_{150}^{(\infty)}
=
\beta_{2, 3}(2, 3) + \rho_{2, 3}(2, 3) - \alpha
=
\frac{ 11 }{ 75 } .
\end{align}
Resetting $\alpha = (77/150) + (11/75) = 33/50$ and $t_{2} = 2$ as in Lines~14 and~15, respectively, i.e., $\bvec{t} = (t_{1}, t_{2}, t_{3}) = (1, 2, 2)$, we go back to Line~4.
As $0 \le \alpha = 33/50 < 1$, we continue the while loop in Lines~4--15 of Algorithm~\ref{alg:main}.

Consider the sixth step of the while loop in Lines~4--15 of Algorithm~\ref{alg:main} with the following parameters: $\alpha = 33/50$ and $\bvec{t} = (t_{1}, t_{2}, t_{3}) = (1, 2, 2)$.
Set $(i, j) = (1, 2)$ as in Line~5, and go to the while loop in Lines~6--12 of Algorithm~\ref{alg:main}.
It can be verified that
\begin{align*}
\lambda_{1, 2}(2, 3)
& =
\sum_{u_{1} = 2}^{2} \sum_{u_{2} = 0}^{2} \sum_{u_{3} = 0}^{3} \varepsilon_{\langle u_{1}, u_{2}, u_{3} \rangle}
=
\varepsilon_{4} + \varepsilon_{12} + \varepsilon_{20} + \varepsilon_{36} + \varepsilon_{60} + \varepsilon_{100} + \varepsilon_{180} + \varepsilon_{300} + \varepsilon_{500} + \varepsilon_{900} + \varepsilon_{1500} + \varepsilon_{4500}
=
\frac{ 1 }{ 3 } ,
\\
\rho_{1, 2}(2, 3)
& =
\sum_{u_{1} = 0}^{1} \sum_{u_{2} = 3}^{2} \sum_{u_{3} = 0}^{3} \varepsilon_{\langle u_{1}, u_{2}, u_{3} \rangle}
=
0 .
\end{align*}
Since $\lambda_{1, 2}(2, 3) > \rho_{1, 2}(2, 3)$, store $(k, l) = (1, 1)$ as in Line~11; reset $(i, j) = (1, 3)$ as in Line~12; and go back to Line~6.
It can be verified that
\begin{align*}
\lambda_{1, 3}(2, 3)
& =
\sum_{u_{1} = 2}^{2} \sum_{u_{2} = 0}^{2} \sum_{u_{3} = 0}^{2} \varepsilon_{\langle u_{1}, u_{2}, u_{3} \rangle}
=
\varepsilon_{4} + \varepsilon_{12} + \varepsilon_{20} + \varepsilon_{36} + \varepsilon_{60} + \varepsilon_{100} + \varepsilon_{180} + \varepsilon_{300} + \varepsilon_{900}
=
\frac{ 11 }{ 50 } ,
\\
\rho_{1, 3}(2, 3)
& =
\sum_{u_{1} = 0}^{1} \sum_{u_{2} = 0}^{2} \sum_{u_{3} = 3}^{3} \varepsilon_{\langle u_{1}, u_{2}, u_{3} \rangle}
=
\varepsilon_{125} + \varepsilon_{250} + \varepsilon_{375} + \varepsilon_{750} + \varepsilon_{1125} + \varepsilon_{2250}
=
\frac{ 19 }{ 150 } .
\end{align*}
Since $\lambda_{1, 3}(2, 3) > \rho_{1, 3}(2, 3)$, store $(k, l) = (1, 1)$ as in Line~11; reset $(i, j) = (1, 4)$ as in Line~12; and go back to Line~6.
As $j = 4 > 3 = m$, the while loop in Lines~6--12 of Algorithm~\ref{alg:main} is finished and we go to Line~13.
It can be verified that
\begin{align*}
\beta_{1, 3}(2, 3)
& =
\sum_{u_{1} = 0}^{1} \sum_{u_{2} = 0}^{2} \sum_{u_{3} = 0}^{2} \varepsilon_{\langle u_{1}, u_{2}, u_{3} \rangle}
\notag \\
& =
\varepsilon_{1} + \varepsilon_{2} + \varepsilon_{3} + \varepsilon_{5} + \varepsilon_{6} + \varepsilon_{9} + \varepsilon_{10} + \varepsilon_{15} + \varepsilon_{18} + \varepsilon_{25} + \varepsilon_{30} + \varepsilon_{45} + \varepsilon_{50} + \varepsilon_{75} + \varepsilon_{90} + \varepsilon_{150} + \varepsilon_{225} + \varepsilon_{450}
=
\frac{ 27 }{ 50 } .
\end{align*}
Since $\lambda_{1, 3}(2, 3) > \rho_{1, 3}(2, 3)$, we get in Line~13 that
\begin{align}
\mu_{\langle 1, 2, 2 \rangle}^{(\infty)}
=
\mu_{450}^{(\infty)}
=
\beta_{1, 3}(2, 3) + \rho_{1, 3}(2, 3) - \alpha
=
\frac{ 1 }{ 150 } .
\end{align}
Resetting $\alpha = (33/50) + (1/150) = 2/3$ and $t_{1} = 2$ as in Lines~14 and~15, respectively, i.e., $\bvec{t} = (t_{1}, t_{2}, t_{3}) = (2, 2, 2)$, we go back to Line~4.
As $0 \le \alpha = 2/3 < 1$, we continue the while loop in Lines~4--15 of Algorithm~\ref{alg:main}.

Consider the seventh step of the while loop in Lines~4--15 of Algorithm~\ref{alg:main} with the following parameters: $\alpha = 2/3$ and $\bvec{t} = (t_{1}, t_{2}, t_{3}) = (2, 2, 2)$.
Set $(i, j) = (1, 2)$ as in Line~5, and go to the while loop in Lines~6--12 of Algorithm~\ref{alg:main}.
It can be verified that
\begin{align*}
\lambda_{1, 2}(3, 3)
& =
\sum_{u_{1} = 3}^{2} \sum_{u_{2} = 0}^{2} \sum_{u_{3} = 0}^{3} \varepsilon_{\langle u_{1}, u_{2}, u_{3} \rangle}
=
0 ,
\\
\rho_{1, 2}(3, 3)
& =
\sum_{u_{1} = 0}^{2} \sum_{u_{2} = 3}^{2} \sum_{u_{3} = 0}^{3} \varepsilon_{\langle u_{1}, u_{2}, u_{3} \rangle}
=
0 .
\end{align*}
Since $\lambda_{1, 2}(3, 3) = \rho_{1, 2}(3, 3)$, store $(k, l) = (2, 1)$ as in Line~8; reset $(i, j) = (2, 3)$ as in Line~9; and go back to Line~6.
It can be verified that
\begin{align*}
\lambda_{2, 3}(3, 3)
& =
\sum_{u_{1} = 0}^{2} \sum_{u_{2} = 3}^{2} \sum_{u_{3} = 0}^{2} \varepsilon_{\langle u_{1}, u_{2}, u_{3} \rangle}
=
0 ,
\\
\rho_{2, 3}(3, 3)
& =
\sum_{u_{1} = 0}^{2} \sum_{u_{2} = 0}^{2} \sum_{u_{3} = 3}^{3} \varepsilon_{\langle u_{1}, u_{2}, u_{3} \rangle}
=
\varepsilon_{125} + \varepsilon_{250} + \varepsilon_{375} + \varepsilon_{500} + \varepsilon_{750} + \varepsilon_{1125} + \varepsilon_{1500} + \varepsilon_{2250} + \varepsilon_{4500}
=
\frac{ 6 }{ 25 } .
\end{align*}
Since $\lambda_{2, 3}(3, 3) < \rho_{2, 3}(2, 3)$, store $(k, l) = (3, 2)$ as in Line~8; reset $(i, j) = (3, 4)$ as in Line~9; and go back to Line~6.
As $j = 4 > 3 = m$, the while loop in Lines~6--12 of Algorithm~\ref{alg:main} is finished and we go to Line~13.
It can be verified that
\begin{align*}
\beta_{2, 3}(3, 3)
& =
\sum_{u_{1} = 0}^{2} \sum_{u_{2} = 0}^{2} \sum_{u_{3} = 0}^{2} \varepsilon_{\langle u_{1}, u_{2}, u_{3} \rangle}
=
\varepsilon_{1} + \varepsilon_{2} + \varepsilon_{3} + \varepsilon_{4} + \varepsilon_{5} + \varepsilon_{6} + \varepsilon_{9} + \varepsilon_{10} + \varepsilon_{12} + \varepsilon_{15} + \varepsilon_{18} + \varepsilon_{25} + \varepsilon_{30}
\notag \\
& \qquad \qquad \qquad
{} + \varepsilon_{36} + \varepsilon_{45} + \varepsilon_{50} + \varepsilon_{60} + \varepsilon_{75} + \varepsilon_{90} + \varepsilon_{100} + \varepsilon_{150} + \varepsilon_{180} + \varepsilon_{225} + \varepsilon_{300} + \varepsilon_{450} + \varepsilon_{900}
=
\frac{ 19 }{ 25 } .
\end{align*}
Since $\lambda_{2, 3}(3, 3) < \rho_{2, 3}(3, 3)$, we get in Line~13 that
\begin{align}
\mu_{\langle 2, 2, 2 \rangle}^{(\infty)}
=
\mu_{900}^{(\infty)}
=
\beta_{2, 3}(3, 3) + \rho_{2, 3}(3, 3) - \alpha
=
\frac{ 7 }{ 75 } .
\end{align}
Resetting $\alpha = (2/3) + (7/75) = 19/25$ and $t_{3} = 4$ as in Lines~14 and~15, respectively, i.e., $\bvec{t} = (t_{1}, t_{2}, t_{3}) = (2, 2, 3)$, we go back to Line~4.
As $0 \le \alpha = 19/25 < 1$, we continue the while loop in Lines~4--15 of Algorithm~\ref{alg:main}.

Consider the eighth step of the while loop in Lines~4--15 of Algorithm~\ref{alg:main} with the following parameters: $\alpha = 19/25$ and $\bvec{t} = (t_{1}, t_{2}, t_{3}) = (2, 2, 3)$.
Set $(i, j) = (1, 2)$ as in Line~5, and go to the while loop in Lines~6--12 of Algorithm~\ref{alg:main}.
Note that $\bvec{t} = \bvec{r} = (r_{1}, r_{2}, r_{3}) = (2, 2, 3)$.
Since $\lambda_{1, 2}(3, 3) = \rho_{1, 2}(3, 3)$, store $(k, l) = (2, 1)$ as in Line~8; reset $(i, j) = (2, 3)$ as in Line~9; and go back to Line~6.
It can be verified that
\begin{align*}
\lambda_{2, 3}(3, 4)
& =
\sum_{u_{1} = 0}^{2} \sum_{u_{2} = 3}^{2} \sum_{u_{3} = 0}^{3} \varepsilon_{\langle u_{1}, u_{2}, u_{3} \rangle}
=
0 ,
\\
\rho_{2, 3}(3, 4)
& =
\sum_{u_{1} = 0}^{2} \sum_{u_{2} = 0}^{2} \sum_{u_{3} = 4}^{3} \varepsilon_{\langle u_{1}, u_{2}, u_{3} \rangle}
=
0 .
\end{align*}
Since $\lambda_{2, 3}(3, 4) = \rho_{2, 3}(3, 4)$, store $(k, l) = (3, 2)$ as in Line~8; reset $(i, j) = (3, 4)$ as in Line~9; and go back to Line~6.
As $j = 4 > 3 = m$, the while loop in Lines~6--12 of Algorithm~\ref{alg:main} is finished and we go to Line~13.
It can be verified that
\begin{align*}
\beta_{2, 3}(3, 4)
& =
\sum_{u_{1} = 0}^{2} \sum_{u_{2} = 0}^{2} \sum_{u_{3} = 0}^{3} \varepsilon_{\langle u_{1}, u_{2}, u_{3} \rangle}
=
\sum_{d|q} \varepsilon_{d}
=
1 .
\end{align*}
Since $\lambda_{2, 3}(3, 4) = \rho_{2, 3}(3, 4)$, we get in Line~13 that
\begin{align}
\mu_{\langle 2, 2, 3 \rangle}^{(\infty)}
=
\mu_{4500}^{(\infty)}
=
\beta_{2, 3}(3, 4) + \rho_{2, 3}(3, 4) - \alpha
=
\frac{ 6 }{ 25 } .
\end{align}
Resetting $\alpha = (19/25) + (6/25) = 1$ and $t_{3} = 4$ as in Lines~14 and~15, respectively, i.e., $\bvec{t} = (t_{1}, t_{2}, t_{3}) = (2, 2, 4)$, we go back to Line~4.
As $\alpha = 1$, we finish the while loop in Lines~4--15 of Algorithm~\ref{alg:main}, and the asymptotic distribution $( \mu_{d}^{(\infty)} )_{d|q}$ is just obtained.


\begin{thebibliography}{99}

\bibitem{alsan3}
M.~Alsan and E.~Telatar,
A simple proof of polarization and polarization for non-stationary memoryless channels,''
\emph{IEEE\ Trans.\ Inf.\ Theory},
vol.~62, no.~9, pp.~4873--4878, Sept.~2016.

\bibitem{arikan}
E.~Ar{\i}kan,
``Channel polarization: A method for constructing capacity-achieving codes for symmetric binary-input memoryless channels,''
\emph{IEEE\ Trans.\ Inf.\ Theory},
vol.~55, no.~7, pp.~3051--3073, July 2009.

\bibitem{mori}
R.~Mori and T.~Tanaka,
``Source and channel polarization over finite fields and Reed--Solomon matrices,''
\emph{IEEE\ Trans.\ Inf.\ Theory},
vol.~60, no.~5, pp.~2720--2736, May 2014.

\bibitem{ergodic1}
R.~Nasser,
``Ergodic theory meets polarization I: A foundation of polarization theory,''
\emph{IEEE\ Trans.\ Inf. Theory},
vol.~62, no.~12, pp.~6931--6952, Dec.~2016.

\bibitem{ergodic2}
---------,
``Ergodic theory meets polarization II: A foundation of polarization theory for MACs,''
\emph{IEEE\ Trans.\ Inf. Theory},
vol.~63, no.~2, pp.~1063--1083, Feb.~2017.

\bibitem{quasigroup}
R.~Nasser and E.~Telatar,
``Polarization theorems for arbitrary DMCs and arbitrary MACs,''
\emph{IEEE\ Trans.\ Inf.\ Theory},
vol.~63, no.~6, pp.~2917--2936, June~2016.

\bibitem{park}
W.~Park and A.~Barg,
``Polar codes for $q$-ary channels, $q = 2^{r}$,''
\emph{IEEE\ Trans.\ Inf.\ Theory},
vol.~59, no.~2, pp.~955--969, Feb.~2013.

\bibitem{sahebi}
A.~G.~Sahebi and S.~S.~Pradhan,
``Multilevel channel polarization for arbitrary discrete memoryless channels,''
\emph{IEEE\ Trans.\ Inf.\ Theory},
vol.~59, no.~12, pp.~7839--7857, Dec.~2013.

\bibitem{itw2016}
Y.~Sakai and K.~Iwata,
``A generalized erasure channel in the sense of polarization for binary erasure channels,''
in \emph{Proc.\ IEEE\ Inf.\ Theory, Workshop} (ITW),
Cambridge, UK, Sept.~2016, 5~pages.
[Online]. Available at \url{https://arxiv.org/abs/1604.04413}.

\bibitem{sasoglu}
E.~{\c{S}}a{\c{s}}o{\u{g}}lu,
``Polar codes for discrete alphabets,''
in \emph{Proc.\ IEEE\ Int.\ Symp.\ Inf.\ Theory} (ISIT),
Cambridge, MA, USA, July~2012, pp.~2137--2141.

\bibitem{sasoglu_etal}
E.~{\c{S}}a{\c{s}}o{\u{g}}lu, E.~Telatar, and E.~Ar{\i}kan,
``Polarization for arbitrary discrete memoryless channels,''
in \emph{Proc.\ IEEE\ Inf.\ Theory\ Workshop} (ITW),
Sicily, Italy, Oct.~2009, pp.~144--148.

\end{thebibliography}
\end{document}